\documentclass[fleqn,usenatbib]{mnras}

\usepackage{newtxtext,newtxmath}

\usepackage[T1]{fontenc}
\usepackage{ae,aecompl}

\usepackage{graphicx}	
\usepackage{amsmath}	
\usepackage{amssymb}	
\usepackage{blindtext}  
\usepackage{natbib}
\usepackage{rotating,times,graphicx,latexsym}
\usepackage{color}
\usepackage{longtable}
\usepackage{lscape}
\usepackage{lipsum} 
\usepackage{array}
\usepackage{flafter}

\usepackage{soul} 
\usepackage{xargs} 
\usepackage[pdftex,dvipsnames]{xcolor}  
\usepackage{amsmath}
\usepackage[colorinlistoftodos,prependcaption]{todonotes}
\newcommandx{\tobedone}[2][1=]{\todo[linecolor=red,backgroundcolor=red!25,bordercolor=red,inline,#1]{#2}}
\newcommandx{\changed}[2][1=]{\todo[linecolor=blue,backgroundcolor=blue!25,bordercolor=blue,inline,#1]{#2}\noindent}
\newcommandx{\thiswillnotshow}[2][1=]{\todo[disable,#1]{#2}}
\newcommandx{\alex}[2][1=]{\todo[linecolor=Orchid,backgroundcolor=Orchid!25,bordercolor=Orchid,inline,#1]{#2}\noindent}
\newcommandx{\dirk}[2][1=]{\todo[linecolor=BurntOrange,backgroundcolor=BurntOrange!25,bordercolor=BurntOrange,inline,#1]{#2}\noindent}
\newcommandx{\justyn}[2][1=]{\todo[linecolor=SpringGreen,backgroundcolor=SpringGreen!40,bordercolor=SpringGreen,inline,#1]{#2}\noindent}
\newcommandx{\jochen}[2][1=]{\todo[linecolor=BurntOrange,backgroundcolor=SpringGreen!40,bordercolor=SpringGreen,inline,#1]{#2}\noindent}
\newcommandx{\jack}[2][1=]{\todo[linecolor=BurntOrange,backgroundcolor=BurntOrange!40,bordercolor=BurntOrange,inline,#1]{#2}\noindent}
\newcommandx{\sally}[2][1=]{\todo[linecolor=BurntOrange,backgroundcolor=BurntOrange!40,bordercolor=BurntOrange,inline,#1]{#2}\noindent}

\newcommand{\hc}{{HOYS}}
\newcommand{\ha}{{H$\alpha$}}
\newcommand{\vc}{{V1490\,Cyg}}
\newcommand{\dg}{{$^\circ$}}
\graphicspath{ {images/} }

\title[\vc]{A survey for variable young stars with small telescopes: II - Mapping a protoplanetary disk with stable structures at 0.15\,AU}

\author[Jack J.Evitts et al.]{Jack J. Evitts$^{1}$,
Dirk Froebrich$^{1}$\thanks{E-mail: df@star.kent.ac.uk},
Aleks Scholz$^{2}$, 
Jochen Eisl\"offel$^{3}$, 
\newauthor
Justyn Campbell-White$^{1,4}$, 
Will Furnell$^{1,5}$, 
Thomas Urtly$^{6}$\thanks{\hc\ Observer}, 
Roger Pickard$^{6}\dagger$, 
\newauthor
Klaas Wiersema$^{7,8}\dagger$, 
Pavol A. Dubovsk\'{y}$^{9}\dagger$, 
Igor Kudzej$^{9}\dagger$, 
Ramon Naves$^{10}\dagger$, 
\newauthor
Mario Morales Aimar$^{10,11}\dagger$, 
Rafael Castillo Garc\'{i}a$^{10,11,12}\dagger$, 
Tonny Vanmunster$^{13,14}\dagger$, 
\newauthor
Erik Schwendeman$^{11}\dagger$, 
Francisco C. Sold\'{a}n Alfaro$^{10,11}\dagger$, 
Stephen Johnstone$^{6, 11}\dagger$, 
\newauthor
Rafael Gonzalez Farf\'{a}n$^{10}\dagger$, 
Thomas Killestein$^{6,7}\dagger$, 
Jes\'{u}s Delgado Casal$^{10}\dagger$, 
\newauthor
Faustino Garc\'{i}a de la Cuesta$^{10,15}\dagger$, 
Dean Roberts$^{16}\dagger$, 
Ulrich Kolb$^{16}\dagger$, 
Lu\'{i}s, Montoro$^{10}\dagger$, 
\newauthor
Domenico Licchelli$^{17}\dagger$, 
Alex Escartin Perez$^{10}\dagger$, 
Carlos Perell\'{o} Perez$^{10,18}\dagger$, 
\newauthor
Marc Deldem,$^{11}\dagger$ 
Stephen R.L. Futcher$^{6, 19}\dagger$, 
Tim Nelson,$^{19}\dagger$ 
Shawn Dvorak,$^{11}\dagger$ 
\newauthor
Dawid Mo\'{z}dzierski$^{20}\dagger$, 
Nick Quinn$^{6}\dagger$, 
Krzysztof Kotysz$^{20}\dagger$, 
Katarzyna Kowalska$^{20}\dagger$, 
\newauthor
Przemys{\l}aw Miko{\l}ajczyk$^{20}\dagger$, 
George Fleming$^{6}\dagger$, 
Mark\,Phillips$^{21}\dagger$, 
Tony Vale$^{6,22,23}\dagger$, 
\newauthor
Franky Dubois$^{24,25}\dagger$, 
Ludwig Logie$^{24,25}\dagger$, 
Steve Rau$^{24,25}\dagger$, 
Siegfried Vanaverbeke$^{24,25,26}\dagger$, 
\newauthor
Barry Merrikin$^{19}\dagger$, 
Esteban Fern\'{a}ndez Ma\~{n}anes$^{10}\dagger$, 
Emery Erdelyi$^{11,27}\dagger$, 
\newauthor
Juan-Luis Gonzalez Carballo$^{10}\dagger$, 
Fernando Limon Martinez$^{10}\dagger$, 
Timothy P. Long,$^{28}\dagger$ 
\newauthor
Adolfo San Segundo Delgado$^{10}\dagger$, 
Jos\'{e}p Luis Salto Gonz\'{a}lez$^{10,29}\dagger$, 
\newauthor
Luis Tremosa Espasa$^{10}\dagger$, 
Georg Piehler$^{30}\dagger$, 
James Crumpton$^{1}$\thanks{Observer Beacon Observatory},
Lord Dover$^{1}\ddagger$,
\newauthor
Samuel J. Billington$^{1}\ddagger$,
Emma D'Arcy$^{1}\ddagger$,
Sally V. Makin$^{1}\ddagger$,
Bringfried Stecklum$^{3}$,
\\
$^{1}$Centre for Astrophysics and Planetary Science, School of Physical Sciences, University of Kent, Canterbury CT2 7NH, UK\\
$^{2}$SUPA, School of Physics and Astronomy, University of St Andrews, North Haugh, St Andrews KY16 9SS, UK\\
$^{3}$Th\"{u}ringer Landessternwarte, Sternwarte 5, 07778 Tautenburg, Germany\\
$^{4}$SUPA, School of Science and Engineering, University of Dundee, Nethergate, Dundee DD1 4HN, UK\\
$^{5}$School of Computing, University of Kent, Canterbury CT2 7NF, UK\\
$^{6}$The British Astronomical Association, Variable Star Section, Burlington House Piccadilly, London W1J 0DU, UK\\
$^{7}$Department of Physics, University of Warwick, Coventry CV4 7AL, UK\\
$^{8}$University of Leicester, University Road, Leicester LE1 7RH, UK\\
$^{9}$Vihorlat Observatory, Mierov\'{a} 4, 06601 Humenn\'{e}, Slovakia\\
$^{10}$Observadores de Supernovas$^{\thanks{\tt \href{https://sites.google.com/view/sn2017eaw/}{Observadores de Supernovas}}}$, Spain\\
$^{11}$AAVSO, 49 Bay State Road, Cambridge, MA 02138, USA\\
$^{12}$Asociacion Astronomica Cruz del Norte, Calle Caceres 18, 28100 Alcobendas, Madrid, Spain\\
$^{13}$Center for Backyard Astrophysics Extremadura, 06340 Fregenal de la Sierra, Spain\\
$^{14}$Vereniging voor Sterrenkunde VVS, 3401 Landen, Belgium\\
$^{15}$Sociedad Astron\'{o}mica Asturiana "Omega", Spain\\
$^{16}$School of Physical Sciences, The Open University, Walton Hall, Milton Keynes MK7 6AA, UK\\
$^{17}$Center for Backyard Astrophysics, Piazzetta del Ges\`{u} 3, 73034, Gagliano del Capo, Italy\\
$^{18}$Agrupacio Astronomica de Sabadell, C/ Prat de la Riba s/n, 08206 Sabadell (Barcelona), Spain\\
$^{19}$Hampshire Astronomical Group, Clanfield, UK\\
$^{20}$Instytut Astronomiczny, Uniwersytet Wroc{\l}awski, Kopernika 11, 51-622 Wroc{\l}aw, Poland\\
$^{21}$Astronomical Society of Edinburgh, Edinburgh, UK\\
$^{22}$Wiltshire Astronomical Society, 2 Oathills, Corsham, SN13 9NL, UK\\
$^{23}$The William Herschel Society, The Herschel Museum of Astronomy, 19 New King Street, Bath BA1 2BL, UK\\
$^{24}$Public Observatory AstroLAB IRIS, Provinciaal Domein De Palingbeek, Verbrandemolenstraat 5, B-8902 Zillebeke, Ieper, Belgium\\
$^{25}$Vereniging voor Sterrenkunde, werkgroep veranderlijke sterren, Oostmeers 122 C, 8000 Brugge, Belgium\\
$^{26}$Center for Mathematical Plasma Astrophysics, University of Leuven, Belgium\\
$^{27}$San Diego Astronomy Association, P.O. Box 23215, San Diego, CA 92193-321, US\\
$^{28}$Tigra Astronomy, 16 Laxton Way, Canterbury, Kent, CT1 1FT, UK\\
$^{29}$Sociedad Malague\~{n}a de Astronom\'{i}a (SMA), M\'{a}laga, Spain\\
$^{30}$Selztal Observatory, D-55278 Friesenheim, Bechtolsheimer Weg 26, Germany\\
}

\date{Accepted XXX. Received YYY; in original form ZZZ}

\pubyear{2019}

\begin{document}
\label{firstpage}
\pagerange{\pageref{firstpage}--\pageref{lastpage}}
\maketitle

\begin{abstract}

The \hc\ citizen science project conducts long term, multifilter, high cadence monitoring of large YSO samples with a wide variety of professional and amateur telescopes. We present the analysis of the light curve of \vc\ in the Pelican Nebula. We show that colour terms in the diverse photometric data can be calibrated out to achieve a median photometric accuracy of 0.02\,mag in broadband filters, allowing detailed investigations into a variety of variability amplitudes over timescales from hours to several years. Using Gaia\,DR2 we estimate the distance to the Pelican Nebula to be 870\,$^{+70}_{-55}$\,pc. \vc\ is a quasi-periodic dipper with a period of 31.447\,$\pm$\,0.011\,d. The obscuring dust has homogeneous properties, and grains larger than those typical in the ISM. Larger variability on short timescales is observed in U and R$_c-$\ha, with U-amplitudes reaching 3\,mag on timescales of hours, indicating the source is accreting. The \ha\ equivalent width and NIR/MIR colours place \vc\ between CTTS/WTTS and transition disk objects. The material responsible for the dipping is located in a warped inner disk, about 0.15\,AU from the star. This mass reservoir can be filled and emptied on time scales shorter than the period at a rate of up to 10$^{-10}$\,M$_\odot$/yr, consistent with low levels of accretion in other T\,Tauri stars. Most likely the warp at this separation from the star is induced by a protoplanet in the inner accretion disk. However, we cannot fully rule out the possibility of an AA\,Tau-like warp, or occultations by the Hill sphere around a forming planet.

\end{abstract}





\section{Introduction}

Young stellar objects (YSOs) were initially discovered by their irregular and large optical variability \citep{1945ApJ...102..168J}. Their fluxes can be affected by a wide variety of physical processes such as changeable excess emission from accretion shocks, variable emission from the inner disk, and variable extinction along the line of sight \citep{2001AJ....121.3160C}. Furthermore, variability in YSOs occurs on a wide variety of time scales - from short term (minutes) accretion rate changes (e.g. \citet{2008A&A...491L..17S}, \citet{2013A&A...557A..69M}) to long term (years to tens of years) outburst or disk occultation events (e.g. \citet{2019MNRAS.486.4590C}, \citet{2016MNRAS.463.4459B}). Thus, observing variable young stars over a wide range of time scales and wavelengths allows us to explore the physical processes, structure and evolution of their environment, and provides key insights into the formation of stars.

Numerous photometric variability surveys have been conducted in the past aiming to address the study of YSO variability in optical and near infrared filters. Often they have either focused on high cadence over relatively short periods (e.g. with COROT -- Convection, Rotation and planetary Transits -- \citet{2009A&A...506..411A}, Kepler \citet{2014AJ....147...82C, 2016ApJ...816...69A} and TESS -- Transiting Exoplanet Survey Satellite, \citet{2015JATIS...1a4003R}) or longer term but lower cadence (e.g. with UKIRT Galactic Plane Survey -- UGPS -- and VISTA Variables in the Via Lactea -- VVV -- \citet{2014MNRAS.439.1829C, 2017MNRAS.465.3011C}). We have recently initiated the Hunting Outbursting Young Stars (\hc) project which aims to perform high cadence, long term, and simultaneous multifilter optical monitoring of YSOs. The project uses a combination of professional, university and amateur observatories \citep{2018MNRAS.478.5091F} in order to study accretion and extinction related variability over the short and long term in a number of nearby young clusters and star forming regions. 

Characterising the structure and properties of the inner accretion disks of YSOs is vital for our understanding of the accretion processes and the formation of terrestrial (inner) planets in those systems. However, investigating the innermost disk structure of YSOs on scales below 1\,AU is currently only possible through indirect methods such as photometric monitoring of disk occultation events. Direct observations of disks with ALMA \citep{2018ApJ...869L..41A} or SPHERE \citep{2018ApJ...863...44A} are limited to about 25\,--35\,mas, which corresponds to 2.5\,--\,3.5\,AU for the nearest (d\,$\approx$\,100\,pc) young stars. Of particular interest are periodic or quasi-periodic occultation events (e.g. in AA\,Tau \citep{1999A&A...349..619B} or UX\,Ori \citep{1999AJ....118.1043H} type objects) as they allow us to identify the exact physical location of the occulting structures in the disks based on the period, and thus to determine the spatial scales of the material directly.

In this paper we aim to show how long term optical photometric data from a variety of telescopes can be calibrated with sufficient accuracy to be useful for this purpose. With our light curves, we investigate the properties of the material causing quasi-periodic occultations in the young star \vc. This paper is organised as follows. In Sect.\,\ref{data} we describe the \hc\ data obtained for the project and detail the internal calibration procedure and accuracy for our inhomogeneous data set. We then describe the results obtained for \vc\ in Sect.\,\ref{results} and discuss the implications for the nature of the source in Sect.\,\ref{discussion}.


\section{Observational Data and Photometric Calibration}\label{data}

\subsection{The \hc\ Project}\label{hc_project}

All data presented in this paper has been obtained as part of the \hc\ project \citep{2018MNRAS.478.5091F}. The project utilises a network of amateur telescopes, several university observatories as well as other professional telescopes, currently distributed over 10 different countries across Europe and the US. At the time of writing the project had 58 participants submitting data, in several cases from multiple amateur observers or multiple telescopes/observing sites. In total, approximately 12500 images have been gathered. In those we have obtained $\approx$95 million reliable photometric measurements for stars in all of the 22 \hc\ target regions. 

To ease and streamline the data submission and processing for all participants we have developed an online portal. The website has been written using the Django\footnote{\tt \href{https://www.djangoproject.com/}{Django Project}} web framework. Django ORM has been used for managing the MariaDB database\footnote{\tt \href{https://mariadb.org/}{MariaDB}} into which the processed data is automatically added. The website also allows users and the public to plot and download light curves\footnote{\tt \href{http://astro.kent.ac.uk/HOYS-CAPS/lightcurve/}{\hc\ Light curve Plotter}} for all objects in the database. 

\subsection{\vc\ Imaging Observations}

In this paper we analyse the data for the star \vc, which is situated in the Pelican Nebula, or IC\,5070, corresponding to the \hc\ target number 118. At the time of writing we have gathered a total of 85, 419, 1134, 932, 249, and 755 images in the U, B, V, R$_c$, \ha\, and I$_c$ filters, respectively for this target field. The target itself has data with sufficient quality (magnitude uncertainty smaller than 0.2\,mag) in 3321 images from 44 different users and 66 different imaging devices - see Sect.\,\ref{Data:Colour-term_correction} for details. A full description of the observatories, the equipment used, the typical observing conditions and patterns, as well as data reduction procedures is given in Appendix\,A in the online supplementary
material. All \hc\ observations included in the paper for \vc\ have been taken over the last 4\,yr.

\subsection{Photometric Data Calibration}

The basic data calibration for all the \hc\ data has been detailed in \citet{2018MNRAS.478.5091F}. The images are submitted to our database server\footnote{\tt \href{http://astro.kent.ac.uk/HOYS-CAPS/}{\hc\ Database Server}} by the participants. They then indicate for each image which target region and imaging device (telescope/detector combination) has been used. We then extract the date/time, filter, and exposure time information from the FITS header and use the Astrometry.net\footnote{\tt \href{http://nova.astrometry.net/}{Astrometry.net}} software  \citep{2008ASPC..394...27H} to accurately determine the image coordinate system. 

\subsubsection{Basic Photometric Calibration}
\label{Data:Relative_photometric_calibration}
    
The initial photometric calibration process is carried out on the data before it is submitted to the \hc\ database. The Source Extractor\footnote{\tt \href{https://www.astromatic.net/software/sextractor}{The Source Extractor}} software \citep{1996A&AS..117..393B} is used to perform aperture photometry for all images. For each region and filter a deep image obtained at photometric conditions has been chosen as a reference image. The U-band reference frames are from the Th\"{u}ringer Landessternwarte (see Appendix\,A2.7 in the online supplementary material), while all the other reference images (B, V, R$_c$, I$_c$) are from the Beacon Observatory (see Appendix\,A2.5 in the online supplementary material). We have determined the calibration offsets into apparent magnitudes for those reference images by utilising the Cambridge Photometric Calibration Server\footnote{\tt \href{http://gsaweb.ast.cam.ac.uk/followup}{Cambridge Photometric Calibration Server}}, which has been set up for Gaia follow up photometry. 

The magnitude dependent calibration offsets $f(m_i)$ for all images into the reference frames have been obtained by fitting a photo-function and 4$^{\rm th}$ order polynomial{\bf , $\mathcal{P}_4(m_i)$} \citep{2005MNRAS.362..542B, 1969A&A.....3..455M} to matching stars with accurate photometry.

\begin{equation}
\label{eq:photo_cal}
f(m_i) = A \cdot log(10^{B \cdot (m_i - C)} + 1) + \mathcal{P}_4(m_i)
\end{equation}
        
\noindent See Sect.\,2.4 in \citet{2018MNRAS.478.5091F} for more details. Note that all \ha\ images are calibrated against the R$_c$-band reference images. Typically the accuracy of this basic relative calibration ranges from a few percent for the brighter stars to 0.20\,mag for the faintest detected stars, depending on the observatory, filter, exposure time, and observing conditions.

\subsubsection{Photometry Colour Correction}
\label{Data:Colour-term_correction}

In \citet{2018MNRAS.478.5091F} we limited the analysis to data taken either with the Beacon Observatory or data taken in the same filters. Now, with a much larger fraction of amateur data using a variety of slightly different filters, in particular from digital single-lens reflex (DSLR) cameras, the calibration of the photometry needs to consider colour terms. We have therefore devised a way to internally calibrate the photometry in the database. The general steps of the correction process are outlined below. 
        
The correction procedure utilises stars in each target region that do not change their brightness over time. Hence, these stars do have a known magnitude and colour. By comparing the photometry (after the basic calibration described above) of these stars in an image to their known brightness, any difference can be attributed to colour terms caused by either the filter used, the sensitivity curve of the specific detector, or the observing conditions (e.g. thin/thick cirrus) which can then be corrected for. This will furthermore correct any systematic errors that have potentially been introduced during the basic calibration step. We thus need a reliable catalogue of non-variable stars for each \hc\ target region.

\subsubsection*{Identifying Non-Variable Stars}
\label{Data:Calibration_catalogue}

\begin{figure}
\centering
\includegraphics[width=\columnwidth]{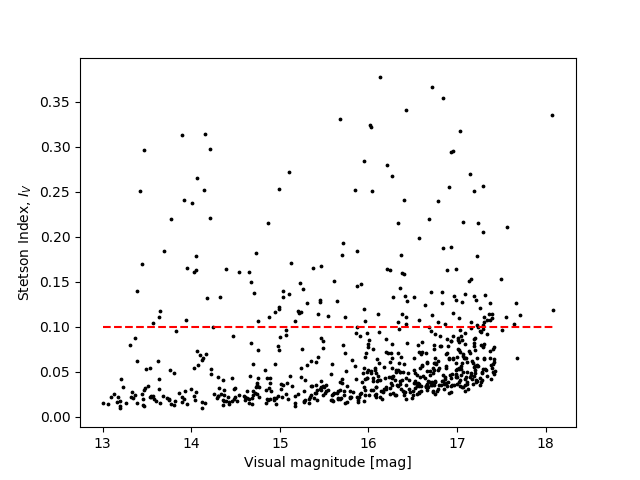}
\caption{Stetson variability index $\mathcal{I}$ against visual magnitude for all stars within the \hc\ target region IC\,5070. The red dashed line shows the cut made, below which stars were deemed to be non-variable.}
\label{fig:Stetsons}
\end{figure}

For each target region and in the V, R$_c$, I$_c$ filters we identify the image with the largest number of accurately measured (Source Extractor flag less than 5 -- see \citet{1996A&AS..117..393B} for details) and calibrated magnitudes. Stars in those images that are detected in all three filters (matched within a 3\arcsec\ radius -- the typical seeing in our images) are selected to generate a master list of stars for the region in our database. For the selected stars the accurately measured photometry in all filters (U, B, V, R$_c$, \ha, I$_c$) is extracted. Stars with fewer than 100 data points in V, R$_c$, and I$_c$ are removed. We then determine the Stetson index $\mathcal{I}$ \citep{1993AJ....105.1813W} for the V, R$_c$, and I$_c$ data. Figure\,\ref{fig:Stetsons} shows the Stetson index for V plotted against visual calibrated magnitude, for all stars within the target region of IC\,5070. For the purpose of this paper we classify stars with a Stetson index of less than 0.1 in all three filters (V, R$_c$, I$_c$) as non-variable. For the non-variable stars we determine the median magnitudes and colours in all of the filters (U, B, V, R$_c$, \ha , I$_c$) as reference brightness for the subsequent calibration. The Stetson index cut ensures that images with small fields of view contain a sufficient number of calibration stars.
 
As can be seen in Fig.\,\ref{fig:Stetsons} there is a slight upward trend of the Stetson index in V for fainter stars, which is also seen in the R$_c$ and I$_c$ filters. This is caused by a small underestimation of the photometric uncertainties for the fainter stars during the basic photometric calibration due to images having different limiting magnitudes. The effect is, however very small and thus has no significant impact on the selection of non-variable sources. 

\subsubsection*{Colour Correction}
\label{Data:Correction_method}

To correct for any systematic magnitude offset caused by colour terms we determine for each image ($N$) a unique function $\mathcal{W}_N(m, c)$, where $m$ is the calibrated magnitude of the stars in the image (in the filter the image is calibrated into) and $c$ the colour of the stars. For the purpose of this paper we use V$-$I$_c$, but any other colour can be chosen if the star is detectable in those filters. The functional form of the correction factor used is a simple 2$^{\rm nd}$ order polynomial for both magnitude and colour, with no mixed terms and a common offset $p_0$, i.e.:

\begin{equation}
\mathcal{W}_N(m, V-I_c) = p_0 + \mathcal{P}^2(m) + \mathcal{P}^2(V-I_c),
\label{eq:functionalform}
\end{equation}

\noindent where $\mathcal{P}^2$ represents a second order polynomial without the offset. Thus, it is necessary to determine the five free parameters for the correction function $\mathcal{W}_N(m, c)$. We hence identify all non-variable stars detected (with Source Extractor flag less than 5) in image $N$ and determine their difference in magnitude from their real magnitude. We remove any stars that show a magnitude difference of more than $\pm$\,0.5\,mag and whose magnitude uncertainty is greater than 0.2\,mag. This is necessary, since stars selected as non-variable in V, R$_c$, and I$_c$ may still change their brightness in U and \ha, especially if they are young and potentially accreting sources. We require at least 10 non-variable stars to be present in the image. We then perform a least-squares optimisation of these magnitude differences to determine the required parameters for $\mathcal{W}_N(m, c)$. Typically, the bright non-variable stars in each image are far outnumbered by fainter ones, so for each star $i$, we introduce a magnitude $m_i$ dependent weighting factor $w_i$ during the fitting process:

\begin{equation}
w_i = \frac{1}{(m_i - {\rm min}(m_i) - 2)^{2}}
\label{eq:weightingfactor}
\end{equation}

\begin{figure*}
\centering
\includegraphics[width=\columnwidth]{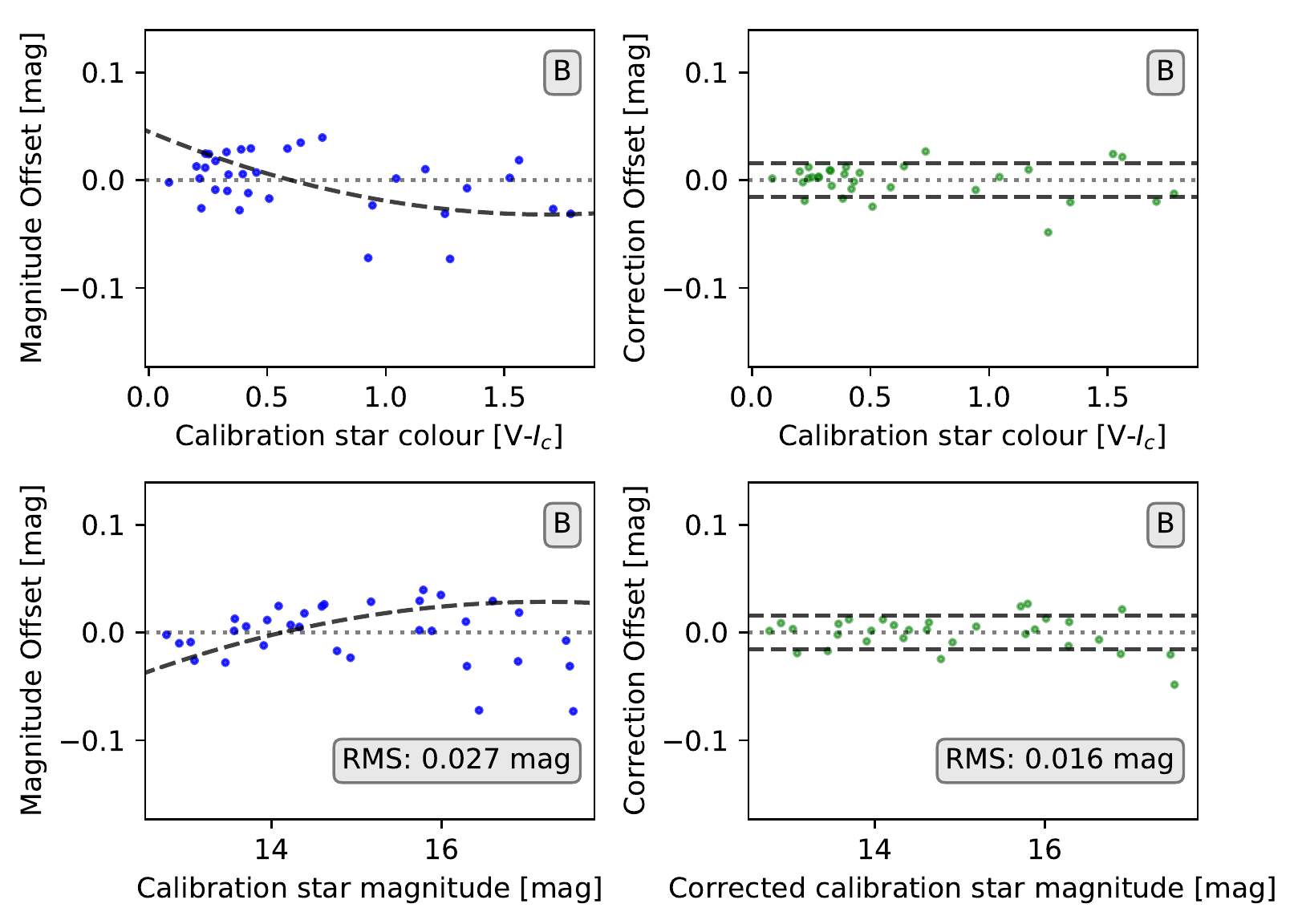} \hfill
\includegraphics[width=\columnwidth]{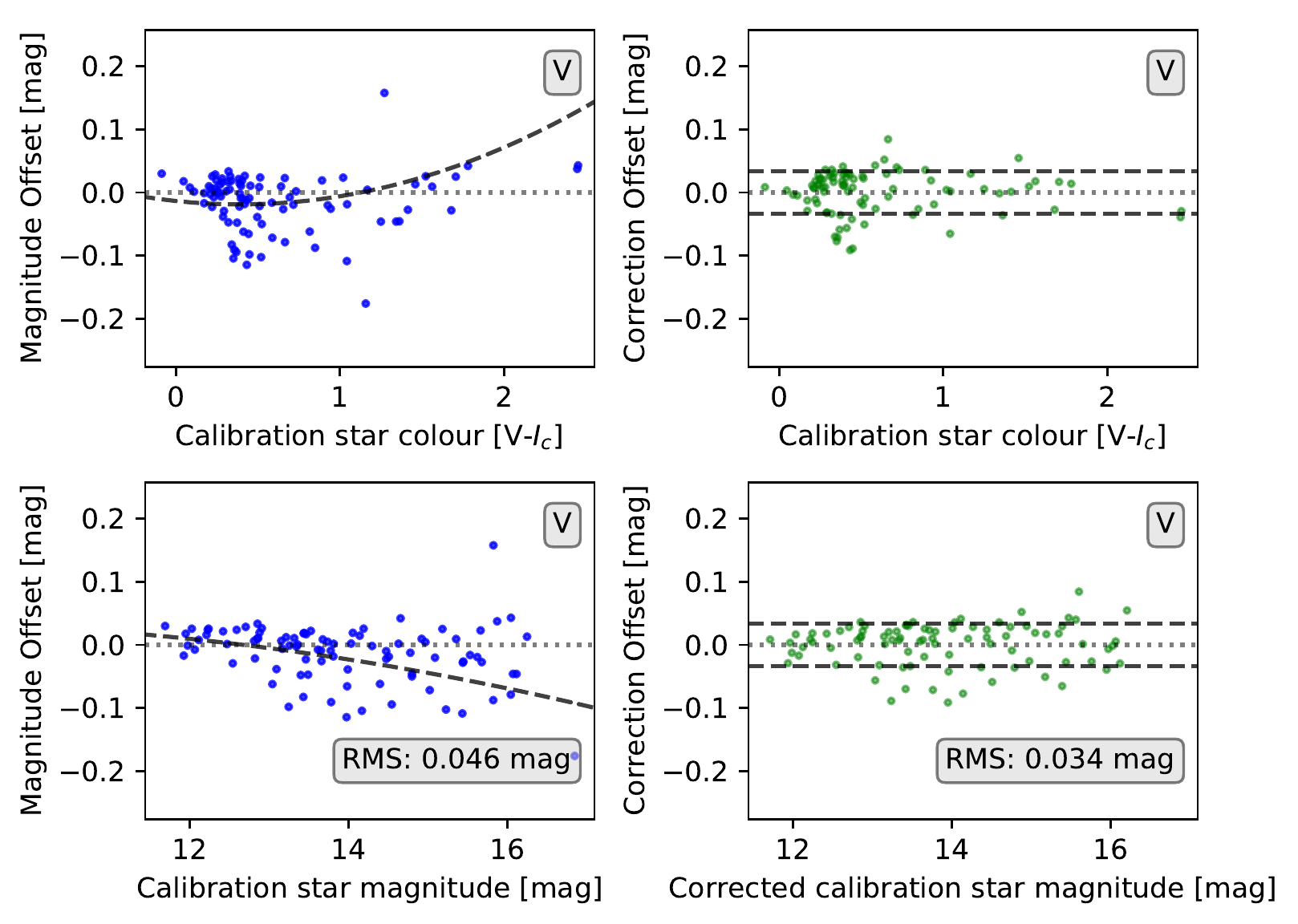} \\ 
\includegraphics[width=\columnwidth]{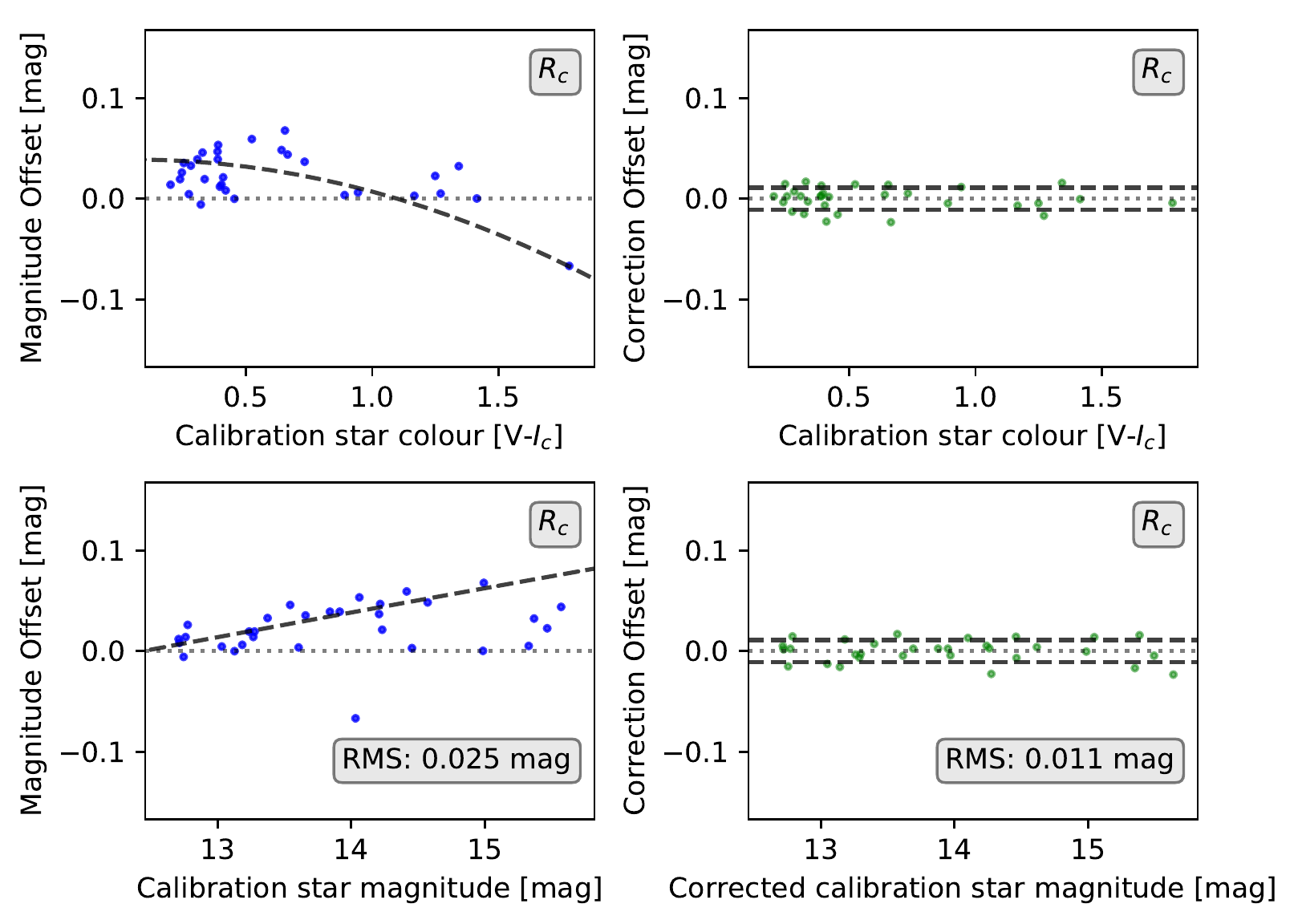} \hfill
\includegraphics[width=\columnwidth]{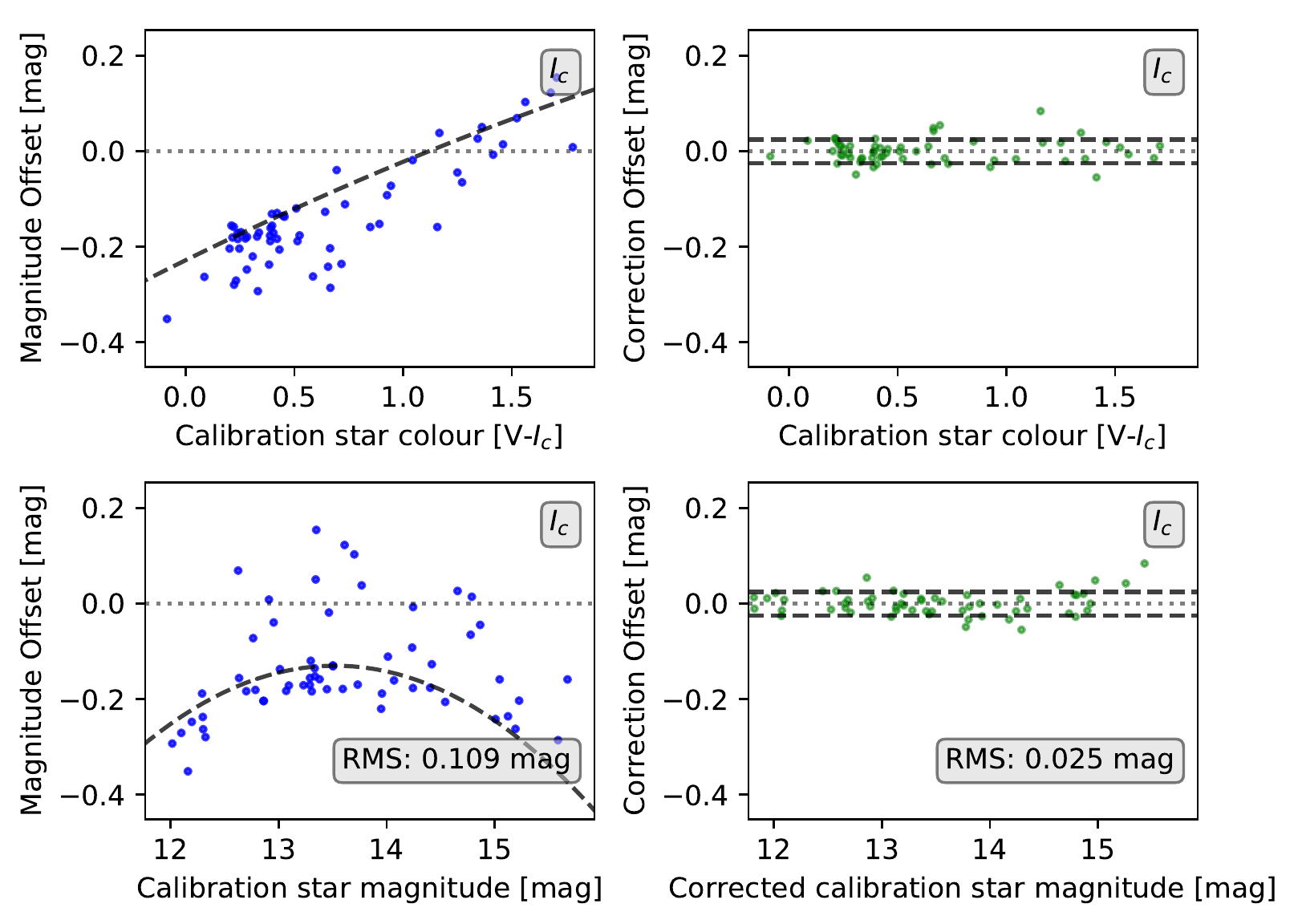} \\
\caption{\label{fig:correction1} Example correction plots for four images in B, V, R$_c$, and I$_c$ (from top left to bottom right). The B and V examples show very small colour terms, while the R$_c$ and I$_c$ images have been selected to show some of the largest colour terms present in our data to demonstrate how the colour correction procedure works. In each set of figures there are four panels. \textbf{Top Left:} The dots show the offset calculated for each calibration star against their median colour (V$-$I$_c$). The dashed line shows $\mathcal{W}_N(m, V-I_c)$ for the median magnitude of the calibration stars. \textbf{Bottom left:} The dots show the offset calculated for each calibration star against their magnitudes. The dashed line shows $\mathcal{W}_N(m, V-I_c)$ for the median colour of the calibration stars. \textbf{Top right:} The dots show the corrected offsets for the calibration stars against their colour. \textbf{Bottom right:} The dots show the corrected offsets for the calibration stars against their magnitudes. All sigma-clipped stars are removed from the right panels. The dashed lines in the right panels indicate the RMS scatter after the correction.}
\end{figure*}

\noindent Here min$(m_i)$ represents the magnitude of the brightest star included in the fitting process. This is the same weighting factor that is used when fitting the photo function and 4$^{\rm th}$ order polynomial during the basic data calibration. This gives brighter stars a larger weight during the optimisation. To ensure the fit is not influenced by misidentified objects, or stars showing variability (e.g. from flares not detected previously), the fitting is done using a three sigma clipping process. In Fig.\,\ref{fig:correction1} we show four examples of how the fitting process reliably removes any systematic colour- and magnitude-dependent photometry offsets.

To correct the magnitude $m_i$ of a particular star $i$ in image $N$ using the determined parameters of $\mathcal{W}_N(m, V-I_c)$, the colour of the star must be known at the time of the observation for image $N$. We determine the median magnitude in V and I$_c$ from all images taken within $\pm\,5$\,days of the observation date to estimate the colour. If there is insufficient data the time range is doubled until a value is found. As can be seen in Fig.\,\ref{fig:correction1} the colour dependence of the correction is weak in all cases, thus estimating the star's colour from the uncorrected photometry will not introduce any considerable systematic offsets, particularly since the majority of the data have been obtained using filters that have a very small colour term. 

We also use the calibration procedure to estimate a more representative uncertainty for the photometry after the correction of systematic offsets. We define the uncertainty as the RMS scatter of the magnitude offsets of all calibration stars in the image which have the same magnitude (within $\pm$\,0.1\,mag) as the star in question. If there are fewer than 10 calibration stars in that range, the magnitude range is increased until there are at least 10 calibration stars from which the RMS can be estimated.

\begin{figure*}
\centering
\includegraphics[width=\textwidth]{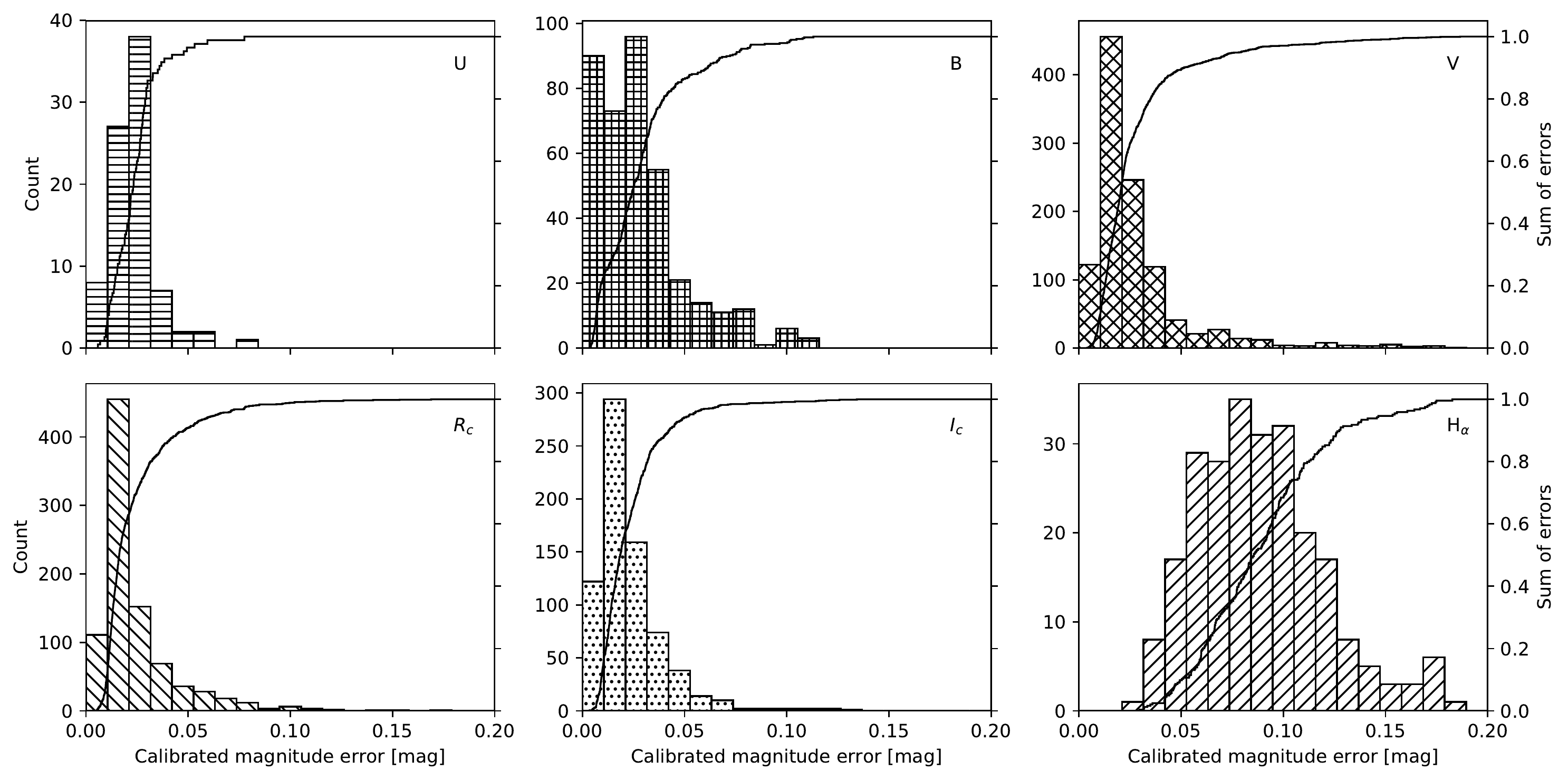}
\caption{\label{fig:histcumfreq} Histograms showing the number of the photometric uncertainties of all brightness measurements of \vc\ after the colour correction. Each panel represents one of the filters. The solid line represents the cumulative distribution. One can see that more than 80\,\% of the V, R$_c$, and I$_c$ measurements have photometric uncertainties of less than 0.04\,mag.}
\end{figure*}

The colour correction procedure only fails for 181 of the 3513 images when applied to the data of \vc. The main reason for these failures is a very small field of view. The typical median calibrated magnitude uncertainties for the U and \ha\ filters are 0.08\,mag and 0.09\,mag, respectively. This decreases to about 0.02\,mag for the B, V, R$_c$, and I$_c$ filters. Figure\,\ref{fig:histcumfreq} shows histograms of all calibrated magnitude uncertainties that are less than $0.2$\,mag for each filter. The cumulative frequency distribution of the same data shows that approximately $80$\,\% of the uncertainties in the broadband filters are less than $0.04$\,mag.

The entire analysis presented in this paper has been conducted using the colour corrected light curve for all data submitted and processed in the \hc\ database before September\,1$^{\rm st}$ 2019, and excludes any measurement with a higher than 0.2\,mag photometric uncertainty after our colour correction procedure.

\subsection{Spectroscopic Data}\label{lco_spectra}

From August\,1$^{\rm st}$ to September\,15$^{\rm th}$ in 2018 we coordinated \hc\ observations in a high cadence photometric monitoring campaign of \vc\ in order to monitor the short term variability of the source. In support of this campaign we attempted to obtain optical spectra of the source every five days. We utilised the FLOYDS spectrograph \citep{2014htu..conf..187S} on the 2\,m Las Cumbres Observatory Global Telescope network (LCOGT) telescope on Haleakala in Hawaii. It has a resolution between R\,=\,400 (blue) and R\,=\,700 (red) and covers a wavelength range from 320\,nm\,--\,1000\,nm. Due to local weather conditions, observations were only carried out on six nights during the above mentioned period. During each observing night we took 3\,$\times$\,600\,s exposures of the target, using a slit width of 1.2\,\arcsec. 

\begin{figure}
\centering
\includegraphics[width=\columnwidth]{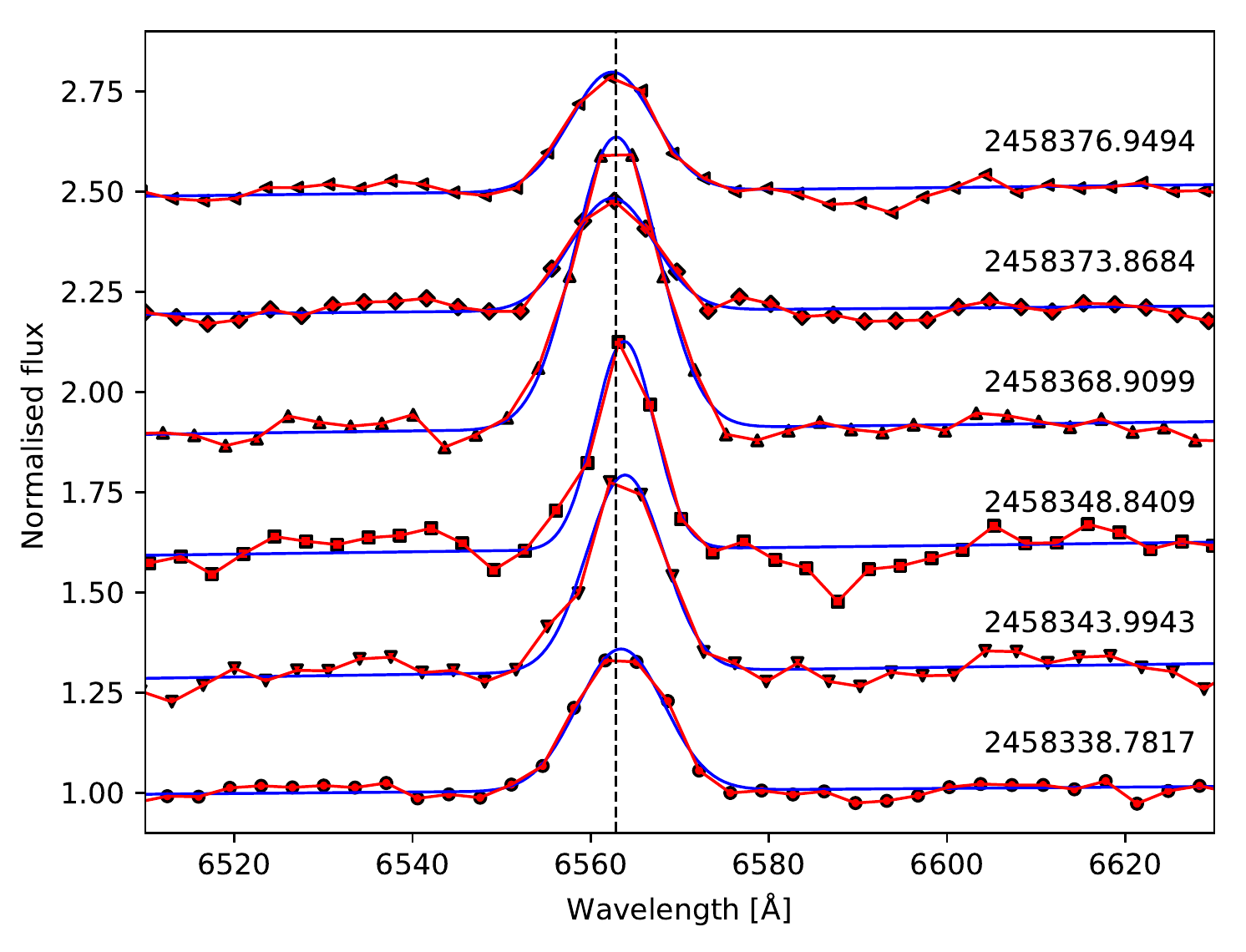}
\caption{\label{halinefig} \ha\ line profiles for \vc\ taken during August and September\,2018, ordered in time from bottom (earliest) to top (latest). The exact Julian dates are shown above each spectrum and the \ha\ equivalent widths are listed in Table\,\ref{tab:ha_ew}. The symbols and connecting red lines represent the data, while the smooth blue lines show the Gaussian fit to the \ha\ line from which the equivalent widths are measured. The dashed vertical line indicates the nominal \ha\ wavelength.}
\end{figure}

We utilised the pipeline reduced spectra and downloaded them from the LCOGT archive. Spectra taken in the same night are averaged. The only feature visible in all the spectra is the \ha\ line. We therefore normalised the spectra to the continuum near the \ha\ line and determined the equivalent width (EW) of this line by fitting a Gaussian profile to it. We show the region around the \ha\ line for all spectra in Fig.\,\ref{halinefig}.

\begin{table}
\caption{\label{tab:ha_ew} Observation dates of LCOGT optical spectra and the measured \ha\ equivalent width for \vc. The uncertainties in the equivalent widths are about 0.3\AA.}
\centering
\begin{tabular}{cc}
Julian Date & EW [\AA] \\ \hline
2458338.7817 & $-$4.31 \\
2458343.9943 & $-$5.30 \\
2458348.8409 & $-$4.36 \\
2458368.9099 & $-$8.63 \\
2458373.8684 & $-$3.29 \\
2458376.9494 & $-$3.49 \\
\end{tabular}
\end{table}

\section{Results}\label{results}

\subsection{The Quasi Periodic Light Curve of \vc}\label{period}

\begin{figure}
\includegraphics[width=\columnwidth]{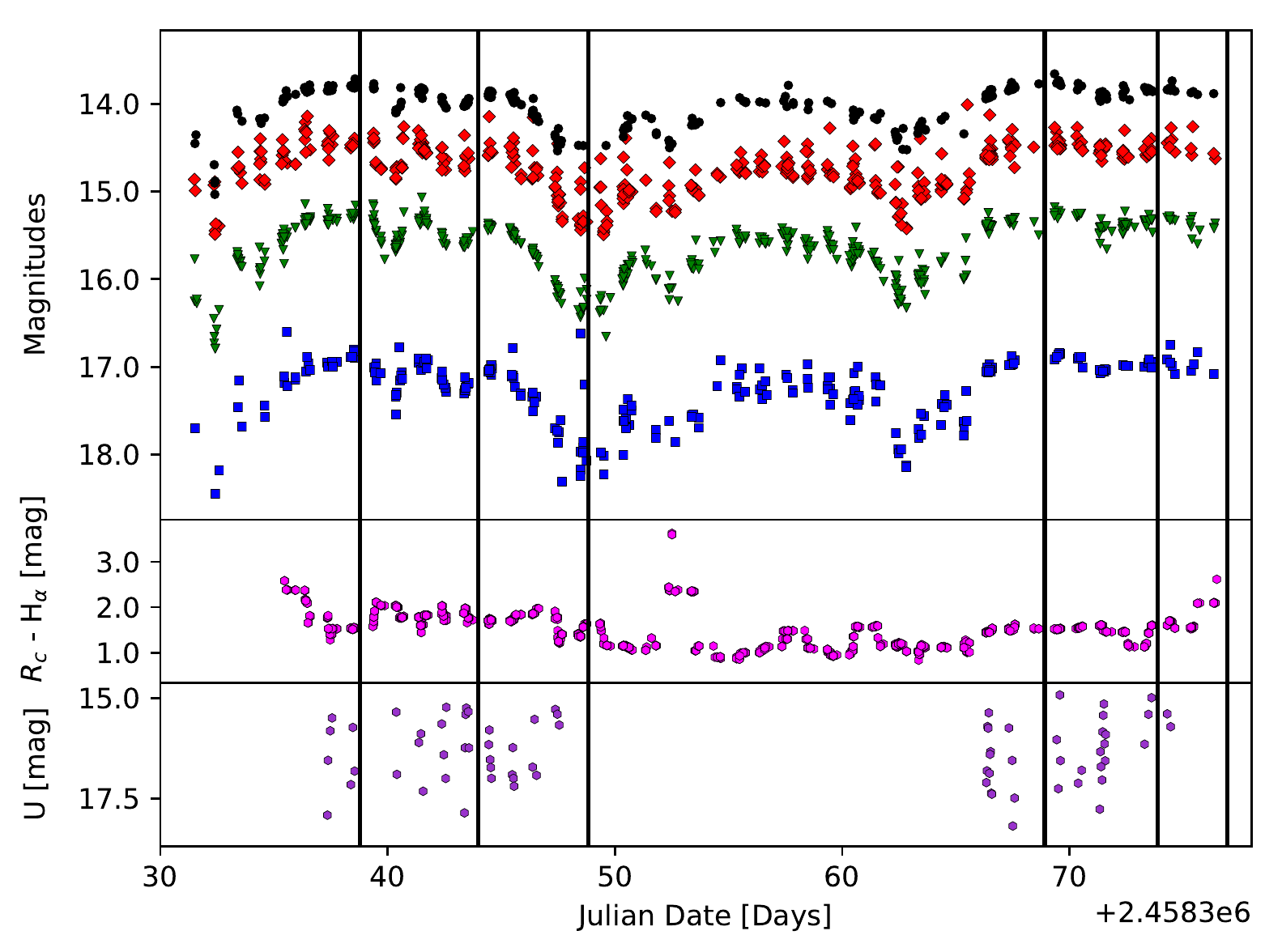}
\caption{\label{BVRI_lc} {\bf Top panel:} \hc\ light curves in B (blue), V (green), R$_c$ (red), and I$_c$ (black) of \vc\ for the high cadence campaign run between August\,1$^{\rm st}$ and September\,15$^{\rm th}$ 2018. {\bf Middle panel:} R$_c-$\ha\ magnitudes for \vc\ for the same time period. {\bf Bottom panel:} U magnitudes for \vc\ for the same time period. The vertical lines in all panels indicate the times when the LCOGT spectra are taken.}
\end{figure}

\vc, also known as 2MASS\,J20505357$+$4421008, is located in the Pelican Nebula (IC\,5070) at the J2000 position RA\,=\,20h50m53.58s, DEC\,=\,$+$44\dg21\arcmin00.88\arcsec\ \citep{2018yCat.1345....0G}. It has been classified as a low mass YSO and emission line object by \citet{2002AJ....123.2597O}. The star is classified as a variable star of Orion Type ($Or^*$) in the General Catalogue of Variable Stars (GCVS, \citet{2003AstL...29..468S}. It is not listed in \citet{2013ApJ...768...93F} who surveyed the area using data from the Palomar Transient Factory. It is included in the ASAS-SN variable stars database\footnote{\tt \href{ASAS-SN database}{https://asas-sn.osu.edu/variables}} (variable number 263219), but no period has been determined. \vc\ was however included in the list of candidates for YSOs published by \citet{2009ApJ...697..787G} and has been further studied by \citet{2018RAA....18..137I}, who do not report any periodicity. \citet{2018RNAAS...2b..61F} investigate the source and determine a period of approximately 31.8\,days in the V-Band. In the analysis of \hc\ data by \citet{2018MNRAS.478.5091F} it was classified as one of the most variable sources, grouped into the extreme dipper category. The colour and brightness changes were estimated to be caused by extinction due to dust grains that are larger than typical ISM grains. 

The 4\,yr long term \hc\ light curves of \vc, observed between September\,2015 and September\,2019 in the U, B, V, R$_c$, \ha, and I$_c$ filters are shown in Figs.\,B1, B2, B3 in Appendix\,B in the online supplementary material. The object shows strong variability, with large amplitude changes ($\Delta {\rm m} > 1$\,mag) in all filters, in agreement with its classification as a variable star by \citet{2003AstL...29..468S}. In Table\,\ref{tab:flux_variations} we list the range of maximum variability in each filter. The brightness variations of \vc\ are seen over the entirety of our observing period.

In Fig.\,\ref{BVRI_lc} we show a closeup of the light curve in U, B, V, R$_c$ and I$_c$, as well as for R$_c-$\ha\ for the time period from August\,1$^{\rm st}$\,--\,September\,15$^{\rm th}$ 2018, where we ran a focused coordinated campaign with all participants in order to investigate the short term variability of the source. This figure clearly shows distinct, short duration changes (down to sub-one day) in the brightness of the source during this 45\,day observing period. In the four broadband filters (B, V, R$_c$, I$_c$) the light curves look very similar, i.e. they show exactly the same behaviour and only the amplitudes are different. The U-band data, even if much more sparse, does not follow this trend and shows variability with $\Delta {\rm U} \sim$\,3\,mag. We will discuss this in more detail in Sect.\,\ref{Analysis:Fingerprint}. The R$_c-$\ha\ lightcurve also does not follow the general trend of the broadband filters, but shows longer term (weeks) and short term (day/s) variability with short, up to 1\,mag bursts. 

\begin{table}
\caption{\label{tab:flux_variations} Ranges of variability for each filter for \vc. Given are the minimum recorded magnitudes, the maximum recorded magnitudes and the difference between them.}
\centering
\begin{tabular}{cccc}
Filter & min [mag] & max [mag] & $\Delta m$ [mag] \\ \hline
U & 18.18 & 14.92 & 3.26 \\
B & 18.70 & 16.60 & 2.10 \\
V & 17.31 & 15.03 & 2.27 \\
R$_c$ & 15.96 & 13.97 & 1.99 \\
\ha\ & 15.29 & 11.00 & 4.29 \\
I$_c$ & 15.26 & 13.65 & 1.61 \\
\end{tabular}
\end{table}

A more detailed investigation of the entire 4\,yr light curve reveals that the variations in the star's brightness occur as dips in brightness lasting from a few days to about two weeks. We used Lomb-Scargle Periodograms \citep{1982ApJ...263..835S} to investigate whether or not these dips are periodic in nature. The periodograms for the B, V, R$_c$, and I$_c$ filters are shown in Figs.\,C1, C2 in Appendix\,C in the online supplementary material. We can identify that \vc\ exhibits a clear quasi-periodic behaviour for the broadband filters V, R$_c$ and I$_c$, with a period of approximately $31.5$\,days, in agreement with \citet{2018RNAAS...2b..61F}. 


Utilising the V, R$_c$ and I$_c$ \hc\ data we find an average period of the dips of 31.423\,$\pm$\,0.023\,d. The uncertainty is the RMS of the periods determined for the individual filters. In order to improve the accuracy we include the V, R$_c$, and I$_c$ data from \citet{2018RAA....18..137I} in our period determinations as this extends the observation period from September\,2010 to September\,2019, i.e. doubling the baseline of the data. We have `colour corrected' the \citet{2018RAA....18..137I} data using common observing dates between their and our data. The mean period with the additional data for the three filters then becomes 31.447\,$\pm$\,0.011\,d, which we will use for the purpose of this paper throughout. The phase zero point (taken to be the point of maximum light) occurs at JD\,=\,2458714.0, which corresponds to 12:00\,UT on August 18$^{\rm th}$ 2019. The phased lightcurves in Figs.\,C1, C2 in the online supplementary material indicate that the object is most likely to be observable in its bright state within about $\pm$\,5\,d (15\,\% of the period) from the nominal phase\,=\,0 point. 

Note that despite the much smaller amount of B-band data in our light curve, the periodogram shows a peak at the above determined period (see Fig.\,C1 in the online supplementary material). The peak is however not as significant as for the other filters.

\subsection{Analysis of the Variability Fingerprints}
\label{Analysis:Fingerprint}

\begin{figure*}
\centering
\includegraphics[width=0.49\textwidth]{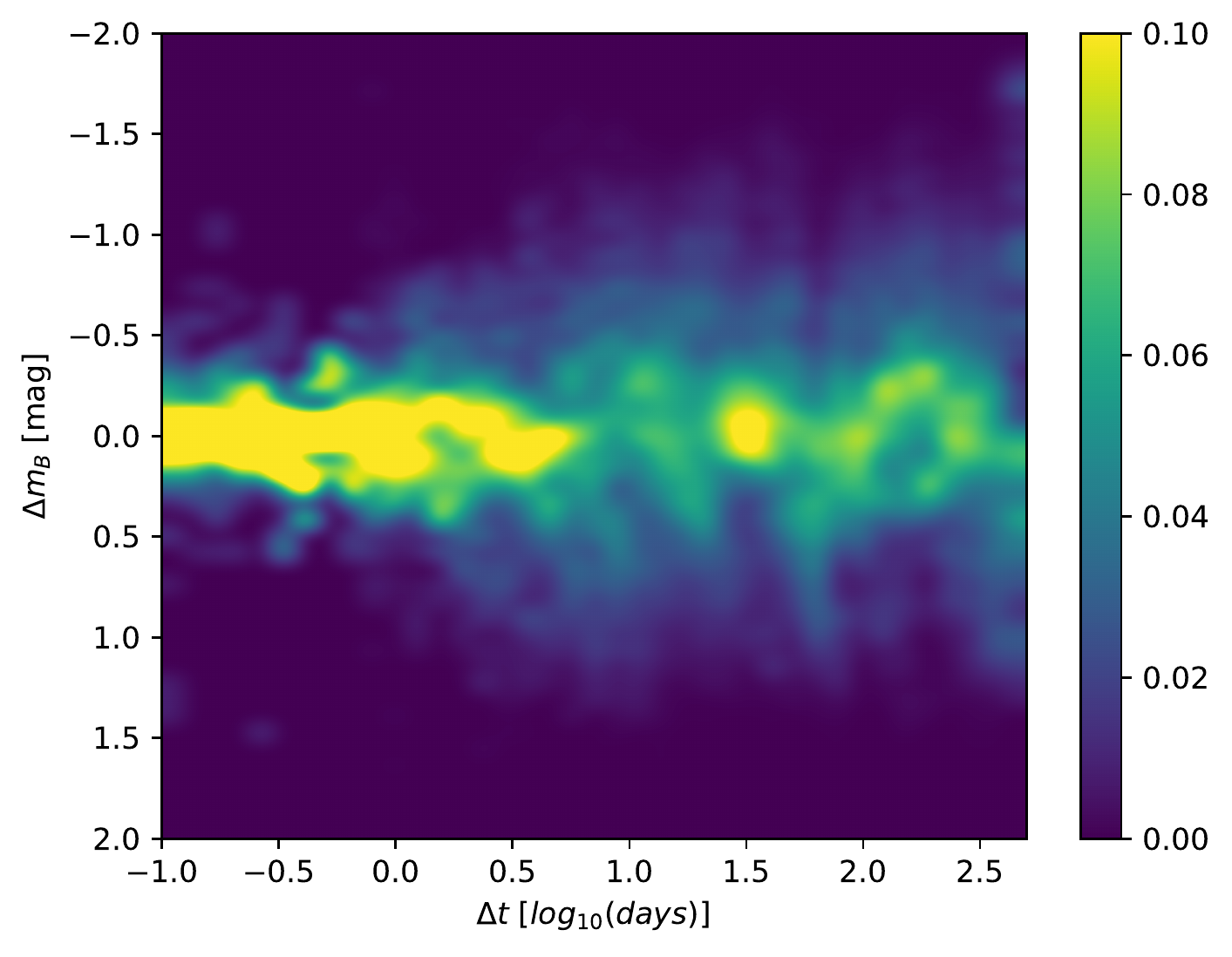} \hfill \includegraphics[width=0.49\textwidth]{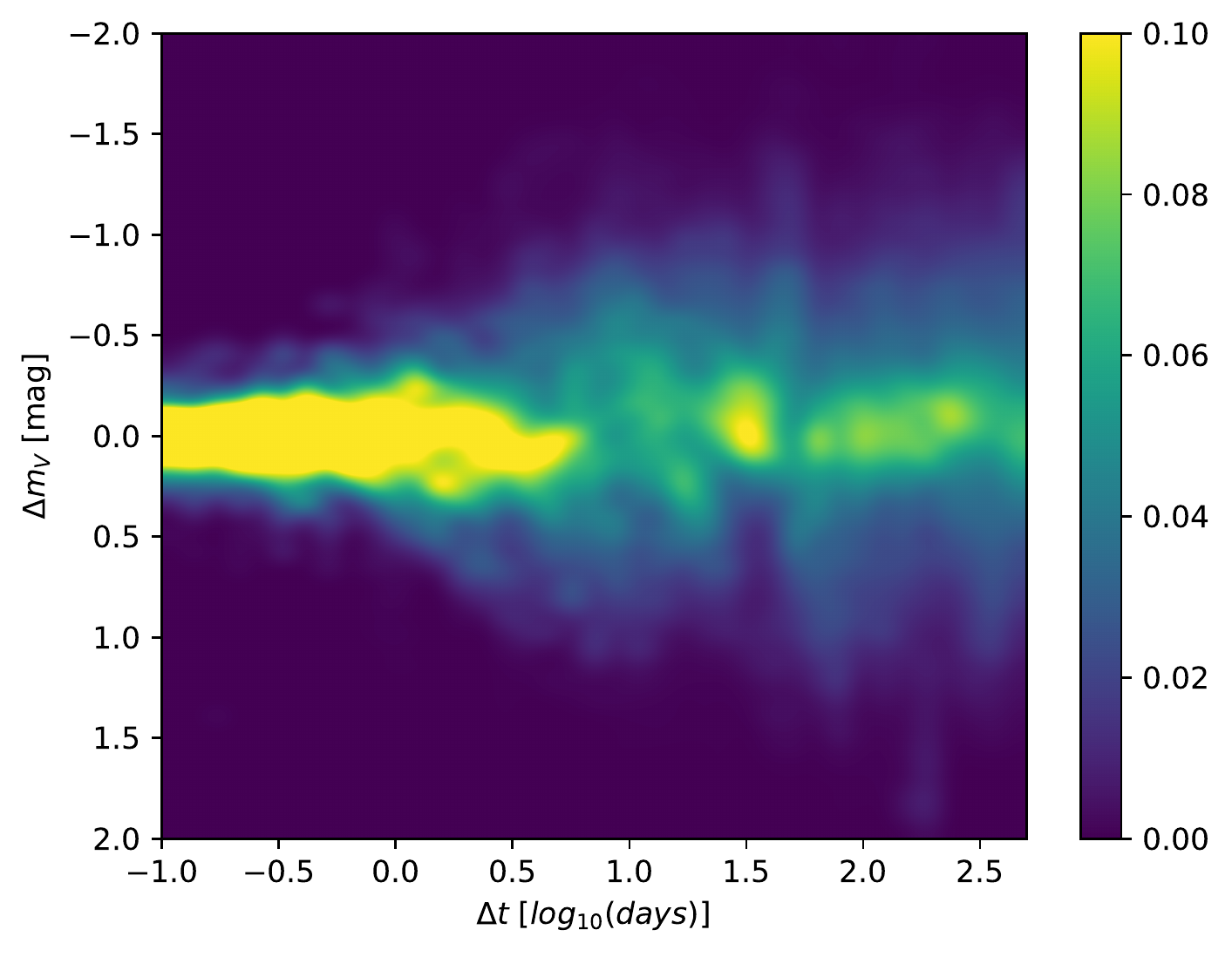} \\
\includegraphics[width=0.49\textwidth]{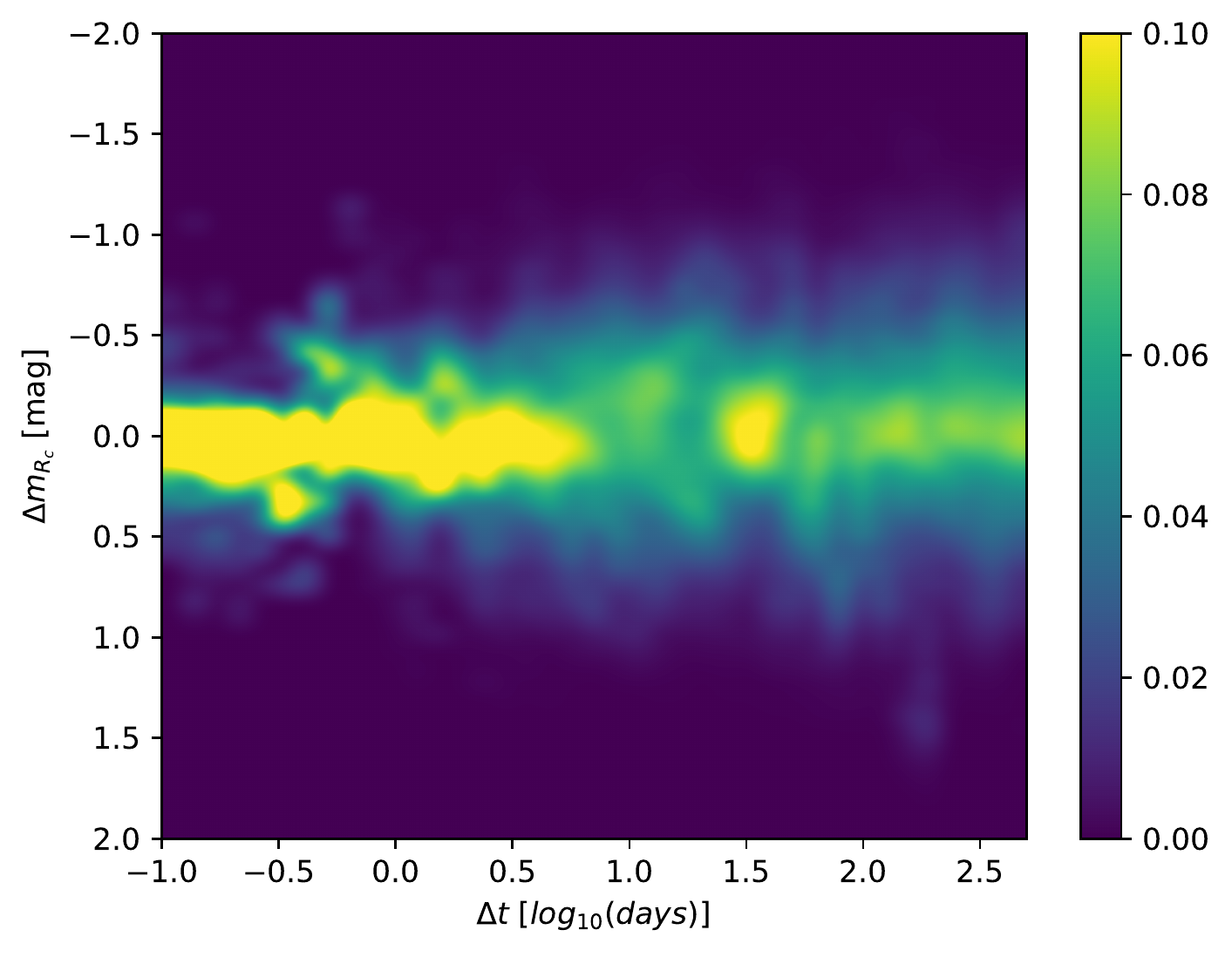} \hfill \includegraphics[width=0.49\textwidth]{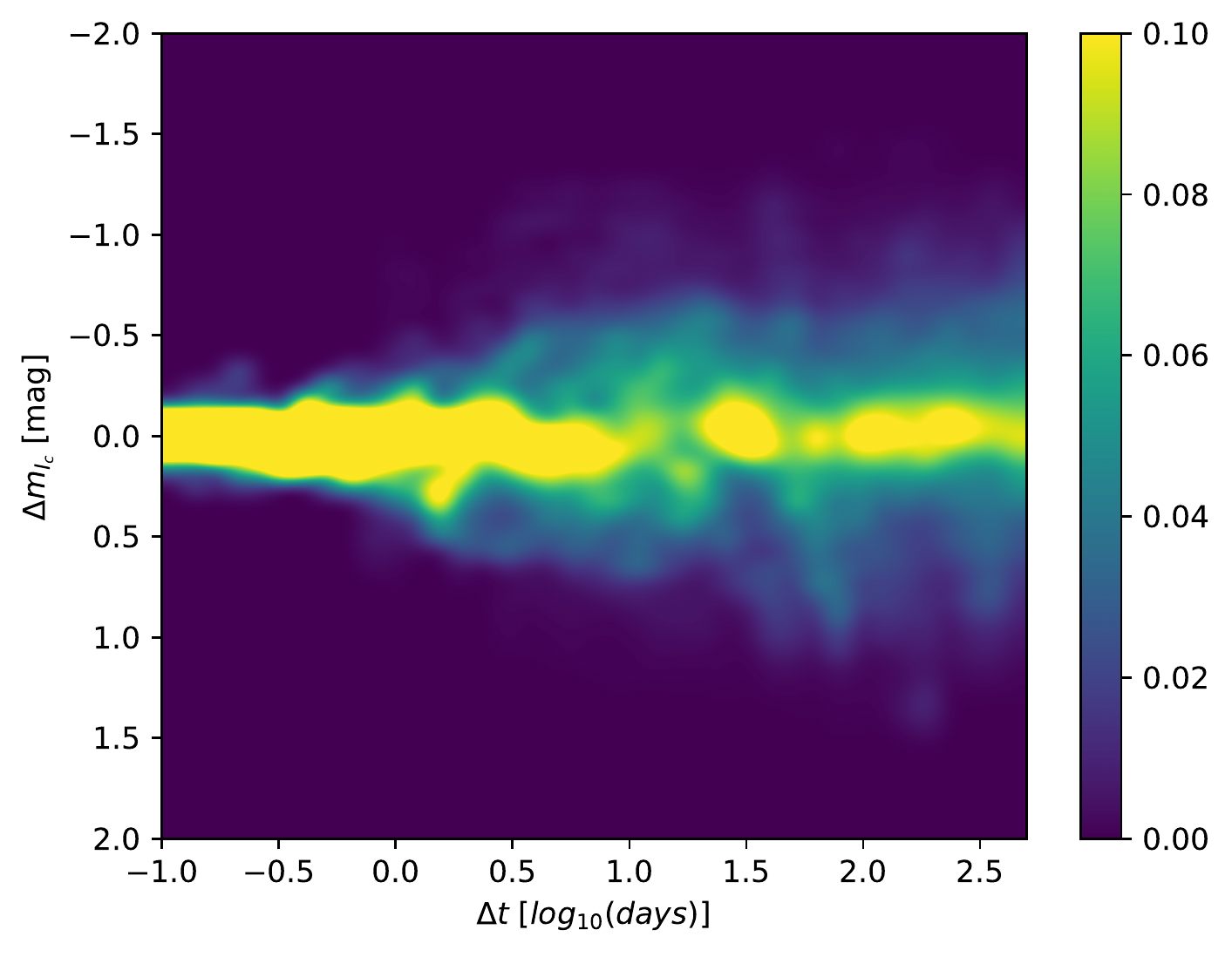} \\
\includegraphics[width=0.49\textwidth]{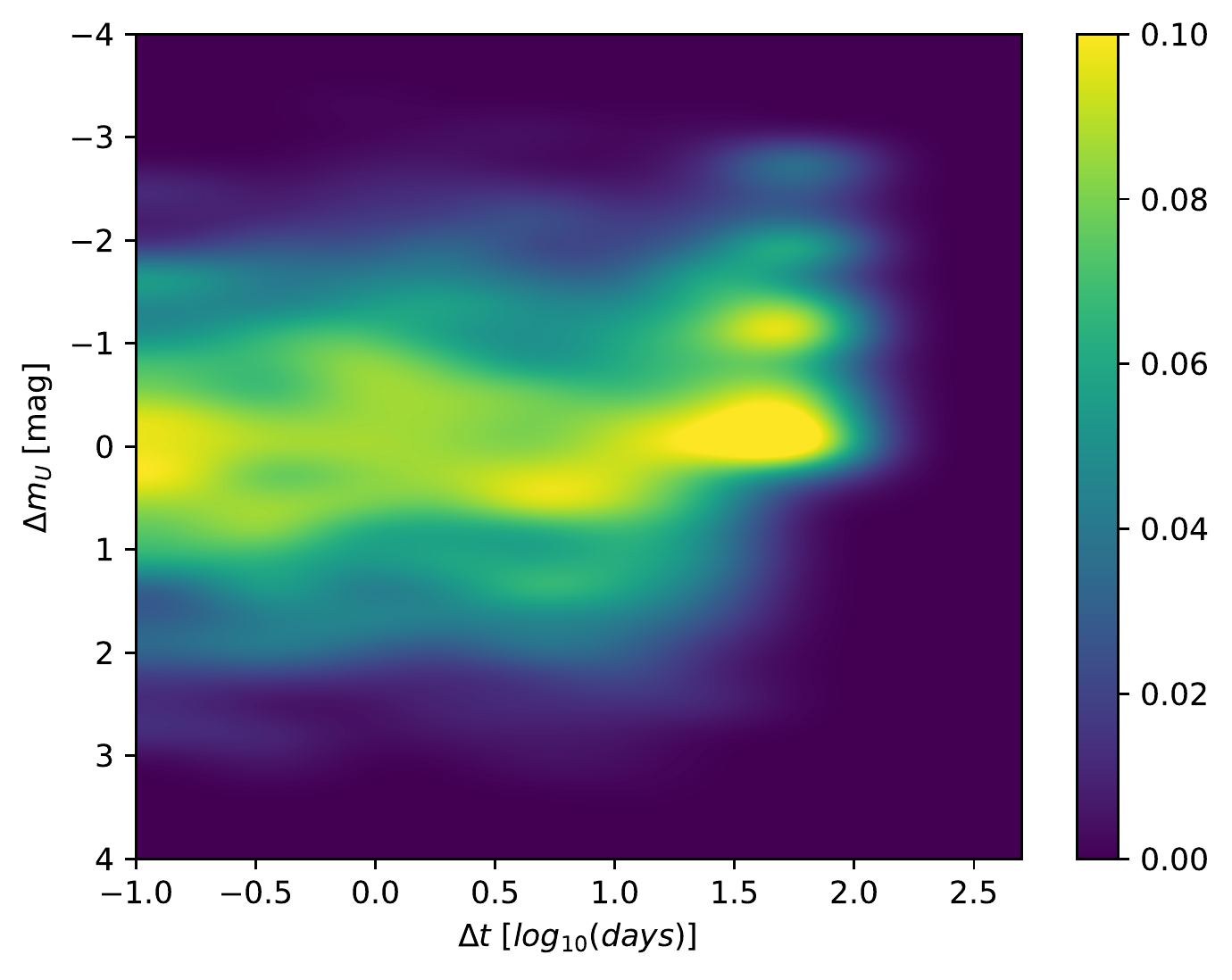} \hfill \includegraphics[width=0.49\textwidth]{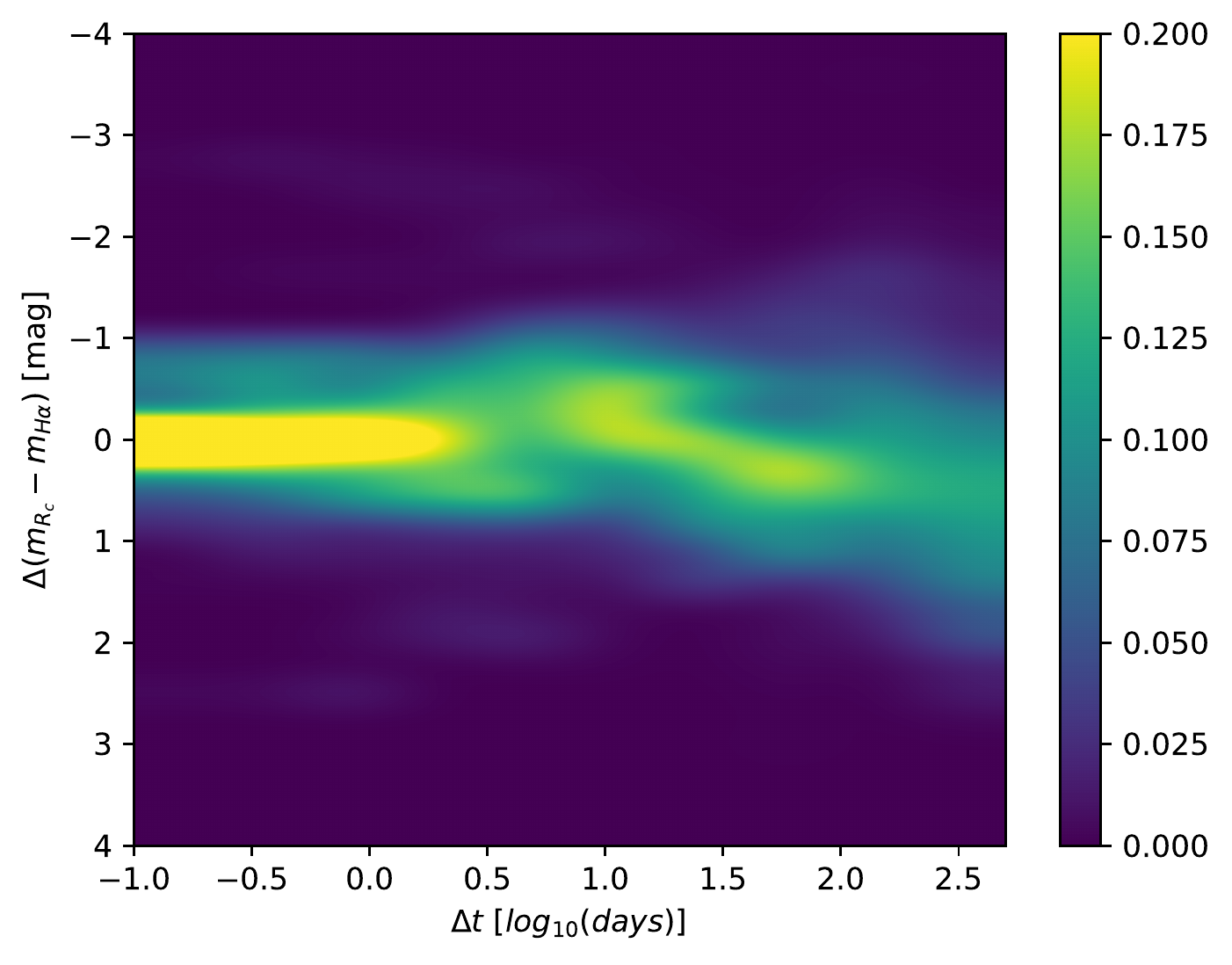} \\
\caption{\label{fig:fingerprints} Variability fingerprint plots of B, V, R$_c$, I$_c$, U and R$_c-$\ha\ data (top left to bottom right) for \vc. The colours indicate the probability of observing the object to undergo a change in magnitude for a given time gap between observations. Clearly observed in the B, V, R$_c$, I$_c$ fingerprints is the periodicity of \vc\ at $\log(\Delta t)\approx1.5$, which equates to the determined period of 31.447\,d. The bin size in the top four plots is 0.08\,mag and a factor of 1.237 for $\Delta t$, in the bottom two plots the bin size is 0.27\,mag and a factor of 2 for $\Delta t$.}
\end{figure*}

In order to further analyse the star's variability, in particular to investigate time scales and amplitudes, we determine a {\it variability fingerprint} of the source in each filter. This follows and improves upon previous work by e.g. \citet{2004A&A...419..249S}, \citet{2015ApJ...798...89F} and \citet{2017MNRAS.465.3889R}. We determine for all data points in a light curve that are taken in the same filter, all possible time ($t$) and magnitude ($m$) differences for two measurements $i, k$ ($\Delta t = t_i - t_k$; $\Delta m = m_i - m_k$) where $t_i > t_k$. All these differences are then used to populate a 2D $\Delta t$ vs. $\Delta m$ histogram, where the bins in $\Delta t$ are log10-spaced. These histograms are normalised to an integral of one for each bin of $\Delta t$. This ensures that the values in these variability fingerprints represent the probability ($p$) that the source shows a change of $\Delta m$ for a given time interval $\Delta t$ between observations. 

In Fig.\,\ref{fig:fingerprints} we show these variability fingerprints for \vc\ for the five broadband filters as well as for R$_c-$\ha. The plots for B, V, R$_c$ and I$_c$ (top four plots) show an extremely similar behaviour, however the B-band data suffers from a smaller number of observations compared to the other filters. For time intervals between observations shorter than a few days the object is most likely not variable and is observed at the same magnitude. The width of the high probability ($p >$\,10\,\%) behaviour corresponds very well to the typical photometric errors measured in the photometry of the object after the calibration procedure (see Fig.\,\ref{fig:histcumfreq} in Sect.\,\ref{Data:Colour-term_correction}). 

Between 10\,d and 30\,d intervals, the probability to find the object changed within its range of variability is almost homogeneously distributed. At a roughly 30\,d interval the object is once again most likely to be observed at an unchanged magnitude, representing the quasi-periodicity of the source. However, there is still a significant probability that the source does not return to precisely the same brightness after one period. This indicates that the internal structure of the dips varies each time it is observed. We will investigate this in more detail in Sect.\,\ref{Analysis:column_density}.

For time intervals longer than 30\,d the fingerprints show that the object most likely does not change brightness but still has a significant probability to show variations within the min/max values found for each filter. Thus, the overall behaviour of the object remains unchanged for time scales beyond one period. Furthermore, one can see that from time intervals of one day onward, there is a non-zero probability that the object starts to vary by more than the photometric uncertainty. In the $\Delta t$ vs. $\Delta m$ space this trend is almost linear from one day to about half the period, after which the variability does not increase any further. This short term variability is also very evident in the detailed light curve presented in Fig.\,\ref{BVRI_lc}.

Compared to the B, V, R$_c$, and I$_c$ fingerprints, the behaviour for U and R$_c-$\ha\ is different (see bottom graphs in Fig.\,\ref{fig:fingerprints}). This difference is not just caused by the reduced number of observations available and the (particularly in the U-band) much shorter total time interval covered by the observations. 

In the U-band the variability for short time intervals is clearly different from the photometric uncertainties and also does not systematically increase towards one period. In essence for U the full range of variability is achieved within time intervals of less than one day. Furthermore, the magnitude of variability is much larger. For example the U-Band magnitude changes by up to 3\,mag in a single night. Variability in the U-Band fluxes is indicative of accretion rate changes \citep{1998ApJ...492..323G, 2008ApJ...681..594H, 1998ApJ...509..802C}. This indicates short term (hours) accretion rate fluctuations in this object by a factor of up to ten or slightly larger. This will be discussed in more detail in Sect.\,\ref{evol_stage}.

The R$_c-$\ha\ fingerprint over short time periods (i.e. less than a few days) is similar the B, V, R$_c$, and I$_c$ filters, in that it is dominated by the photometric uncertainties, which are of course higher for the \ha\ images. But there is a non-negligible probability that the source varies by up to 1\,mag even at those short time intervals. This is caused by short bursts of variability, one of which is evident in the detailed lightcurve in Fig.\,\ref{BVRI_lc}. However, there is no strong indication that the R$_c-$\ha\ magnitude follows the same periodic behaviour that can be seen in the broadband filters. Generally, the variability in R$_c-$\ha\ increases slightly with an increase in time interval between observations. The apparent trend of a slight long term decrease in R$_c-$\ha\ visible in the fingerprint plot is not real, as it depends upon a single \ha\ data point. 

\subsection{Column density distribution of consecutive dips} 
\label{Analysis:column_density}

\begin{figure}
\includegraphics[width=\columnwidth]{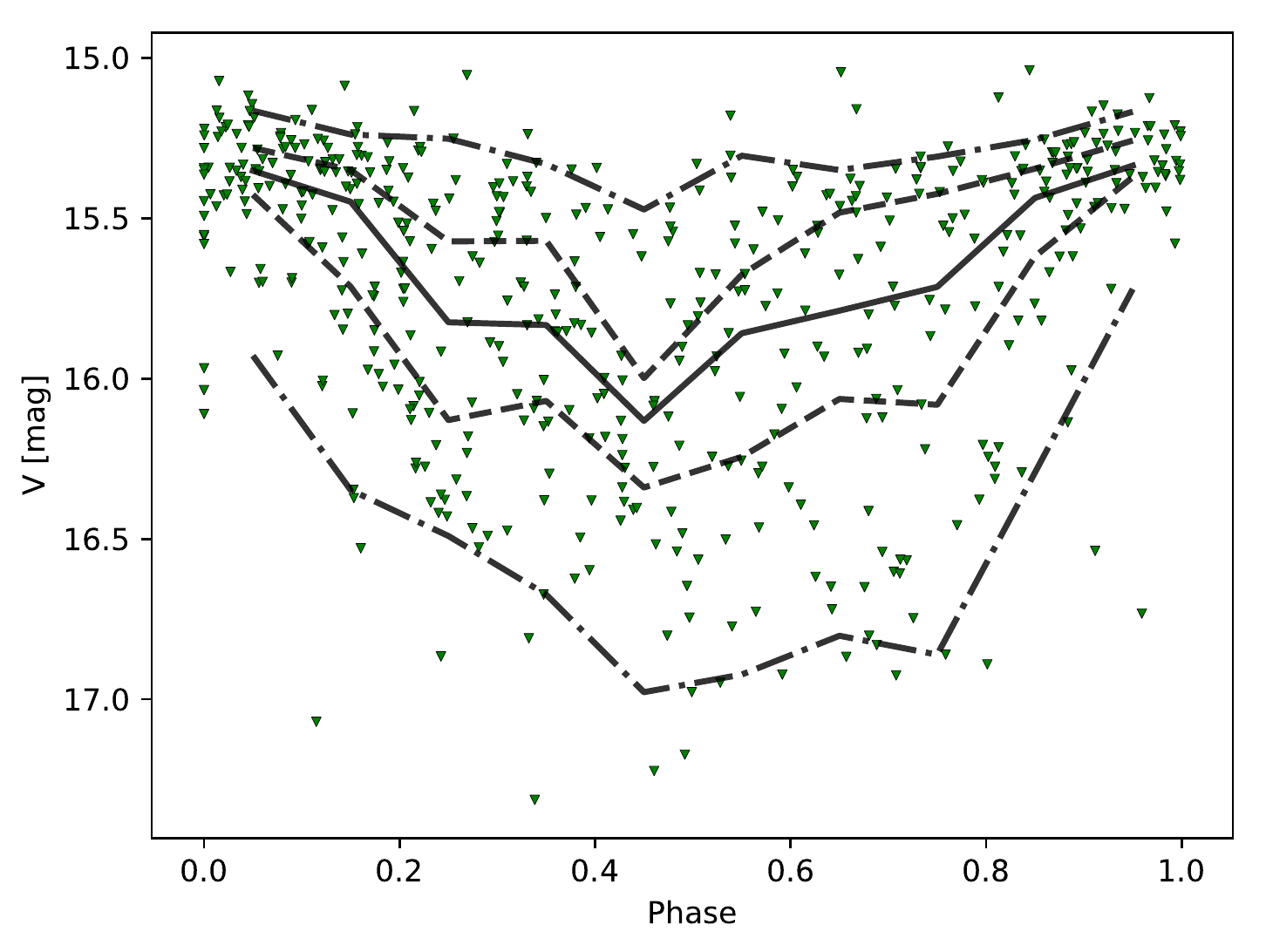}
\caption{\label{fig:phased_V} Phased light curve of V-band data for \vc. The solid line indicates the running median magnitude. The two most extreme dash-dotted lines indicate the range within which 95\,\% of the data points can be found. The other two dashed lines indicate the 68\,\% range for the data points. Note that the high cadence data (shown in Fig.\,\ref{BVRI_lc}) has been removed from this plot as it otherwise would dominate the typical shape due to the large amount of data in that 6\,week period.}
\end{figure}

The quasi-periodic dips in the light curve of \vc\ are potentially caused by orbiting material in the accretion disk. The different time scales of the variability in the U and \ha\ filters strongly suggest that accretion variability is not the cause of the longer term variations in the B, V, R$_c$, and I$_c$ continuum. In the following, we aim to constrain the properties of the occulting material.

The quasi-periodic nature of \vc\ found in Sect.\,\ref{period} indicates orbiting material held in place within the inner disk for at least the duration of our lightcurve, i.e. more than 4\,yr. As evident in the variability fingerprint plots shown in Fig.\,\ref{fig:fingerprints}, there is a significant probability that the source does not return to the exact same brightness after one full period. This indicates variation in the column density distribution of the material along its orbital path. In this section we will investigate if we can identify any systematic changes in the column density distribution, which can hint at the time scales and/or mechanisms by which the material is either moved into and out of the orbiting structure, or redistributed within it.

In Fig.\,\ref{fig:phased_V} we show the folded V-Band light curve of \vc\ with the high cadence data from Aug.\,--\,Sep.\,2018 (shown in Fig.\,\ref{BVRI_lc}) removed as this would otherwise dominate the analysis. The plot has the running median overlaid which indicates the typical column density distribution of material along the orbit. As one can see, the median occultations are up to about 0.7\,mag deep and have a broad minimum. In the figure we also show the typical one and two sigma deviations from the median, which are represented by the dashed and dash-dotted lines, respectively. They indicate that in some cases there is almost no detectable occultation, while the dips can, in extreme cases, be up to 1.7\,mag deep in the V-Band, during a larger part of the period. 

Our \hc\ data now covers about 40 complete periods of \vc. We are hence able to investigate how the column density distribution along the orbital path varies as a function of the time interval between observations (in units of the period of the source). In essence we can construct the structure function of the column density. For this we determine the difference $\Delta$($m$) in the depth of the dip for all pairs of V-band measurements which are $N$ times the period ($\pm$\,1\,day) apart from each other, whereby N runs from 1 to 10. We find that there are no significant trends in the structure function and hence refrain from showing it here. The values for $\Delta$($m$) scatter less (for all values of $N$) when the phase is close to 0 or 1, compared to phase values near 0.5. This is of course expected from Fig.\,\ref{fig:phased_V}. There are no systematic trends for the value of the structure function with $N$, either for a particular part of the phase space, or when averaged over the entire period. 

\begin{figure}
\includegraphics[width=\columnwidth]{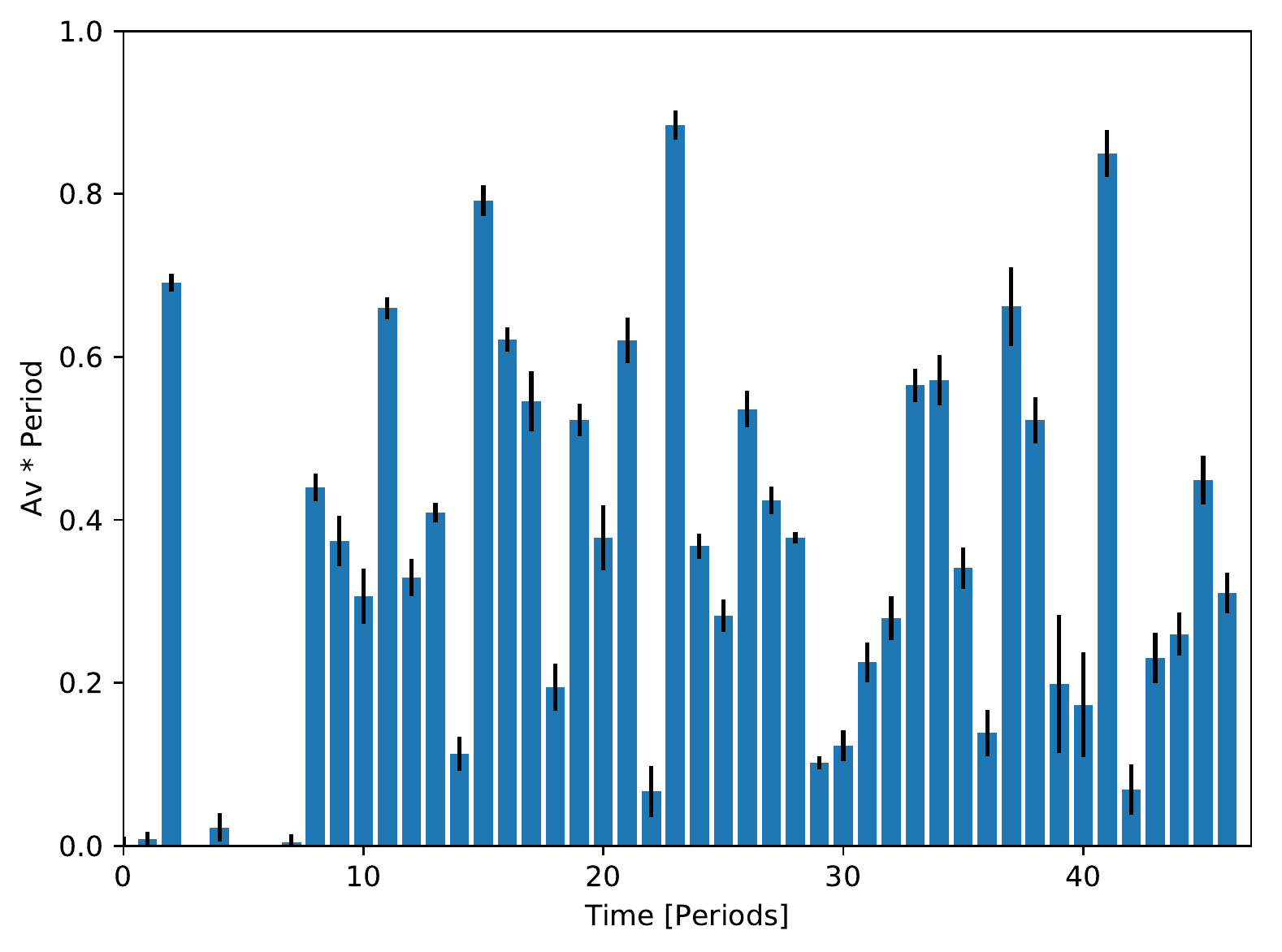}
\caption{\label{fig:dip_mass} This plot shows the integrated column density for each orbital period, i.e. a measure of the mass of material contained in the orbiting structure around \vc\ for the observed dips. The dips are numbered arbitrarily starting at our first available data. The value of one on the y-axis corresponds to about 1\,$\times$\,10$^{-11}$\,M$_\odot$ or 0.03\,\% of the Lunar mass.}
\end{figure}

Given the lack of correlation found above of column density from dip to dip, we have tried to investigate the total amount of material along the orbital path during each dip. Under the assumption of a constant line of sight column density, the mass is proportional to the integrated depth of the V-Band lightcurve for each dip. We show the results of this calculation in Fig.\,\ref{fig:dip_mass}. We use a trapezium interpolation between V-band data points to determine the values displayed in the figure. The error bars are solely based on the photometric uncertainty and do not consider gaps in the data. As one can see, the amount of mass in the occulting structure varies by up to a factor of 10. This suggests that the material in the line of sight is moving in and out of the structure on time scales of the order of, or shorter than the period of the occultations, and the mass flow rate varies by typically a factor of a few when averaged over one period. This is in agreement with the R$_c-$\ha\ and U-band variability discussed in the previous sections, which indicate variable accretion through the disk.

We can make some simple assumptions about the star and the geometry of the occulting structure to obtain an order of magnitude estimate of the mass in the dips. Given the classification of the source as low mass YSO \citep{2002AJ....123.2597O} we assume that the object has a central mass of 0.5\,M$_\odot$. The period of about 31.5\,d then means that the occulting material is located about 0.15\,AU from the central star. We further assume dimensions of the structure in the two directions perpendicular to the orbital path of 0.05\,AU, and dust scattering properties in agreement with $R_{\rm V}$\,=\,5.0 (see below in Sect.\,\ref{Analysis:Dust_Scattering_Properties}). This converts the value of one in Fig.\,\ref{fig:dip_mass} to approximately 1\,$\times$\,10$^{-11}$\,M$_\odot$ (or 0.03\,\% of the Lunar mass). Given the period, the typical mass flow (accretion rate) of material through the structure in \vc\ is hence of the order of 10$^{-10}$\,M$_\odot$/yr. This is consistent with low levels of accretion as seen in other T\,Tauri stars (e.g. \citealt{2017A&A...600A..20A}).

\subsection{Colour variations during dips}
\label{Analysis:Dust_Scattering_Properties}

In order to investigate the scattering properties of the occulting material in the line of sight during the dips, we investigate the star's behaviour in the V vs. V$-$I$_c$ parameter space. 

\begin{figure}
\includegraphics[width=\columnwidth]{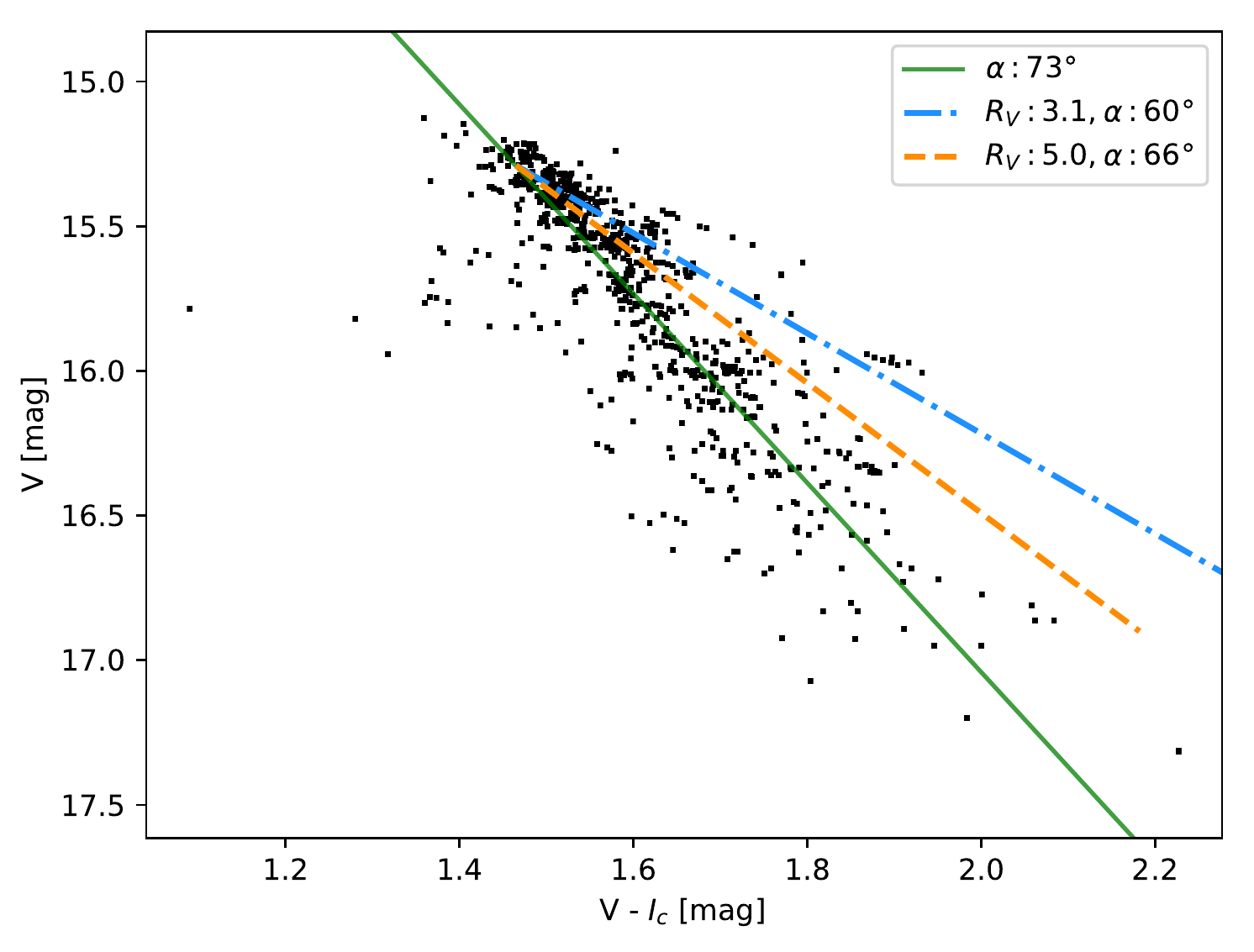}
\caption{\label{fig:alpha} The V vs. V$-$I$_c$ colour magnitude plot for the \hc\ data of \vc. The blue dash-dotted line corresponds to behaviour expected for occultations by material with $R_{\rm V}$\,=\,3.1, while the orange dashed line is for $R_{\rm V}$\,=\,5.0. The solid green line indicates the determined angle for \vc\ of 73\dg.}
\end{figure}

To characterise the movement of the star in this diagram we follow \citet{2018MNRAS.478.5091F} and determine the angle $\alpha$, measured counterclockwise from the horizontal. Using a standard reddening law by e.g. \citet{1990ARA&A..28...37M}, disk material composed of normal ISM dust would have $\alpha$ values between 60\,\dg\ (R$_{\rm V}$\,=\,3.1) and 66\,\dg\ (R$_{\rm V}$\,=\,5.0). We show the V vs. V$-$I$_c$ diagram for \vc\ in Fig.\,\ref{fig:alpha}. Note that the exact values for $\alpha$ do also depend on the filter transmission curves used in the observations. However, there will be a general trend of increasing $\alpha$ values with increasing typical grain size. Occultations by very large grains compared to the observed wavelengths, or by optically thick material, should generate colour independent dimming ($\alpha$\,=\,90\dg). If occultations cause a change in the dominant source of the light we observe, i.e. from direct light to scattered light, other colour changes including bluing during deep dimming events can occur, see e.g. \citet{1999AJ....118.1043H}.

Since the determined value of $\alpha$ can be very sensitive to small changes in the V and V\,$-$\,I$_c$ values, we only include high  signal-to-noise (S/N) measurements in this analysis. In particular we only include V and I$_c$ magnitudes that have a determined uncertainty of less than 0.05\,mag after the colour correction (see Sect.\,\ref{Data:Correction_method}). Furthermore, since the object is constantly changing its brightness the V and I$_c$ data need to be taken as simultaneously as possible. Thus, we only utilise pairs of V and I$_c$ data that were obtained within one day. 

In order to determine $\alpha$ we need to establish the baseline brightness in V and I$_c$ which represents the bright state of the source. We select all data that are taken at a phase within 0.1 from the bright state of the source (see Sect.\,\ref{period}) and use their median magnitude as the baseline brightness for the source. This equates to 15.29\,mag in V and 13.86\,mag in I$_c$. The uncertainty in $\alpha$ is determined from error propagation of the individual photometry errors obtained during the colour correction.

\begin{figure}
\includegraphics[width=\columnwidth]{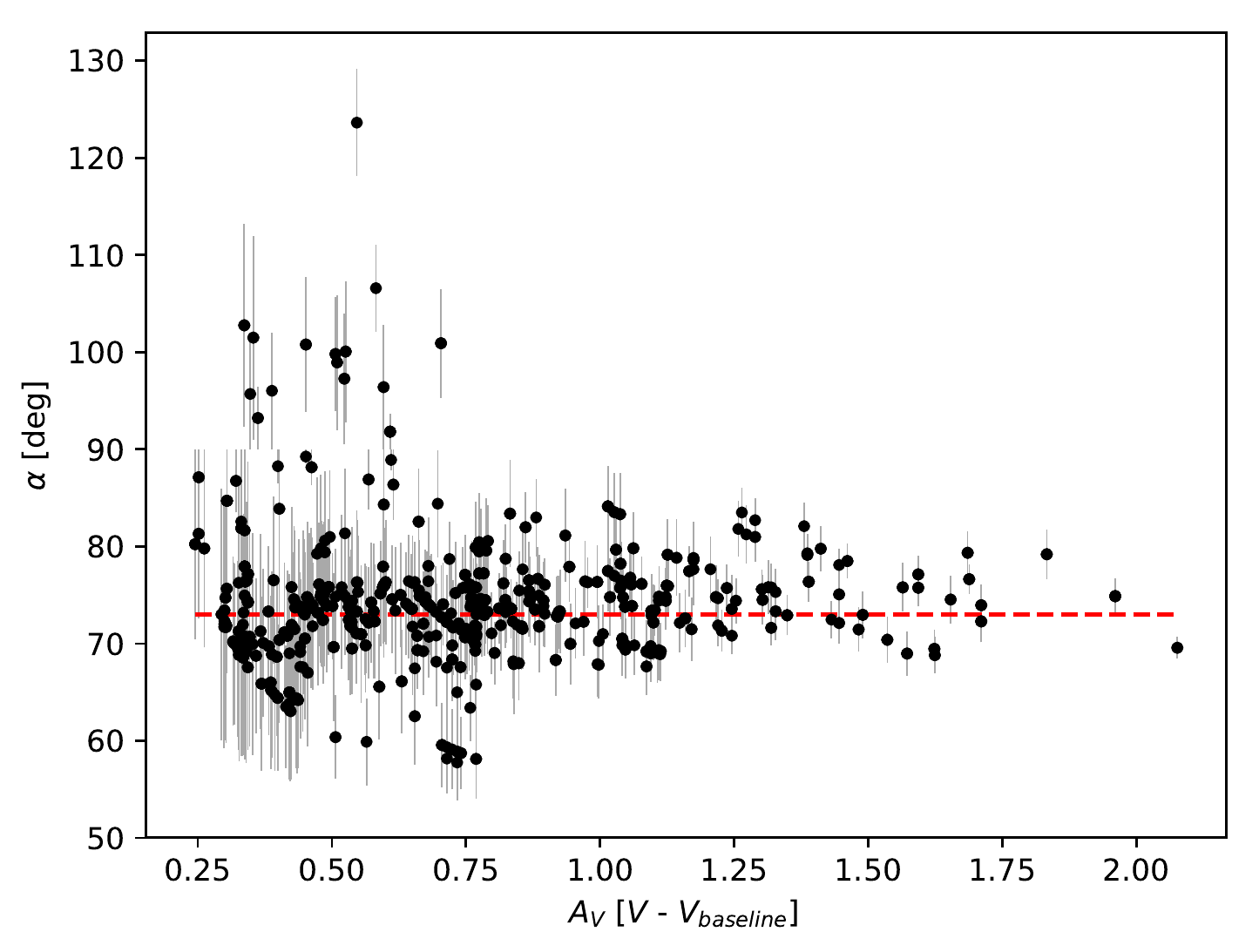}
\caption{\label{fig:alpha_av} $\alpha$ values in the V vs. V$-$I$_c$ parameter space during dips as a function of dip depth, i.e. $A_{\rm V}$. Points were removed within $5\sigma$ of the V-Band baseline of 15.29\,mag. The red dashed line indicates the median $\alpha$ value of (73\,$\pm$\,4)\,\dg.}
\end{figure}

In Fig.\,\ref{fig:alpha_av} we show the $\alpha$ values plotted against the depth of the occultation in the V-Band, i.e. the $A_{\rm V}$ of the occulting material. We only include points that correspond to measurements taken at an occultation depth of more than 5 times the nominal maximum uncertainty of the V-Band data, i.e. when the dip is deeper than 0.25\,mag. The median value for $\alpha$ is 73\dg\ with a scatter of 4\dg. This angle is systematically higher than can be expected for normal interstellar dust-grain-dominated scattering. Hence, the disk material in \vc\ does show signs of the onset of general grain growth in the higher column density material.

There is no significant systematic trend of $\alpha$ with $A_{\rm V}$, i.e. the occulting material has the same scattering properties (within the measurement uncertainties) independent of the column density of the material or time, for an extinction above 0.25\,mag in V. There are a few outlying points which are significantly further away from the other data. These points occur at `random' places in the light curve and thus are not caused by single, or multiple dip events with material with different scattering properties. Hence, the outliers are most likely caused by erroneous data where the initial magnitudes have been influenced by e.g. cosmic ray hits, or where the up to one day time gap between the V and I$_c$ observations cause unrealistic values for $\alpha$. As one can see in Fig.\,\ref{BVRI_lc}, significant brightness variations in the source can occasionally happen on time scales of less than one day.

As noted above, small uncertainties in the V and I$_c$ magnitudes can lead to large uncertainties in the $\alpha$ value. The V$-$I$_c$ colour is used in the calibration of the magnitudes (see Sect.\,\ref{Data:Correction_method}). Thus, if initially not exactly correct, the calibration will give systematically different magnitudes and hence might cause systematic and/or random offsets in $\alpha$. As evident from the light curves and Fig.\,\ref{fig:alpha}, the V$-$I$_c$ colour of the source varies by a maximum of about 0.5\,mag between the bright and faint state. To test the influence on the determination of $\alpha$ we have run the following experiment. We have rerun the entire calibration procedure with a systematically overestimated V$-$I$_c$ colour by 0.5\,mag (the worst case scenario). This of course has a systematic effect on the newly calibrated magnitudes of the star. Note that the colour dependence of the calibration is, however, rather weak (see Fig.\,\ref{fig:correction1}). We have then redetermined all the $\alpha$ values and compared them to the original numbers. We find that there are no significant changes to the scatter and uncertainties for $\alpha$ other than a systematic shift by 1\dg\ (to 72\dg) of the median value. Thus, our results for the scattering properties are robust, and the $\alpha$ values do not suffer any systematic uncertainties bigger than 1\dg\ caused by the calibration procedure.

A detailed look at Fig.\,\ref{fig:alpha} reveals that for dip depths of less than 0.25\,mag in V, the scattering behaviour of the material does not follow the same slope as determined for the high $A_{\rm V}$ material. We performed a linear regression of all low extinction points and found that in the V vs. V$-$I$_c$ diagram they are consistent with $R_{\rm V}$\,=\,5.0 scattering material. Thus, this suggests that the material in the occulting structure consists of low column density material with roughly ISM dust properties. Embedded in this envelope are denser, small-scale structures that are most likely composed of larger dust grains. The scattering properties of this material are consistent and do not change over time.

\section{Discussion}\label{discussion}

\subsection{Distance of \vc}

As \vc\ is projected onto IC\,5070, it is assumed to be part of this large star forming region, whose distance was estimated as about 600\,pc \citep{2008hsf1.book...36R, 2009ApJ...697..787G}. With the availability of Gaia\,DR2 we can reevaluate the distance of the source. \vc\ is identified as Gaia\,DR2\,2163139770169674112 with Gmag\,=\,15.0771\,mag, a proper motion of $-$0.539\,$\pm$\,0.377\,mas/yr in RA and $-$2.418\,$\pm$\,0.370\,mas/yr in DEC and a parallax of 0.4560\,$\pm$\,0.2377\,mas. This indicates a much larger distance of 2.2\,kpc with a very high uncertainty.

\begin{figure}
\centering
\includegraphics[width=\columnwidth]{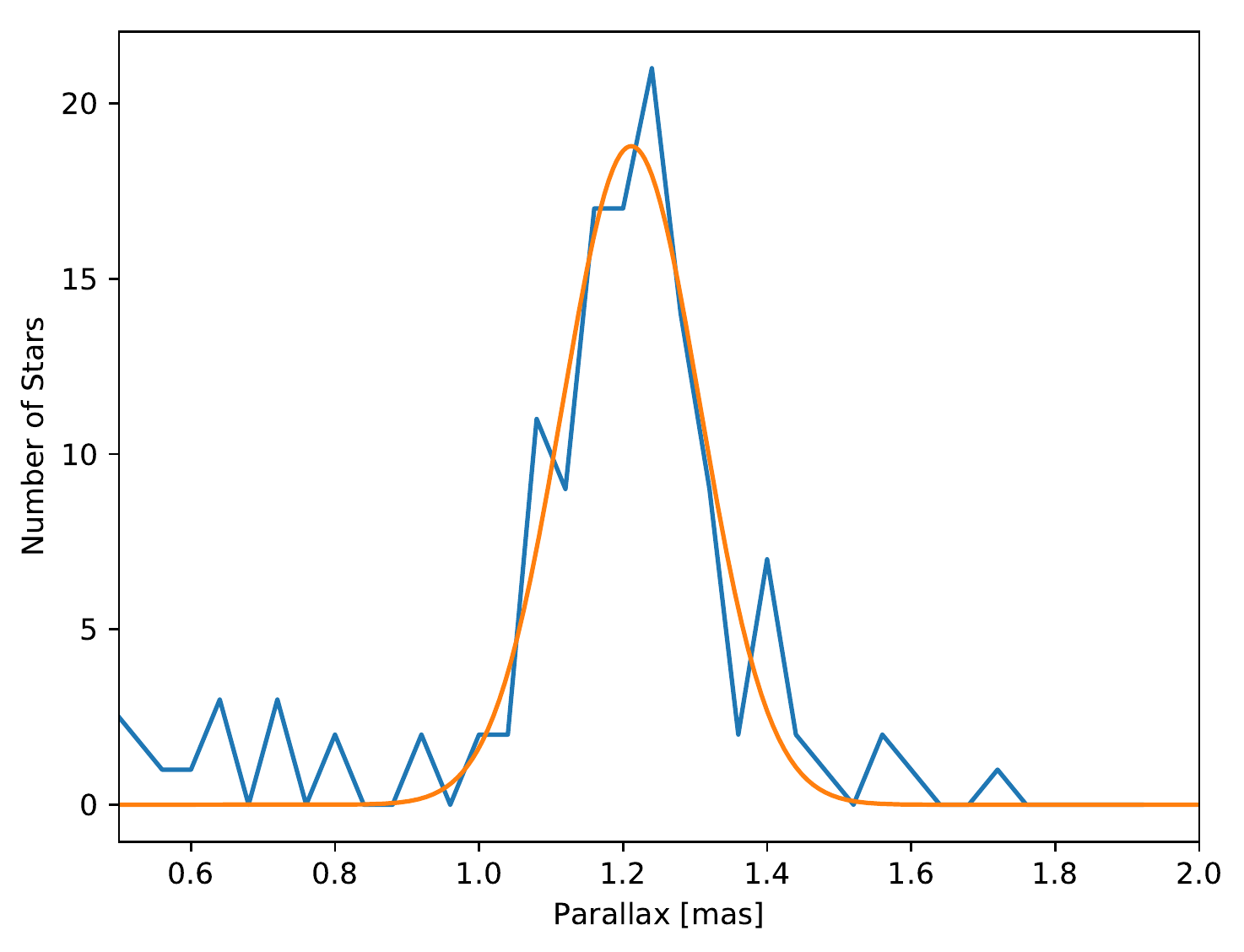}
\caption{\label{gaiaplxfig} Distribution of Gaia DR2 parallaxes in the IC\,5070 region with a parallax S/N of better than 10 and proper motions corresponding to IC\,5070 member stars. Over-plotted is a Gaussian fit to the data, indicating a mean parallax of 1.20\,mas with a scatter of 0.09\,mas.}
\end{figure}

In order to verify the distance we downloaded all Gaia\,DR2 sources within 20\arcmin\ of the object. If one selects only stars with a S/N of three or higher for the parallax, then the IC\,5070 region stands out as a cluster in proper motion vs. parallax plots (see Fig.\,D1 in Appendix\,D in the online supplementary material). The cluster is not identifiable in proper motion space alone, but IC\,5070 members have proper motions in the range from $-$0.35 to $-$2.00\,mas/yr in RA and from $-$2.00 to $-$5.00\,mas/yr in DEC. The cluster over-density seems to extend from parallaxes of 1.0 to 1.5\,mas. A Gaussian fit to the parallax distribution (including only objects with a S/N better than 10 for the parallax) gives a mean parallax of $\approx$1.2\,mas with a standard deviation of 0.09\,mas (see Fig.\,\ref{gaiaplxfig}). If we apply the suggested zero point correction of $-$0.0523\,mas \citep{2019MNRAS.tmp.2167L}, this indicates a distance for IC\,5070 of the order of 870\,$^{+70}_{-55}$\,pc, which we will adopt throughout this paper for this region and \vc. 

To understand why the parallax of \vc\ differs from the cluster value and also has a very low S/N ratio, we determine the typical parallax error of all stars in the same field, with proper motions indicating they are potentially part of IC\,5070 and with a Gmag value within 1\,mag of \vc\ -- there are 49 such stars. For those stars, the median parallax error is 0.046\,mas/yr with a scatter of 0.021\,mas/yr. Hence the parallax error of 0.2377\,mas/yr represents a $\simeq$9\,$\sigma$ outlier. The source does not stand out in terms of the number of observations when compared to stars in the same field, thus there is no general issue with crowding in this field. However, the astrometric excess noise is much higher for this object, as well as all other quality indicators. This indicates that the object's parallax measurements could be influenced by: i) crowding for this source, caused by the scan angles used so far; ii) the red colour and variability; iii) the source being a binary or that the object's photo-centre position changes due to the variability. Thus, the Gaia DR2 parallax of the source cannot be trusted and we use the above determined IC\,5070 distance. We will briefly discuss the binary interpretation in Sect.\,\ref{binarity}. There is of course a possibility that the source is indeed a background object.

\subsection{Potential binarity of \vc}\label{binarity}

The large Gaia parallax error indicates that the star could be a binary, unresolved in our optical images. We have investigated the highest resolution imaging data available of this source, which comes from the UKIDSS GPS survey (UGPS, \citealt{2008MNRAS.391..136L}). The NIR images reveal three faint, red NIR objects around the star, at separations of 5.6, 2.7 and 2.5\arcsec. However they are fainter by 2.5, 4.5, 7.0\,mag in K, and fainter in J by 3.6 and 5.6\,mag -- the closest source has no J detection. \vc\ itself has the maximum possible value of  {\tt pstar}\,=\,$+$0.999999, indicating its PSF is consistent with a single unresolved source. The ellipticities in all three filters are between 0.07 and 0.10, indicating the source is not elongated at a level above 0.1 of the full width half maximum of the PSF. The seeing in the images is about 0.56\arcsec. Thus, any companion not detectable in UGPS would have to be closer than 0.05\arcsec\ to the source. At our adopted distance of 870\,pc, this corresponds to about 44\,AU maximum separation.

However, the periodic variability of the source is not in agreement with a wide (30\,--\,40\,AU) binary with comparable luminosities. The period of about 31.5\,d indicates material at sub-AU distance from the star. The depth of some occultations reaches up to 2\,mag in V. For equal luminosity objects with one partner being occulted, one should not obtain such deep dips. The maximum dip should only be 0.75\,mag. If the stars are unequal luminosity then the dips could be deeper. In those cases the colour of the occulted system should eventually be dominated by the colour of the fainter object. However, our analysis in Sect.\,\ref{Analysis:Dust_Scattering_Properties} has shown that there is no indication that the dimming is caused by anything other than interstellar dust grains of homogeneous scattering properties. In particular, for deep dips there is no deviation in the changes of colour from the prediction of extinction from dust. Thus, it seems highly unlikely that if there is a wide companion it is contributing a sizeable fraction of the system luminosity. Any closer companion would most likely disturb or remove the sub-AU material in the accretion disk that we see in the system. Thus, \vc\ is most likely single.

\subsection{Literature NIR and MIR data for \vc}

\begin{table}
\caption{\label{vc_phot_table} Summary of literature NIR/MIR data for \vc. We list the Julian date of the observations, the filter/band the observations are in, the magnitude and uncertainty, the survey the data is taken from, and the phase and its uncertainty for the observation. The top part of the table lists data with available observing date, while in the bottom half time-averaged WISE and Spitzer photometry is listed for completeness. The NEOWISE data is averaged over the 1--3 days for each sequence of visits. The phase has been determined following the period determined in Sect.\,\ref{period}. The uncertainty in the phase has been propagated from the nominal 15\,min uncertainty in the period. All magnitudes measured within 0.15 of phase zero are highlighted in bold, to indicate measurements that are most likely less influenced by the unknown variable circumstellar extinction.}
\renewcommand{\tabcolsep}{3pt}
\begin{tabular}{c|c|c|c|c|c|c}
\hline
Date [JD] & Filter & mag & $\Delta$mag & Survey & Phase & $\Delta$Phase \\
\hline
2451707.8263 & J & 11.850 & 0.023 & 2MASS & 0.206 & 0.078 \\
2451707.8263 & H & 11.008 & 0.032 & 2MASS & 0.206 & 0.078 \\
2451707.8263 & K & 10.637 & 0.026 & 2MASS & 0.206 & 0.078 \\
2457589.9499 & J & 12.0472 & 0.0006 & UGPS & 0.256 & 0.013 \\
2457589.9555 & H & 11.0429 & 0.0004 & UGPS & 0.256 & 0.013 \\
2457589.9597 & K & 10.7314 & 0.0006 & UGPS & 0.256 & 0.013 \\
2455861.7314 & K & 10.8222 & 0.0006 & UGPS & 0.299 & 0.032 \\
2456809.4858 & W1 & 9.9188 & 0.0622 & NEOWISE & 0.437 & 0.021 \\
2456809.4858 & W2 & 9.4490 & 0.0258 & NEOWISE & 0.437 & 0.021 \\
2456987.9896 & W1 & {\bf 10.225} & 0.118 & NEOWISE & 0.113 & 0.019 \\
2456987.9896 & W2 & {\bf 9.6490} & 0.0705 & NEOWISE & 0.113 & 0.019 \\
2457171.2939 & W1 & {\bf 10.057} & 0.03 & NEOWISE & 0.942 & 0.017 \\
2457171.2939 & W2 & {\bf 9.5810} & 0.0243 & NEOWISE & 0.942 & 0.017 \\
2457346.0040 & W1 & 10.061 & 0.015 & NEOWISE & 0.498 & 0.015 \\
2457346.0040 & W2 & 9.5360 & 0.0346 & NEOWISE & 0.498 & 0.015 \\
2457349.4515 & W1 & 10.006 & 0.022 & NEOWISE & 0.608 & 0.015 \\
2457349.4515 & W2 & 9.5000 & 0.0182 & NEOWISE & 0.608 & 0.015 \\
2457535.4401 & W1 & 10.191 & 0.057 & NEOWISE & 0.522 & 0.013 \\
2457535.4401 & W2 & 9.6460 & 0.0279 & NEOWISE & 0.522 & 0.013 \\
2457538.7336 & W1 & 10.040 & 0.032 & NEOWISE & 0.627 & 0.013 \\
2457538.7336 & W2 & 9.4930 & 0.0171 & NEOWISE & 0.627 & 0.013 \\
2457708.6578 & W1 & {\bf 9.9273} & 0.0701 & NEOWISE & 0.030 & 0.011 \\
2457708.6578 & W2 & {\bf 9.4918} & 0.0551 & NEOWISE & 0.030 & 0.011 \\
2457902.4970 & W1 & 9.9078 & 0.0227 & NEOWISE & 0.194 & 0.009 \\
2457902.4970 & W2 & 9.4613 & 0.0146 & NEOWISE & 0.194 & 0.009 \\
2458069.3243 & W1 & 9.9929 & 0.0441 & NEOWISE & 0.500 & 0.007 \\
2458069.3243 & W2 & 9.4577 & 0.0568 & NEOWISE & 0.500 & 0.007 \\
2458266.6522 & W1 & 9.8730 & 0.0221 & NEOWISE & 0.774 & 0.005 \\
2458266.6522 & W2 & 9.3924 & 0.0196 & NEOWISE & 0.774 & 0.005 \\
2458429.9023 & W1 & {\bf 9.8158} & 0.0257 & NEOWISE & 0.966 & 0.003 \\
2458429.9023 & W2 & {\bf 9.3902} & 0.0408 & NEOWISE & 0.966 & 0.003 \\
\hline
  & W1 & 9.929 & 0.023 & WISE all sky & & \\
  & W2 & 9.412 & 0.021 & WISE all sky & & \\
  & W3 & 7.300 & 0.028 & WISE all sky & & \\
  & W4 & 5.209 & 0.036 & WISE all sky & & \\
  & W1 & 9.937 & 0.023 & AllWISE & & \\
  & W2 & 9.443 & 0.021 & AllWISE & & \\
  & W3 & 7.427 & 0.029 & AllWISE & & \\
  & W4 & 5.163 & 0.037 & AllWISE & & \\
  & 3.6 & 9.956 & & IRAC & & \\
  & 4.5 & 9.581 & & IRAC & & \\
  & 5.8 & 9.036 & & IRAC & & \\
  & 8.0 & 8.211 & & IRAC & & \\
  & 24  & 5.387 & & MIPS & & \\
\hline
\end{tabular}
\end{table}

In order to classify the evolutionary stage of the source, we have collected literature near- and mid-infrared photometry of the source. This data is summarised in Table\,\ref{vc_phot_table}. We have extracted the NIR photometry and observing dates from 2MASS \citep{2006AJ....131.1163S}, UGPS \citep{2008MNRAS.391..136L} data release DR11, as well as the MIR observations from NEOWISE \citep{2011ApJ...731...53M,2014ApJ...792...30M} in the W1- and W2-bands from the WISE satellite \citep{2010AJ....140.1868W}. For the latter we have averaged all the measurements taken over the usually one to three day repeated visits for the source and used the RMS as the uncertainty. The individual NEOWISE visits are too short to observe any changes related to the dipping behaviour. For completeness we have further added WISE photometry released from the WISE all sky catalogue \citep{2012yCat.2311....0C} and the ALLWISE catalogue \citep{2013yCat.2328....0C}, as well as the Spitzer IRAC and MIPS photometry presented in \citet{2009ApJ...697..787G} and \citet{2011ApJS..193...25R}. 

As \vc\ is variable, only data with a known observing date should be used for classification purposes (top part of the table). This is particularly important for the shorter wavelength data where the extinction is higher. Since we do not know how high the extinction was during past dipping events (if we do not have contemporary optical data), we should only use photometry taken during the phase of the light curve where the object is most likely at its maximum brightness. Considering the folded light curves in Figs.\,C1, C2 in the online supplementary material and the period of 31.447\,$\pm$\,0.011\,d determined in Sect.\,\ref{period}, we have estimated the phase and its uncertainty for all NIR and MIR data with known observing dates. The star is considered to be in its bright state if the phase is within 0.15 of zero/one. These measurements are highlighted in Table\,\ref{vc_phot_table} in bold face.

\subsection{Evolutionary Stage of \vc}\label{evol_stage}

In order to estimate the evolutionary status of the source we need to estimate the NIR and MIR magnitudes, not influenced by the variable circumstellar extinction. As is evident from Table\,\ref{vc_phot_table}, none of the available NIR data has been taken near the nominal bright state of the source. The variations of the 4 potential observations near the maximum brightness in W1/W2 are of the order of several tenths of magnitudes. As evident in the folded light curves in Figs.\,C1, C2 in the online supplementary material, the source can vary quite significantly even close to the nominally bright phase. Hence we will use the brightest of the JHK/W1/W2 magnitudes for the classification, but note that the resulting colours are potentially uncertain by a few tenths of a magnitude. Similarly we choose the brightest of the W3/W4 measurements from the WISE all sky and ALLWISE catalogues. 

Thus, we find H$-$K\,=\,0.37\,mag, W1$-$W2\,=\,0.43\,mag, W3$-$W4\,=\,2.1\,mag. According to \citet{2014ApJ...791..131K} this places the source at the blue end of the classification as a CTTS in the W1$-$W2 vs. H$-$K diagram, or just outside, considering the potential uncertainties in the colours. In the W3$-$W4 vs. W1$-$W2 diagram the source sits on the border line between CTTS and transition disk objects. Using the WISE data and following \citet{2013Ap&SS.344..175M} we determine the slope ($\alpha_{\rm SED}$) of the spectral energy distribution as $-$0.67. This places the source in the CTTS category.

Our spectra from LCOGT (see Sect.\,\ref{lco_spectra}) allow us to use the \ha\ equivalent width to classify the source. In the six spectra obtained the equivalent width of the \ha\ line varies between $-$3.2\,\AA\ and $-$8.6\,\AA. The most widely used dividing line between CTTS and WTTS is $-$10\,\AA\ (e.g. \citet{1998AJ....115..351M}). However, this is not a fixed value due to the variability of the line. 

The source is, however, still accreting and all available accretion rate indicators show variability. The \ha\ EW (see Fig.\,\ref{halinefig} and Table\,\ref{tab:ha_ew}) is clearly variable by at least a factor of two. Furthermore, the R$_c-$\ha\ magnitudes also vary by at least one magnitude, i.e. more than factor of two. Finally, as can be seen in Fig.\,\ref{fig:fingerprints}, the U-band is highly variable by at least $\pm$\,2\,mag even on very short (hours) time scales. Since the U-band excess is generally acknowledged as one of the best tracers of accretion rate, both empirically \citep{1998ApJ...492..323G, 2008ApJ...681..594H} and theoretically \citep{1998ApJ...509..802C}, this indicates strongly variable accretion in \vc.

Thus, we conclude that \vc\ is most likely a CTTS, with (currently) low, but variable accretion rate, and is potentially at the start of the transition into a WTTS or transition disk object.

\subsection{The Nature of \vc}

Our analysis presented in Sects.\,\ref{period} -- \ref{evol_stage} shows that \vc\ is showing quasi-periodic occultation events of dust in the inner accretion disk at 0.15\,AU from the source. The occulting material is made of material with ISM properties at low $A_{\rm V}$ and shows grain growth at high column densities. The source is still accreting and is at the borderline between CTTS/WTTS or transition disk objects. Below we briefly discuss three possible explanations for the nature of the source, in order of decreasing probability: i) a protoplanet-induced disk warp; ii) A magnetically induced disk warp; iii) The Hill sphere of an accreting protoplanet. Attempting to verify these explanations will require high resolution spectroscopy over several orbital periods which we strongly encourage.

\subsubsection{Protoplanet induced Disk Warp}
\label{Discussion:Protoplanet-induced_Disk_Warp}

Recent Atacama Large Millimeter/sub-millimeter Array (ALMA) observations have discovered several warped protostellar disk systems \citep{2019Natur.565..206S}. For some of these systems, observations rule out the influence of a secondary star, potentially suggesting unseen protoplanets to be the cause of the warping \citep{2018MNRAS.481...20N}. A protoplanet that is capable of driving warping features in the disk would be required to maintain an orbit that is inclined to the disk plane over long time scales. This is, however, in contradiction with current planet formation theory which assumes a flat protoplanetary disk. A co-planar protoplanet could become inclined or eccentric during or after formation, whereby planet-planet interactions are able to move a protoplanet to an inclined orbit \citep{2008ApJ...678..498N}. Measurements of disk inclination in objects such as TW\,Hya hint at a small warp or misalignment, at distances less than 1\,AU for even minor deviations in inclination from the disk plane \citep{2004ApJ...616L..11Q, 2008ApJ...684.1323P, 2011ApJ...727...85H}. The azimuthal surface brightness asymmetry moving with constant angular velocity in TW\,Hya has also been attributed to a planet-induced warp in the inner disk of that system \citep{2017ApJ...835..205D}. Simulations by e.g. \citet{2019MNRAS.484.4951N} show that misaligned planetary orbits are indeed capable of generating such warps.

In Sect.\,\ref{Analysis:Dust_Scattering_Properties}, we have shown that the material in the occulting structure appears to consist of low column density material, with roughly ISM dust properties. Embedded in this envelope are denser, small-scale structures that are most likely composed of larger dust grains. This, combined with the fact that the source is situated in a $\sim$\,3\,Myr star forming region \citep{2008hsf1.book..308B} suggests that planet formation should be ongoing. The source exhibits dips across the majority of the observed periods, demonstrating semi-stability in the occulting structure. Orbital resonance between the disk and an inclined orbiting protoplanet could cause a build up of material at the distances observed. Such a protoplanet could be closer to, or further from the source in relation to the observed orbiting structure.

\subsubsection{Magnetically induced Disk Warp}
\label{Discussion:Magnetically-induced_Disk_Warp}

The observed regular dips of \vc\ suggests it could be an AA\,Tau type source. However, the period of 31.447\,d is particularly long compared to other AA\,Tau type objects. Average rotational periods of CTTSs are 7.3\,d \citep{1995A&A...299...89B}, with AA\,Tau itself having a period of 8.2\,d \citep{2003A&A...409..169B}. Hence, typical values range between 5\,--\,10\,d. 

In AA\,Tau type objects, the periodic dips are caused by a warp of the inner disk due to a misalignment between the rotation axes of both the disk and the star. The warp is therefore located at the co-rotation radius of the disk \citep{1999A&A...349..619B, 2007A&A...463.1017B, 2014AJ....147...82C, 2015A&A...577A..11M}. Assuming a dipole magnetic field aligned with the star's rotation axis, material in the disk would be magnetically displaced from the disk's plane and into the line of sight. If \vc\ is seen at a high inclination, the inner disk warp will occult the stellar photosphere periodically, causing flux dips in the star's light curve. This obscuration could then explain such dimming behaviour as seen in Sect.\,\ref{period}. However, for \vc\ to have a warped disk due to misalignment, the source would have to be a very slow rotator. This would enable the co-rotation radius, and hence the disk warp, to occur further out than is seen for AA\,Tau. Note that 30\,d or slower rotation periods in young clusters are extremely rare (see e.g. \citet{2014prpl.conf..433B}), and that the magnetic field would have to be very strong and reach far out from the star to generate a warp at the observed distance of 0.15\,AU.

\subsubsection{Hill Sphere of Accreting Protoplanet}
\label{Discussion:Hill_Sphere_of_Accreting_Protoplanet}

It is also conceivable that the observed quasi-periodic variability observed in \vc\ comes from material held in orbit not by a magnetic structure, but by a more massive object. A protoplanet located at distances close to the central star within the disk will gravitationally exert influence on the disk material around it. The occultations observed could then be caused by material in the Hill sphere around this protoplanet. The Hill sphere presents as an oblate spheroid of material, gravitationally bound to the protoplanet within the disk. Modelling by \citet{2005A&A...433..247P} suggests that for a protoplanetary mass of $\sim$\,0.1\,M$_\text{J}$, a rapid accretion phase begins. This is a similar mass to that for which either significant perturbation to the protoplanetary disk through local mass accretion or disk-planet interaction begins \citep{2000MNRAS.318...18N}.

On its own, structures of the mass observed in Sect.\,\ref{Analysis:column_density} will not survive for more than one orbit due to shearing. The periodic dips in \vc\ are stable in phase, if not in structure, for more than 40 orbits. We do consider this explanation as the least likely, since the duration of dips observed for the source span more than half the period in many cases and the deepest part of the occultations move significantly in phase. This is not in good agreement with predictions for this scenario as the size of the Hill sphere will not exceed 0.1\,--\,0.2 of the orbital circumference, i.e. the maximum duration of the dips should be no longer than approximately 6\,d. 

\section{Conclusions}

In this work we present results from our long term, high cadence, multifilter optical monitoring of young, nearby star clusters and star forming regions obtained as part of the \hc\ program \citep{2018MNRAS.478.5091F}. The data set consists of images taken with a wide variety of telescopes, detectors and filters, with images also obtained under a variety of observing (light pollution) and weather conditions (from photometric to thin or even thick cirrus). The availability of a large amount of images ($\sim$\,3300) obtained over a long period ($\sim$\,4\,yr) has enabled us to develop an internal photometric calibration procedure to remove systematic magnitude offsets due to colour terms in the photometry. This utilises non-variable sources that are identifiable in the data. The procedure achieves for the star \vc\ (V\,$\approx$\,15.5\,mag) a median uncertainty in the photometry of 0.02\,mag in all broadband filters (U, B, V, R$_c$ and I$_c$) and 0.09\,mag in \ha. 

Our analysis of the Orion variable \vc\ in the Pelican Nebula (IC\,5070) shows that it is actually a quasi-periodic dipper, most likely caused by occultations of the star by material in a warped inner disk. A mean period of 31.447\,$\pm$ 0.011\,d was determined in the V, R$_c$ and I$_c$ filters using a Lomb-Scargle periodogram. This variability is quasi-periodic, with varying depth from orbit to orbit. But the period is stable over the 4\,yr of continuous observations available from \hc. The B, V, R$_c$ and I$_c$ light curves follow an extremely similar pattern on all time scales, caused by the variations in the column density along the line of sight. However, the U and R$_c-$\ha\ data do not follow the same variations and in part show more extreme variation on time scales of hours to days. This suggests that the variability in those accretion rate indicators, is indeed dominated by changes in the mass accretion rate in the system. The U-band data in particular shows variability of up to a factor of 10 on time scales of hours.

We investigate the behaviour of the source in the V vs. V$-$I$_c$ parameter space. We find that the material in the occulting structure consists in part of low column density material ($A_{\rm V}$), with roughly ISM dust properties ($R_{\rm V}$). Embedded in this envelope are denser, higher column density, small-scale structures, that are most likely composed of dust grains that are larger than in the ISM. The scattering properties of this material are consistent and do not change over time, indicating a mixing of material in the disk before it is moved into the observable structure. 

An analysis of the column density distribution of consecutive dips, i.e. the structure function of the material, has been performed. We find no significant or systematic trends in the structure function. This suggests that the material in the line of sight is moving in and out of the occulting structure on time scales of the order of, or shorter than the period of the occultations ($\sim$\,30\,d). We determine the amount of mass in the occulting structure for each orbit and find it to vary by up to a factor of 10 for both mass increase and decrease. Thus, the mass flow rate through the occulting structure varies by typically a factor of a few when averaged over one period. This converts to a minimum mass accretion rate through the occulting structure onto \vc\ of the order of 10$^{-10}$\,M$_\odot$/yr. This is consistent with low levels of accretion as seen in other T\,Tauri stars.

The Gaia\,DR2 parallax of \vc\ is highly uncertain, most likely due to the variability and red colour. We hence use Gaia\,DR2 data from the stars in the same field to determine an accurate distance to IC\,5070. We find a distance of 870\,$^{+70}_{-55}$\,pc for the Pelican Nebula. 

The NIR and MIR data taken from literature, our measured \ha\ equivalent widths, as well as the U-band and R$_c-$\ha\ variability, indicates that \vc\ is most likely a CTTS, with a currently low, but variable accretion rate, that is potentially at the start of the transition into a WTTS or transition disk object.

Our data and analysis show that \vc\ seems to have a warped inner accretion disk which enables us to observe the structure periodically. Assuming the central star is about half a solar mass, this places the orbiting material at a distance of $\sim$\,0.15\,AU from the central star. The most likely interpretation for the cause of the warp is a protoplanetary object with an inclined orbit, located somewhere in the inner accretion disk. There is also the possibility that the warp is magnetically induced similar to AA\,Tau-like object. However, this would require \vc\ to be a very slow rotator. Finally, we also study the possibility that the orbiting structures are associated with the Hill sphere of an accreting protoplanet. However, the long duration of the observed occultations seem the refute this explanation. Longer term high resolution spectroscopy of the object is encouraged to identify the true nature of \vc.

\section*{Acknowledgements}


We would like to thank all contributors of data (even if they decided not to have their name on the author list of the paper) for their efforts towards the success of the \hc\ project.
We thank Nigel Hambly for advise on interpreting Gaia data uncertainties.
J.J.\,Evitts acknowledges a joint University of Kent and STFC scholarship (ST/S505456/1). 
A.\,Scholz acknowledges support through STFC grant ST/R000824/1. 
J.\,Campbell-White acknowledges the studentship provided by the University of Kent. 
S.V.\,Makin acknowledges an SFTC scholarship (1482158). 
K.\,Wiersema thanks Ray Mc\,Erlean and Dipali Thanki for technical support of the UL50 operations.
K.\,Wiersema acknowledges funding by STFC. K.\, Wiersema acknowledges support from Royal Society Research Grant  RG170230 (PI: R.\,Starling).
D.\,Mo\'{z}dzierski acknowledges the support from NCN grant 2016/21/B/ST9/01126.
T.\,Killestein thanks the OpenScience Observatories team at the Open University for allowing the use of and operating the COAST facility.
This work makes use of observations from the LCOGT network.
This work was supported by the Slovak Research and Development Agency under the contract No. APVV-15-0458.
This publication makes use of data products from the Near-Earth Object Wide-field Infrared Survey Explorer (NEOWISE), which is a project of the Jet Propulsion Laboratory/California Institute of Technology. NEOWISE is funded by the National Aeronautics and Space Administration.
This research has made use of the SIMBAD database, operated at CDS, Strasbourg, France.
We acknowledge the use of the Cambridge Photometric Calibration Server (http://gsaweb.ast.cam.ac.uk/followup), developed and maintained by Lukasz Wyrzykowski, Sergey Koposov, Arancha Delgado, Pawel Zielinski, funded by the European Union's Horizon 2020 research and innovation programme under grant agreement No 730890 (OPTICON).


\bibliographystyle{mnras}
\bibliography{bibliography}


\clearpage\newpage

\appendix

\section{Description of Observatories and Data Reduction}\label{observatory_descriptions}

\subsection{Description of Amateur Observatories and Data}

In this section we describe the equipment used by the various amateur astronomers. Each subsection also contains a basic description of the observing and data reduction procedures. In order to protect the privacy and for safety reasons the exact locations of the amateur observatories are not published. We have an online map of the observatory locations\footnote{\tt \href{https://www.google.co.uk/maps/@51,0,6000000m/data=!3m1!1e3!4m2!6m1!1s10hvfem7JcCjRodjofvwzGZBUrSJsJvpw}{\hc\ Observing sites}} that shows their world-wide distribution. For the same reason the markers for the observatories are usually placed on a nearby ($\sim$\,1\,km radius) road junction or landmark.

\subsubsection{Amanecer de Arrakis Observatory}

The observatory is located south of Seville, Andalusia, Spain. It uses an SC 8\arcsec\ telescope with an ICX285AL CCD chip. The optics results in a 1.92\arcsec\ per pixel resolution at 2\,$\times$\,2 binning. A filter set with B, V, R$_{\rm c}$ and I$_{\rm c}$ filters is available. Observations are typically taken with 50\,s\,--\,60\,s exposure time. Science frames are reduced with bias, dark and flat frames. MaximDL software is used for all data and calibration frame acquisition. On average about 3 or 4 nights per week are used for observations. Observing conditions are clear for most of the year.

\subsubsection{Anne and Lou Observatory}

The observatory is located North of Roanoke, Virginia, USA in a roll off roof shed. It has a pier mounted AstroPhysics Mach1GTO mount and a GSO 250mm f/10 Ritchey-Chr\'{e}tien carbon fiber tube telescope. It is further equipped with an AstroPhysics CCDT67 telecompressor, an Atik 428EX monochrome CCD with a Sony ICX674 Sensor. The optics provides a 22\arcmin\,$\times$\,17\arcmin\ field of view at a 0.7\arcsec\ per pixel plate scale. The Atik filter wheel is equipped with Astrodon B, V and I$_{\rm c}$ filters. Depending on the target, typically 30\,s and 60\,s exposures are obtained. The images are bias, dark and flat field corrected and stacked, using Astroart 6 software with ASCOM drivers.

\subsubsection{AstroLAB IRIS Observatory}

The public observatory is located in Zillebeke, South of Ypres in Belgium and host a 684 mm aperture Keller F4.1 Newtonian New Multi-Purpose Telescope (NMPT). It utilises a Santa Barbara Instrument Group (SBIG) STL 6303E CCD operated at -20\dg C. A 4\arcsec\ Wynne corrector feeds the CCD at a final focal ratio of 4.39, providing a nominal field of view of 20\arcmin\,$\times$\,30\arcmin. The 9\,$\mu$m physical pixels project to 0.62\arcsec\ and are read out binned to 3$\times$3 pixels, i.e. 1.86\arcsec\ per combined pixel. The filter wheel is equipped with B, V, and R filters from Astrodon Photometrics. Typical exposures times are 20\,s for \hc\ imaging. Dark and bias correction as well as stacking are done using Lesvephotometry\footnote{\tt \href{http://www.dppobservatory.net/AstroPrograms/Software4VSObservers.php}{Lesvephotometry}}. No flat-field correction is applied to the data. 

\subsubsection{Belako Observatory}

The observatory is located near Muniga (North-East of Bilbao) in Spain. It uses a Meade LX200 SCT 10\arcsec\ telescope (254mm diameter) with ACF GPS f10 (2540mm focal length). The CCD is a SBIG ST2000-XM double chip with a 7.4\,$\mu$m pixel size and 1600\,$\times$\,1200\,pixels. The Mead 0.63 focal reducer results in a 0.84\arcsec\ per pixel resolution. The filter wheel, a SBIG CFW-10, is equipped with optical trichromia R, G, B, L and \ha\ filters from SBIG, as well as photometric B, V, Rc and Ic Johnson Coussin filters. Typical seeing is 2\arcsec\,--\,5\arcsec\ and exposure times up to 3\,min are used for observations. Data capture and reduction including stacking are done using MaximDL, KStars \& INDI\footnote{\tt \href{https://www.indilib.org/}{INDI Software}}, and Deep Sky Stacker\footnote{\tt \href{http://deepskystacker.free.fr/english/index.html}{Deep Sky Stacker}}.

\subsubsection{Bowerhill Observatory}\label{obs_bowerhill}

The Observatory is located East of Bath in the UK. It uses a Skywatcher Startravel telescope with 102\,mm aperture, f/4.9 and a Canon 600D DSLR camera. The field of view is about 2\dg\ with 1.9\arcsec\ per pixel resolution. This telescope is placed on a Skywatcher EQ 5 equatorial mount. Typically up to 120\,$\times$\,60\,s exposures are obtained at ISO 800 depending on darkness and weather conditions during the run. Data reduction is carried out using the IRIS software. The raw files from the DSLR are first decoded, then reduced with dark and flat frames and stacked.

\subsubsection{Cal Maciarol m\`{o}dul 8 Observatory}

The observatory is located in Parc Astron\`{o}mic del Montsec, Starlight reserve in \`{A}ger,  Catalonia, Spain. The typical sky brightness is 21.5\,mag/square arcsec towards the zenith and the typical seeing is 2\arcsec\,--\,4\arcsec\ at the observatory's location. It uses a Meade LX200/R 12\arcsec\ (305\,mm aperture), f/8.9 (2720\,mm focal length) telescope and a full-frame Moravian G9000 (KAF-09000 sensor, 36.7\,mm\,$\times$\,36.7\,mm, 12\,$\mu$m pixel size) CCD camera. The field of view is 46.35\arcmin\,$\times$\,46.35\arcmin\ and the resolution per pixel is 0.91\arcsec\ at 1\,$\times$\,1 binning. The camera is equipped with a Moravian EFW-4L-7 filter wheel (up to 7 filters) with Astrodon 52\,mm g\arcmin\, r\arcmin\, i\arcmin\ Sloan and V Johnson-Cousins filters placed into it. Typical \hc\ observations are taken as 3\,--\,5 frames with 300\,s exposures for each filter. The images are calibrated with bias, dark and twilight flatfield frames (either evening or morning twilight depending on conditions). Image capture is performed with the KStars/Ekos software and data reduction and stacking with MaxIm\,DL.

\subsubsection{CBA Extremadura Observatory}

The Center for Backyard Astrophysics (CBA) Extremadura Observatory has an excellent location in a dry and dark part of Spain, just North of Fregenal de la Sierra. The site has on average about 280 clear nights per year. It is part of the e-EyE complex\footnote{\tt \href{www.entreencinasyestrellas.es}{e-EyE observatory complex}}, which is the largest telescope hosting place in Europe, providing high-end modules of individual observatories allowing astronomers from all over the world to remotely control their telescopes. The telescope is a 0.40\,m f/5.1 Newton with a KAF-16200, ASA DDM-85 mount and a Starlight Xpress SX Trius SX-46 CCD camera. The filter wheel houses Clear, B, V, R and I filters. Observations are typically done with 3\,$\times$\,3 binning at a pixel scale of 1.82\arcsec\ per pixel and a field of view of 46\arcmin\,$\times$\,37\arcmin. A typical observing sequence for \hc\ targets consists of 3\,$\times$\,120\,s exposures in B, V and R. Post processing is done using Lesvephotometry.

\subsubsection{Chicharronian Tres Cantos Observatory}

The observatory is situated about 15\,miles North of Madrid, Spain. It has a 254\,mm aperture f4.8 Skywatcher Newtonian telescope mounted on a Skywatcher EQ6 mount. The system is controlled by Astroberry Kstars and EKOS for scheduling and imaging. It uses an SBIG ST8XME mono CCD camera, cooled to 35\dg C below the ambient temperature. The pixel scale is 1.55\arcsec/pixel with a field of view of 39\arcmin\,$\times$\,28\arcmin.  The telescope is also equipped with a SBIG CFW9 motorised filter wheel with Baader V, J-C and RGB filters. Typically 5\,--\,20 images are stacked, with individual exposure times ranging from 120\,s to 300\,s, depending on object, filter and sky conditions.

\subsubsection{Clanfield Observatory}

The observatory is located in Clanfield, North of Portsmouth, UK and is run by the Hampshire Astronomical Group\footnote{\tt \href{www.hantsastro.org.uk}{Hampshire Astronomical Group}}. There are several telescopes in the observatory that have been used for \hc\ imaging. i) An Astro-Physics 7-inch 'Starfire' F9 Apochromatic refractor with a Starlight Xpress SX-46 CCD Camera, Baader L, R, G, B, \ha, SII, and OIII filters, Plate Scale: 0.773\arcsec/pixel and a field of view of 58.53\arcmin\,$\times$\,46.93\arcmin. ii) A 24\arcsec\ (612\,mm) f7.9 Ritchey-Chr\'{e}tien reflector with a Moravian G4-9000 CCD Camera, Baader L, \ha, SII, and OIII filters, Astrodon R, V and B photometric filters, Plate Scale: 0.515\arcsec/pixel and a field of view of 26.21\arcmin\,$\times$\,26.21\arcmin. iii) A Meade LX200 12 inch Schmidt-Cassegrain with f10 to f6.3 focal reducer and Starlight Xpress SX-46 CCD Camera, Baader L, R, G, B, \ha, SII, and OIII filters, Plate Scale: 0.644\arcsec/pixel and a field of view of 48.72\arcmin\,$\times$\,39.06\arcmin. iv) A 12\arcsec\ (0.305\,m) Newtonian Guided Reflector using an Atair Hypercam 183C 20mp Cooled Colour CMOS Camera with a plate scale of 0.26\arcsec/pixel and TR, TG, TB filters. 

Several \hc\ participants use the various telescopes. Typically observations range from 5\,$\times$\,30--300\,s per filter on the CCD cameras, and 60\,$\times$\,30\,s exposures for the CMOS camera, for up to several targets per night. Images are captured using Astro Photography Tool, MaxIm\,DL or Moravian's SIPS for the G4-9000 camera, and are stacked and calibrated for bias, dark and flat frames using PixInsight or MaxIm\,DL.

\subsubsection{El Guijo Observatory}

The observatory is located North West of Madrid, Spain. It uses a 300\,mm f/4 GSO Newtonian astrograph telescope mounted on a Celestron mount and controlled by TheSky6 Pro and CCDSoft for image capture. It utilises a SBIG ST-7 XME Kaf-0402 mono CCD camera, cooled to $-$15\dg C. It has an image scale of 1.53\arcsec/pixel and a field of view of 19.6\arcmin\,$\times$\,13.1\arcmin. All \hc\ images are taken with 20\,$\times$\,120\,s in B and 15\,$\times$\,120\,s in V, R and I filters.

\subsubsection{Emsworth Observatory}

The observatory is situated North East of Portsmouth, UK. It consists of two telescopes: i) A SW80+SX814 Skywatcher Evostar 80 ED DS Pro refractor with Starlight Xpress SX-814 CCD Camera and Baader Clear, \ha, R, V, B, I and U photometric filters, Plate Scale: 1.269\arcsec/pixel and a field of view of 71.63\arcmin\,$\times$\,57.34\arcmin. ii) A C8+SX814+FR Celestron NEXSTAR 8SE 8\arcsec\ Schmidt-Cassegrain f10 to f6.3 focal reducer with Starlight Xpress SX-814 CCD Camera and Baader Clear, \ha, R, V, B, I and U photometric filters, Plate Scale: 0.982\arcsec/pixel and a field of view of 74.32\arcmin\,$\times$\,59.59\arcmin.

MaxIm\,DL\,6 is used to capture and process images, which are reduced using a library of flats, darks and bias frames taken on each of the telescope/camera combinations.

\subsubsection{Forthimage Observatory}

Forthimage Observatory, in a semi-rural area on the western edge of Edinburgh, about 10\,miles from the City centre. It uses a 250\,mm f4.8 Orion Optics Newtonian telescope, a permanently pillar mounted Skywatcher EQ6-R mount, controlled by EQMOD and Cartes du Ciel, and autoguided using PHD2. It is housed in a motorised 2.2\,m dome. The telescope is equipped with an Atik 460EX mono CCD camera, which is cooled to 25\dg C below ambient. The pixel scale is 0.79\arcsec/pixel resulting in a field of view of 36\arcmin$\times$29\arcmin. A Starlight Xpress motorised filter wheel with Baader tri-colour RGB and \ha\ filters is used. For \hc\ observations typically 3--5 images are stacked, with individual exposure times ranging from 120\,s to 300\,s, depending on the target region, filter and sky conditions. Seeing in Edinburgh is typically around 1.5\arcsec\,--\,3\arcsec. During May, June and July, all night twilight interferes, but useful data can still be gathered, even under civil twilight conditions. Camera control and imaging are done by APT (AstroPhotography Tool) along with plate solving to frame the object accurately, flat fielding, bias and dark frame acquisition. Images are calibrated and stacked in Nebulosity\,4.

\subsubsection{Griffon Educational Observatory}

The observatory is located near El Bosque in the South of Spain. It uses exactly the same equipment and observing procedures as the Bowerhill Observatory (see Sect.\,\ref{obs_bowerhill}) with the exception of a Skywatcher EQ6 mount.

\subsubsection{Horndean Observatory}

The observatory is situated North of Portsmouth in the UK. It uses a Williams Optics Zenith Star SD Doublet APO 66\,mm telescope (Focal Length 388\,mm) with a Canon EOS 600D(Mod) camera (pixel scale 2.29\arcsec/pixel), on a Skywatcher Star Adventurer Non Guided, Pulse dithered mount. The sky quality is Class\,4 Bortle. A typical \hc\ observing session consists of 60\,$\times$\,30\,s exposures to avoid trailing and minimize sky glow. Images are processed with bias, darks and flats in Pixinsight. Usually 1 or 2 \hc\ targets are observed per night.

\subsubsection{Karen Observatory}

The observatory is located in the North West of Warrington, UK. It uses a C11 Sct (working at f7.5) telescope with a 278\,mm apperture and 2780\,mm focal length. It is equipped with a Starlight xpress SXV-H694 Trius CCD with a plate scale of 0.67\arcsec/pixel and field of view of 15.3\arcmin\,$\times$\,12.3\arcmin. The telesope uses off axis guiding with Starlight xpress X2. The motorised filter wheel contains a B, V, R and I, Baader 1/2\,mm filter set. Typical \hc\ observations are taken with 10, 20, or 30\,s exposure times and 10 or 20 are stacked depending on the target region and filter.

\subsubsection{KSE Observatory}

The Observatory is located North of San Diego, US. It uses a Meade LX200, 12\arcsec\ (30.5\,cm) telescope with F0.63 focal reducer. It is equipped with a Santa Barbara Instruments ST-7E CCD camera and a Johnson V filer. 
The field of view is 12.3\arcmin\,$\times$\,8.2\arcmin. Typical \hc\ imaging sessions consist of 6\,$\times$\,60\,s exposures per target. Standard data reduction (bias, dark, flat correction) and stacking is performed with AIP4WIN Version 2.4.8.

\subsubsection{La Vara, Valdes Observatory}

The observatory is located in the North of Spain, West of Oviedo. It uses a RCX400 MEADE telescope with a SBIG STXE camera. The filter wheel is equipped with B, V, R and I filters. At 2\,$\times$\,2 binning the pixel scale is 1.74\arcsec/pixel and the field of view is 22.2\arcmin\,$\times$\,14.8\arcmin. Typical exposure times for \hc\ observations are 3\,min with 3--6 images taken per filter. The typical seeing is about 4\arcsec. Dark, bias and flatfield corrections are applied to the images. Obervations are conducted with CCD Soft v.5.

\subsubsection{Les Barres Observatory}

The observatory is located in the South of France, half way between Avignon and Marseille. It uses a Celestron Schmidt-Cassegrain SCT 203\,mm telescope with focal reducer (f/8.1) and is equipped with a SBIG ST-8XME (KAF-1603ME) CCD. \hc\ data is obtained using an Astrodon Johnson/Cousins V filter. The pixel scale is 1.13\arcsec/pixel and the field of view 28\arcmin\,$\times$\,19\arcmin. Typically total integration times range from 30\,min to 60\,min with 2\,min sub-exposures. Image calibration (flat, dark, bias correction and stacking) is performed with the Prism software\footnote{\tt \href{https://www.hyperion-astronomy.com/pages/prism-landing}{Prism Software}}.

\subsubsection{Mount Oswald Observatory}

The observatory is situated some 2\,km south of Durham city, UK. It is mainly used for the BAA VSS programme of variable stars and the \hc\ project has been added to the list of targets. The telescope is a Skywatcher 190MN DS-Pro with a ZWO ASI1600MM Cooled CMOS Camera on a NEQ6 mount, the ZWO filter wheel holds Bessel B, V and R filters. The plate scale is 0.78\arcsec/pixel and a field of view of 60\arcmin\,$\times$\,45\arcmin. Images are taken with the V-Band filter with integration times of 250\,s with sub-exposures of 10\,seconds managed semi-automatically using the Sequence Gen Pro software. The B and R filters have been recently acquired and will be used in the next imaging cycle. Calibration (bias, dark, flat-frame correction and image stacking) is performed in AstroImageJ.

\subsubsection{Movil Observatory}

The observatory is situated North of Le\'{o}n in Spain. It uses a RC 12\arcsec\ telescope and QHY9 (AF8300 chip) CCD and a LodestarX2 off axis guider. The optics provides a resolution of 0.458\arcsec/pixel with a field of view of 25.63\arcmin\,$\times$\,19.36\arcmin. \hc\ observations are typically done as 10\,$\times$\,180\,s exposure in a CV filter. Standard date reduction is applied. 

\subsubsection{Observatorio de Sant Celoni}

The observatory is located North -East of Barcelona, Spain. It uses a SCT 280\,mm / 2116\,mm telescope and is equipped with a KAF8300 CCD. All \hc\ data are taken in photometric V and R filters. All data is reduced following standard procedures using  MaxIm\,DL V6.

\subsubsection{Observatorio de Sencelles}

The observatory is located South of Inca on Mallorca, Spain. It uses a Meade LX200 10\arcsec\ f/10 SC telescope with f/4.3 reducer and a ST-7XME CCD camera equipped with an Astrodon Johnson V filter. The pixel scale is 1.70\arcsec/pixel and the field of view 21.52\arcmin\,$\times$\,14.34\arcmin. The telescope is autoguided with a 200\,mm, f/2.8 telephoto and Orion CCD. Data gathering and analysis is perfomed with a variety of software packages (MaximDL, TheSky, Fotodif, Elbrus, Astrometrica). Typical \hc\ images are taken with exposure times ranging from 120\,s to 900\,s. Dark, bias and flatfield correction is applied to all images.

\subsubsection{Observatorio El Sue\~{n}o}

The Observatory is situated in Vinyols i els Arcs, West of Tarragona, Spain. It uses a newton 300/1500\,mm GSO telescope and ST8-XE SBIG CCD camera with AO. It is equipped with V, R$_{\rm c}$, Johnson and Cousins filters. The optics results in a scale of 1.22\arcsec/pixel and a field of view of 20.7\arcmin\,$\times$\,31.0\arcmin. Typically \hc\ observations are taken with 1200\,s exposures. All images are dark and flatfield corrected using MaximDL (Windows).

\subsubsection{Observatorio Mazariegos}

The observatory is located North of Valladolid, Spain. It uses a Celestron XLT 8\arcsec\ (2032 FL) and Atik 314L$+$ camera (Sony ICX-285AL CCD), equipped with V and R filters Johnson/Cousins. The pixel scale is 0.85\arcsec/pixel. Typically \hc\ observations was taken as 120\,s exposures. All images are dark and flatfield corrected following standard procedures using the MaximDL software.

\subsubsection{Observatorio Montcabrer}

The Observatory is situated North East of Barcelona, Spain. It uses a Meade ACF 305/3000\,mm telescope and Moravian G4-900 CCD Camera, equipped with V, R$_{\rm c}$ Johnson-Cousins and g, r and i Sloan filters. The image scale is 085\arcsec/pixel and the field of view 43\arcmin$\times$\,43\arcmin. Typically \hc\ observations are taken as 600\,s exposures. All images are dark and flatfield corrected following standard procedures using the Kstars software.

\subsubsection{Observatorio Nuevos Horizontes}

The observatory is located in Camas, West of Seville, Spain. It uses a SC 9.25\arcsec\ telescope with an ICX285AL CCD chip. The optics results in a scale of 1.96\arcsec/pixel at 2\,$\times$\,2 binning. The camera is equipped with a B, V and R filter set. \hc\ observations are typically taken with 60\,s exposure time. Science frames are reduced with bias, dark and flat frames following standard procedures. The MaximDL software is used for all data and calibration frame acquisition. On average about 4 or 5 nights per week are used for observations. Observing conditions are clear for most of the year.

\subsubsection{Rolling Hills Observatory}

The observatory is located West of Orlando, Florida, USA. It uses a 35\,cm aperture, f/10 Schmidt-Cassegrain telescope and a SBIG STT-8300M CCD with Astrodon B and V filters. The pixel scale (at 2\,$\times$\,2 binning) is 0.64\arcsec/pixel and the field of view 17.4\,\arcmin$\times$\,13.1\arcmin. Typical \hc\ observations are done as three images per filter with individual exposures of 75\,s and 180\,s in V and B, respectively. Dark images at the same camera temperature and exposure length were subtracted and then a sky flat is applied. 

\subsubsection{R.P. Feynman Observatory}

The observatory is located in Gagliano del Capo in the South of Italy. The telescope used for \hc\ observations is a 12\arcsec\ f/5.3 Orion Optics newtonian reflector with an Atik460Ex monochrome camera and Custom Scientific B, V, SR and SI filters. Using 2\,$\times$\,2 binned pixels, this provides a plate scale of 1.18\arcsec/pixel and a field of view of 27\arcmin\,$\times$\,21.6\arcmin. Typical seeing in the images is around 3\arcsec\,--\,4\arcsec. Integration times for the images range from 60\,s to 240\,s depending on target and filter. Image calibration (dark, flat-field correction and stacking) is carried out with the AstroArt software. 

\subsubsection{Sabadell Observatory}

The observatory is situated North of Barcelona, Spain. It uses an Newton 500/2000\,mm  telescope and a Moravian G2-1600 camera equipped with Johnson-Cousins filters. The pixel scale is 0.92\arcsec/pixel an the field of view is 23.5\arcmin\,$\times$\,15,7\arcmin. Typically \hc\ observations are taken with 60\,s exposures and set of 10\,--\,15 images are stacked. All images are  dark and flatfield corrected using a number available software packages (Cartes du Ciel, ArtroArt, Astrometrica, Focas).

\subsubsection{Selztal-Observatory}

The observatory has already been described in \citet{2018MNRAS.478.5091F}. For completeness we reproduce here the text used in that earlier publication.

The observatory is located in Friesenheim, approximately 20\,km South of Mainz in Germany. The telescope is a 20\arcsec\ Newton, with f\,=\,2030\,mm and an ASA corrector and an ASA DDM 85 Pro mount. The CCD used is a STL 11000M with anti-blooming gate and a set of RGB filters is available. Twilight flats are taken to correct for variations in pixel sensitivity and image processing is performed with the MaxIm\,DL software. Typical exposure times are 120\,--\,300\,s and observations are guided with an accuracy of less than one pixel and a seeing of about 3\arcsec. Due to surrounding street lights, there are some gradients left in the images not corrected for by the flatfield, but they do not influence the photometry.

\subsubsection{Shobdon Observatory:} 

The observatory has already been described in \citet{2018MNRAS.478.5091F}. For completeness we reproduce here the text used in that earlier publication.

The observatory is situated in Herefordshire about 8\,km from the UK/Wales Border. It houses a Meade LX200 35\,cm SCT (f/7.7) operating at a focal length of 2500\,mm with a Starlight XPress SXV-H9 CCD and a set of Johnson-Cousins B, V, R and I filters. Integration times are typically 60\,s and darks and flats are applied using AIP4WIN software.   

\subsubsection{Steyning Observatory}

The observatory has already been described in \citet{2018MNRAS.478.5091F}. For completeness we reproduce here the text used in that earlier publication.

The observatory is situated in Steyning, West Sussex, UK. The telescope is an 8\arcsec (200\,mm) Ritchey Chretien (f/8.0) operating at a focal length of 1600\,mm with a Santa Barbara Instrument Group (SBIG) STF-8300M mono camera, and a 'green' filter from a tri-colour imaging set made by Astronomik. Using 2\,$\times$\,2 binned pixels, this provides a plate scale of about 1.4\arcsec/pixel with a field of view of 39\arcmin\,$\times$\,29\arcmin. Integration times for the images range from 60\,s to 240\,s. Image calibration (darks, flat-fields and stacking) is carried out with the AstroArt software.

\subsubsection{Tigra Automatic Observatory}

The observatory is one of a pair located in Monkton Nature Reserve on the Isle of Thanet, Kent, UK. It uses a 305\,mm f/10 Schmidt Cassegrain telescope (Meade LX200) mounted on an equatorial fork within a domed observatory. The CCD camera is manufactured by Santa Barbara Instruments Group (SBIG) and has a Kodak KAF-6303E non-antiblooming sensor with 3072\,$\times$\,2048 pixels of 9\,$\mu$m. Guiding is aided by an SBIG AO-X adaptive optics unit. Filters available are Baader L, R, G, B, C and \ha, OIII and SII. Imaging for \hc\ is normally performed with the sensor cooled to $-$35\dg C at 2\,$\times$\,2 binning, for a measured image scale of 1.175\arcsec/pixel. Exposures are typically 300\,s which are dark subtracted, flat fielded and stacked to produce an image for submission to \hc. The observatory is robotic and  is scheduled and orchestrated by Astronomer's Control Program (ACP)\footnote{\tt \href{http://www.dc3.com}{ACP Observatory Control Software}}. Image processing and camera control is provided by MaxIm\,DL from Diffraction Limited\footnote{\tt \href{http://diffractionlimited.com/}{Diffraction Limited}}. Device control is performed using a number of ASCOM\footnote{\tt \href{https://ascom-standards.org}{Astronomy Common Object Model}} drivers developed by Tigra Astronomy\footnote{\tt \href{http://tigra-astronomy.com/}{Tigra Astronomy}}.

\subsubsection{Uraniborg Observatory}

The observatory is located in the South of Spain, between Seville and C\'{o}rdoba. It uses a SC Celestron C11 telescope at f6.3 and an Atik 414ex monochrome CCD camera. The pixel scale is 0.779\arcsec/pixel and the field of view 18\arcmin\,$\times$\,13\arcmin. Images are taken either at the full resolution or using a 2\,$\times$\,2 binning. Typically 40\,s\,--\,60\,s exposures are obtained for \hc\ and 10\,--\,20 are stacked, depending on the target region. Dark, bias and flatfield corrections are applied using MaxIm\,DL.

\subsubsection{Warsash Observatory}

The observatory is located in Warsash, between Portsmouth and Southampton in the UK. It uses a Williams Optics 110\,mm Apochromatic Refractor with William Optics 0.8x Focal reducer/flattener and a Starlight Xpress SX 694 mono CCD camera for guided exposures. The camera has an image scale of 1.53\arcsec/pixel and a field of view of 71\arcmin\,$\times$\,57\arcmin. \hc\ images are typically taken through a Photometric V filter (10\,$\times$\,120\,s), a Baader Red filter (10\,$\times$\,30\,s or 10\,$\times$\,120\,s) and Baader Blue filter (10\,$\times$\,120\,s). All images are calibrated and stacked following standard procedures using the SIPS software.

\subsection{Description of University and Professional Observatories}

In this section we describe the utilised University and professional telescopes. Each subsection also contains a basic description of the observing and data reduction procedures. Like for the amateur observatories, the locations are available on our online map.

\subsubsection{Bia{\l}k\'{o}w Observatory}

The observatory is located at 51.474248\dg N, 16.657821\dg E, to the North West of Wrco{\l}aw in Poland. The data in Bia{\l}k\'{o}w were gathered with the 60\,cm Cassegrain telescope equipped with an Andor Tech iKon-L DW432-BV back-illuminated CCD camera covering 13\arcmin\,$\times$\,12\arcmin\ field of view in the B, V, Rc and Ic passbands of the Johnson-Kron-Cousins photometric system. The CCD has 1250\,$\times$\,1152 pixels with a pixel size of 22.5\,$\mu$m and a scale of 0.619\arcsec/pixel. Exposure times range from 100\,s to 140\,s. The typical seeing is 2.5\arcsec. Observations were calibrated in the standard way, which included dark and bias subtraction and flat-field correction. Custom made software and iraf package routines are used for data reduction.

\subsubsection{Las Cumbres Observatory Global Telescope Network}

Some of the projects participants used access to the range of telescopes from the Las Cumbres Observatory Global Telescope Network (LCOGT). The observatory has already been described in \citet{2018MNRAS.478.5091F}. For completeness we reproduce here the text used in that earlier publication.

LCOGT provides a range of 2\,m, 1\,m and 0.4\,m telescopes located at various sites around the Earth to allow complete longitudinal coverage. The two 2\,m telescopes are the Faulkes telescopes built by Telescope Technologies Ltd. which are f/10 Ritchey-Cretien optical systems. The 1\,m telescopes are also Ritchey-Cretien systems with f/7.95, while the 0.4\,m telescopes are Meade 16\arcsec\ RCX telescopes. Data included in this work has been taken on Haleakala Observatory (0.4\,m, 2\,m), Siding Spring Observatory (0.4\,m, 1\,m) and Tenerife (0.4\,m). All data from LCOGT are returned reduced with dark and flat-field corrections applied. Integration times are typically 60\,s but depend on the target and telescope size.

\subsubsection{OpenScience Observatories - COAST Observatory}\label{obs_coast}

The observatory is located at the Observatorio del Teide, Tenerife, Spain (same site as the PIRATE observatory - see Sect.\,\ref{obs_pirate}). It is currently operated by the Open University, and a fully autonomous, queue-scheduled system. The telescope is a Celestron 14\arcsec\ Schmidt-Cassegrain (f/10) on a 10Micron GM4000 mount. It uses a FLI ProLine KAF-09000 CCD with photometric Johnson B, V and R filters. The field of view is 33\arcmin\,$\times$\,33\arcmin\ at a pixel scale of 0.65\arcsec/pixel.

Image calibration has been performed with COAST pipeline, and fully calibrated images have been retrieved. Dark and bias subtraction has been done with library frames and dawn sky flats are used for flat-fielding. Typical \hc\ observations consist of single integrations of 40\,s\,--\,60\,s, repeated roughly 2\,--\,3 times a week.

\subsubsection{OpenScience Observatories - PIRATE (Open University)}\label{obs_pirate}

The observatory \citep{2018RTSRE...1..127K} is sited at Teide Observatory (Latitude: 28.299286\dg N, Longitude: 16.510297\dg W, Altitude: 2370\,m). It uses a 17\arcsec\ (432\,mm) Aperture Planewave CDK17 telescope with Cassegrain optics (2939\,mm Focal Length, Focal ratio f/6.8) on a 10Micron GM4000 HPS mount. It is equipped with a FLI ProLine PL16803 Camera with a KAF-16803 CCD and a 10 position filter wheel (U, B, V, R, I, \ha, OIII, SII, Clear). The field of view is 43\arcmin\,$\times$\,43\arcmin\ with 0.63\arcsec/pixel resolution.

Dark, bias and flatfield frames are taken at dusk and dawn every day. Data are reduced using a custom built pipeline that is loosely based off AstroImageJ and follows a standard calibration technique, it also removes the overscan region of the CCD. Seeing conditions are typically better than 1\arcsec, and during the summer, 50\,\% of the time the seeing is better than 0.54\arcsec. All images were taken with 100\,s exposure times. The typical \hc\ observing pattern includes taking two exposures in B, V, R and \ha\ filters every night.

\subsubsection{University of Kent Beacon Observatory}\label{beacon}

The observatory has already been decribed in \citet{2018MNRAS.478.5091F}. For completeness we reproduce here the text used in that earlier publication.

The Beacon Observatory consists of a 17\arcsec\ {\em Planewave} Corrected Dall-Kirkham (CDK) Astrograph telescope situated at the University of Kent (51.296633\dg\ North, 1.053267\dg\ East, 69\,m elevation). The telescope is equipped with a 4k\,$\times$\,4k Peltier-cooled CCD camera and a B, V, R$_c$, I$_c$ and \ha\ filter set. The pixel scale of the detector is 0.956\arcsec, giving the camera a field of view of about 1\dg\,$\times$\,1\dg. Due to the optical system of the telescope the corners of the detector are heavily vignetted. Hence the usable field of view of the detector is a circular area with a diameter of approximately 1\dg.

The observatory has, despite its location, a good record for observations. Over the first four years of operations an average of 12 nights per month were used for science observations, with an average of 60\,hrs per month usable, i.e. about 60\,\% of the time is used in each night with clear skies. The typical seeing in the images is about 3\arcsec\,--\,4\arcsec.

Images taken by the observatory for the \hc\ project are typically taken in the following sequence: 120\,s integrations are done in V, R, and I and this sequence is repeated 8 times. Including filter changes and CCD readout, this sequence takes one hour. All individual images are dark and bias subtracted and flat-fielded using sky-flats. All images taken of a particular target during a sequence are median averaged using the Montage software package\footnote{\tt \href{http://montage.ipac.caltech.edu/}{Montage Software}}. 

\subsubsection{University of Leicester Observatory (UL50)}

The observatory has already been decribed in \citet{2018MNRAS.478.5091F}. For completeness we reproduce here the text used in that earlier publication.

The University of Leicester runs a 0.5\,m telescope (the University of Leicester 50\,cm, or UL50). This is a 20\arcsec\ {\em Planewave} CDK telescope with a SBIG  ST2000XM camera. It is equipped with a Johnson-Cousins B, V, R and I filter set. Data were reduced using dark, bias and flat-frames taken the same night as science observations, using an {\sc IRAF} pipeline. In addition the observatory now operates a Moravian Instruments G3-11000 CCD camera with Johnson-Cousins B, V, R and I filters.

\subsubsection{Th\"{u}ringer Landessternwarte}\label{tls}

The observatory has already been decribed in \citet{2018MNRAS.478.5091F}. For completeness we reproduce here the text used in that earlier publication.

The Th\"{u}ringer Landessternwarte is operating its Alfred-Jensch 2\,m telescope\footnote{\tt \href{http://www.tls-tautenburg.de/TLS/index.php?id=25&L=1}{Th\"{u}ringer Landessternwarte}} near Tautenburg (50.980111\dg\ North, 11.711167\dg\ East, 341\,m elevation) Germany. For \hc\ the telescope is used in its Schmidt configuration (clear aperture 1.34\,m, mirror diameter 2.00\,m, focal length 4.00\,m). It is equipped with a 2k\,$\times$\,2k liquid nitrogen-cooled CCD camera and with a U, B, V, R, I and \ha\ filter set. The employed SITe CCD has 24\,$\mu$m\,$\times$\,24\,$\mu$m pixels, leading to a field of view of 42\arcmin\,$\times$\,42\arcmin.  Single exposures of 20\,s to 120\,s integration time -- depending on the filter -- are obtained, and several consecutive frames may be co-added. Dark frames and dome-flats are used for image calibration.

\subsubsection{Vihorlat Observatory}

The observatory is located in eastern Slovakia at 48.935000\dg N, 22.273889\dg E. The Vihorlat National Telescope is a 1000\,mm aperture and 9000\,mm focal length telescope equipped with a FLI PL 1001E camera. At 2\,$\times$\,2 binning the pixel scale is 1.11\arcsec/pixel and the field of view 9.47\arcmin\,$\times$\,9.47\arcmin. The filter wheel contains a B, V, R$_{\rm c}$, I$_{\rm c}$, and Clear Johnson Cousins filter set. For \hc\ targets we use 2\,min or 3\,min exposure times, with at least 5 images in every filter. Typical seeing conditions are around 3\arcsec. The data reduction is following standard procedures with bias, dark and sky flat corrections and image stacking. For image acquisition we use MaximDL. The data reduction is performed by recently developed CoLiTecVS software\footnote{\tt \href{http://www.neoastrosoft.com/colitecvs$\_$en/}{CoLiTecVS software}}.

\subsubsection{iTelescope Network}

Several of the amateur observers have used access to the iTelescope network\footnote{\tt \href{https://www.itelescope.net/}{iTelescope Network}} to support \hc\ observations. In particular the T5 and T7 telescopes were used.

The T5 telescope is situated at the New Mexico Skies Observatory near Mayhill in the US. It uses a 0.25\,m f/3.4 reflector and SBIG ST-10XME CCD eqipped with Red Green Blue, Ha, SII, OIII, Clear and Johnson's Cousin's Photometric B, V, and I filters. The field of view is 60.6\arcmin\,$\times$\,40.8\arcmin\ at a scale of 1.65\arcsec/pixel.

The T7 telescope is situated at Astro Camp Nerpio, west of Murcia in Spain. It uses a 0.43\,m f/6.8 reflector and SBIG STL-11000M CCD equipped with R, V, B, Ha, OIII, SII and I filters. The field of view is 28.2\arcmin\,$\times$\,42.3\arcmin\ at a scale of 0.63\arcsec/pixel.

\clearpage\newpage

\onecolumn 

\section{Long term Light Curves of \vc}\label{longtermlight-curves}

\begin{figure}
\centering
\vspace{-10cm}
\includegraphics[width=0.75\textwidth]{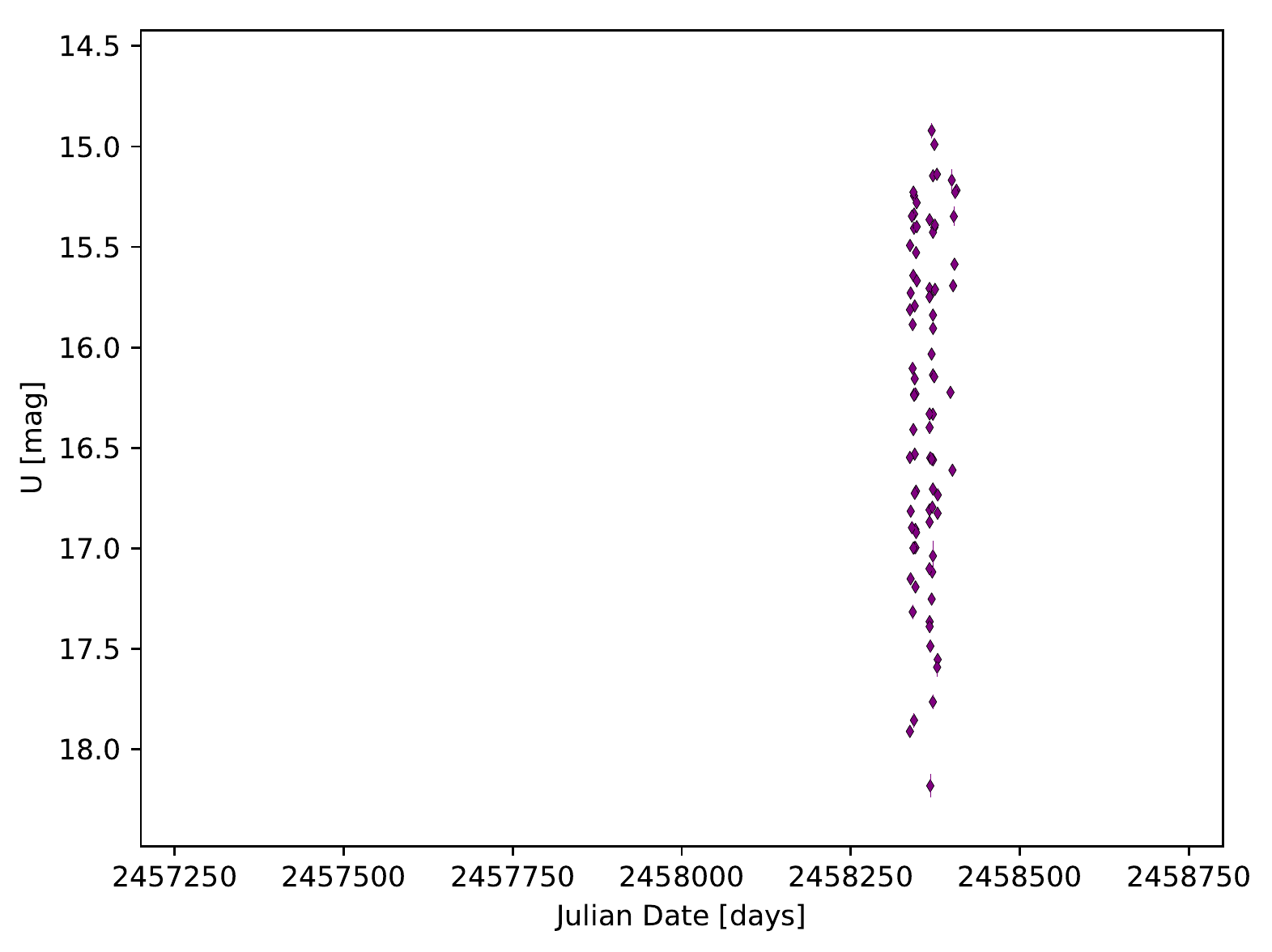} \\
\includegraphics[width=0.75\textwidth]{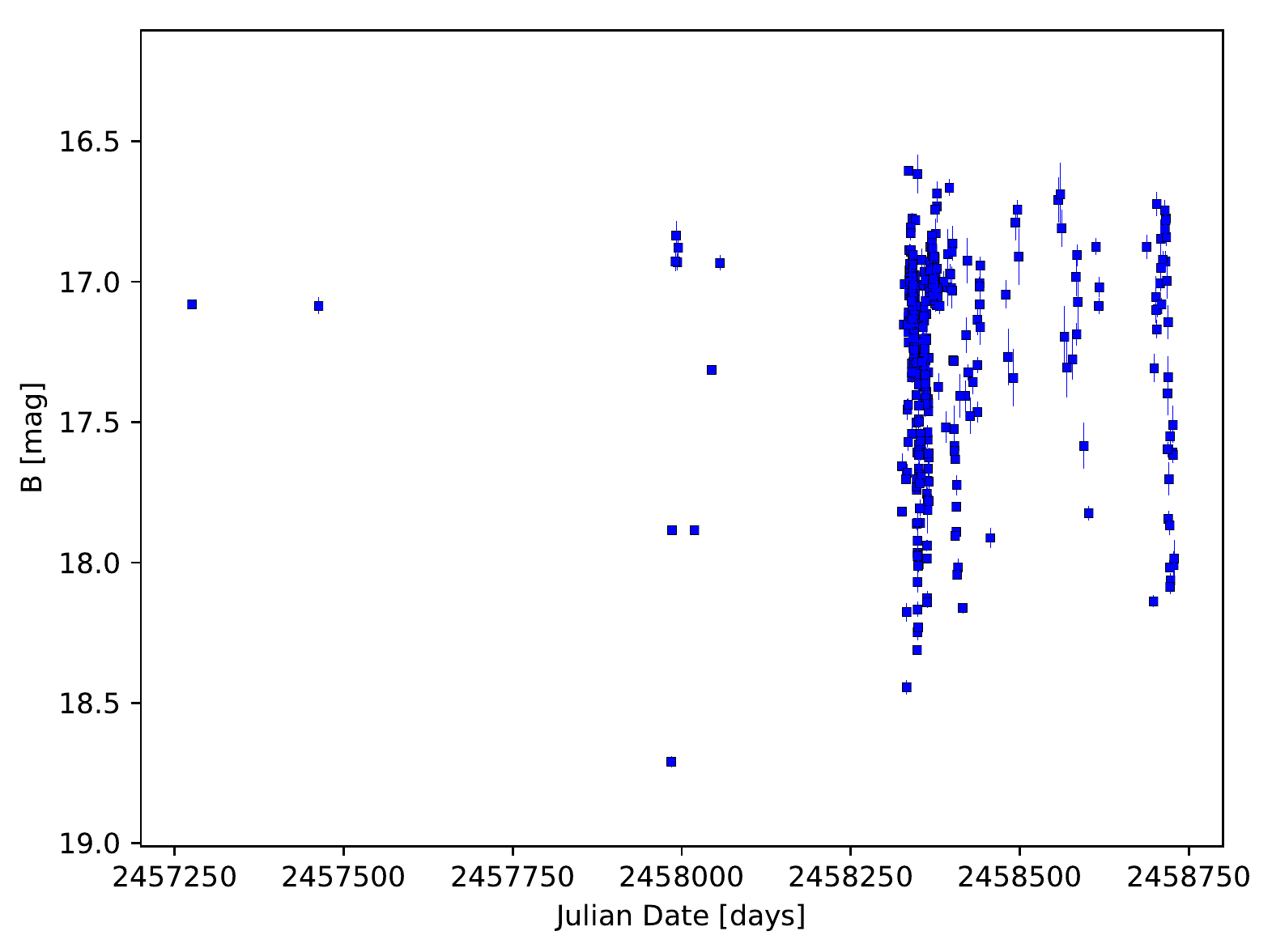} \\
\caption{\label{UB_lc} Long term \hc\ U and B light curves of \vc.}
\end{figure}

\begin{figure}
\centering
\includegraphics[width=0.75\textwidth]{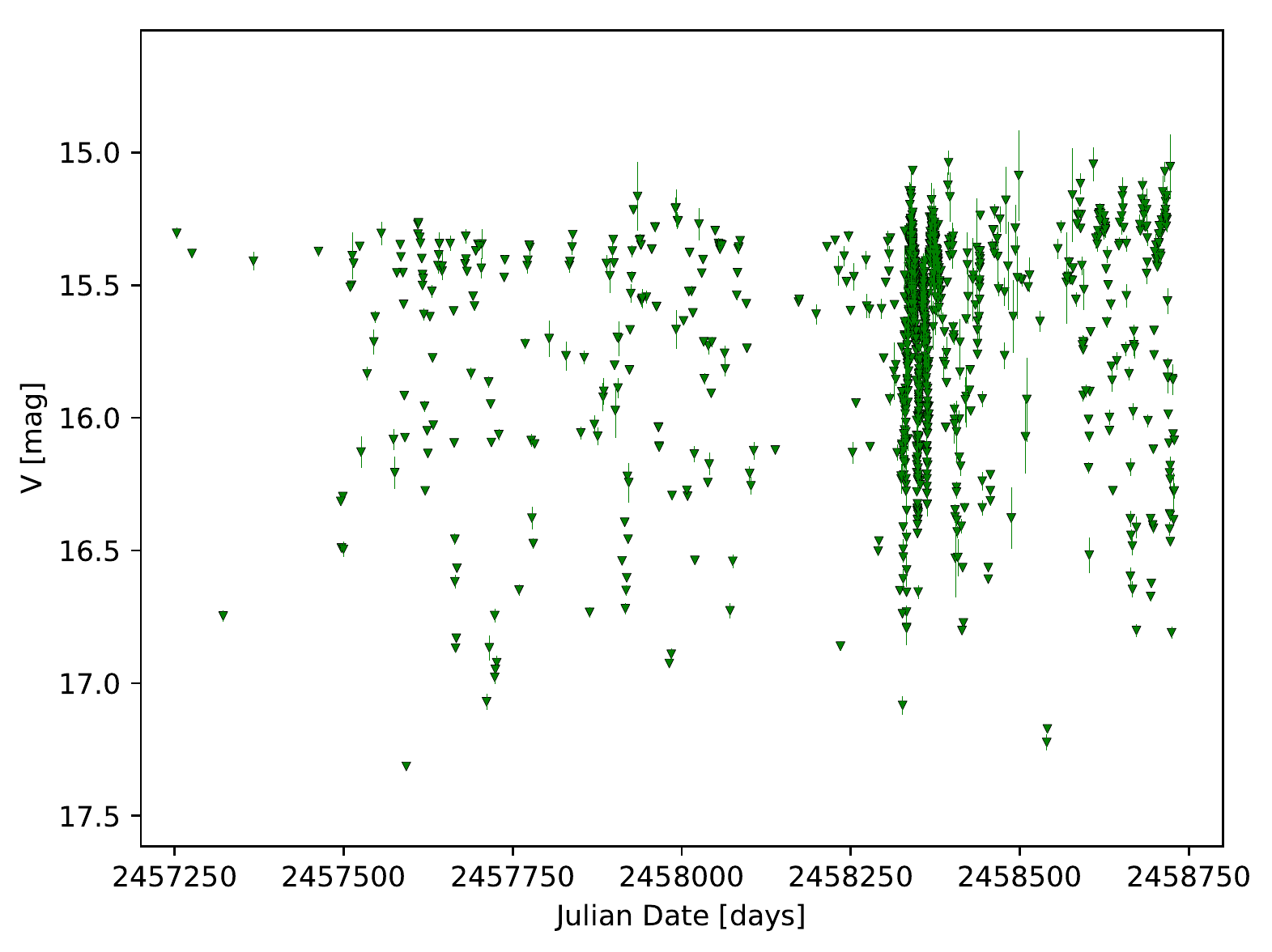} \\
\includegraphics[width=0.75\textwidth]{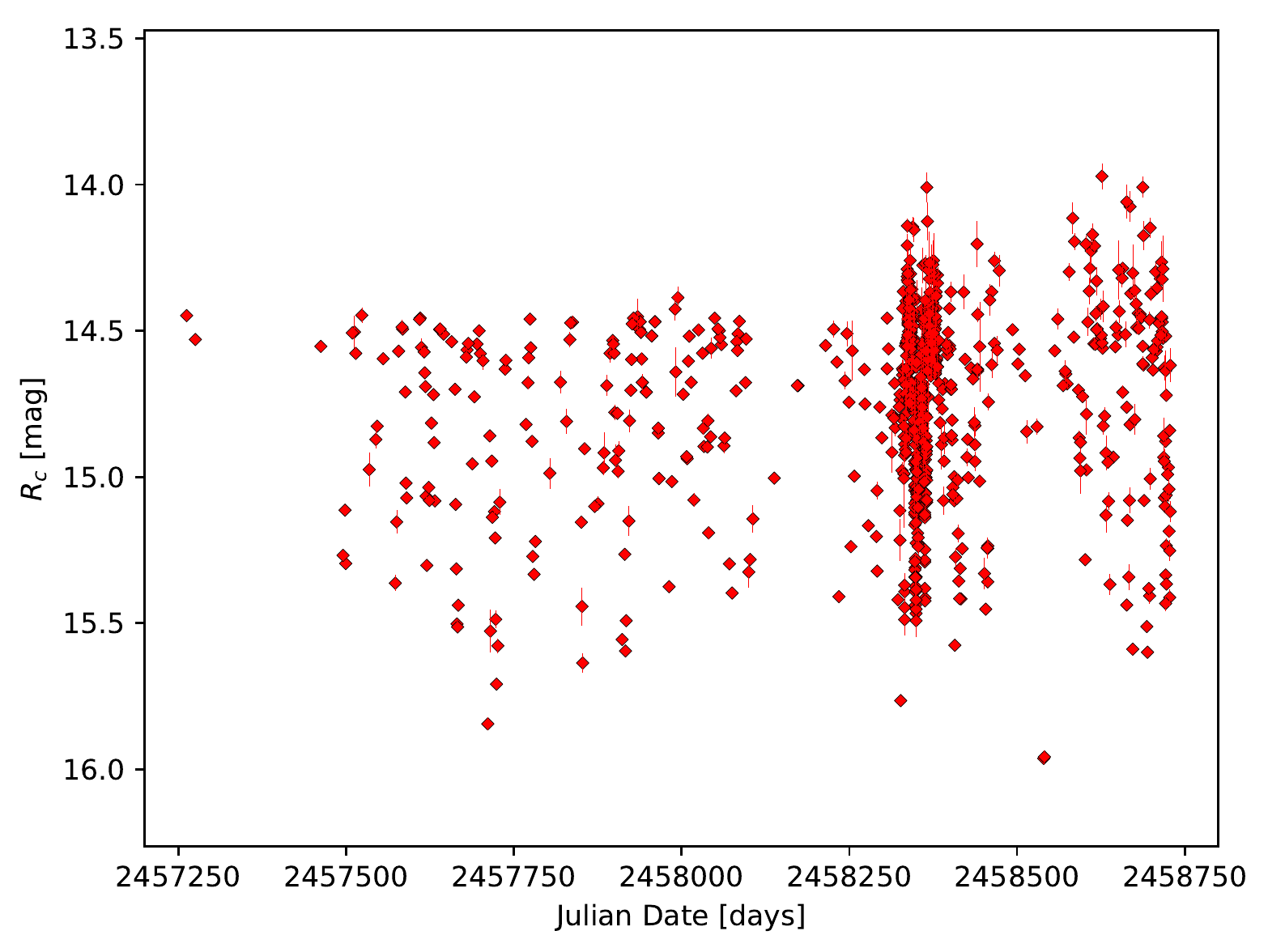} \\
\caption{\label{VR_lc} Long term \hc\ V and R$_c$ light curves of \vc.}
\end{figure}

\begin{figure}
\centering
\includegraphics[width=0.75\textwidth]{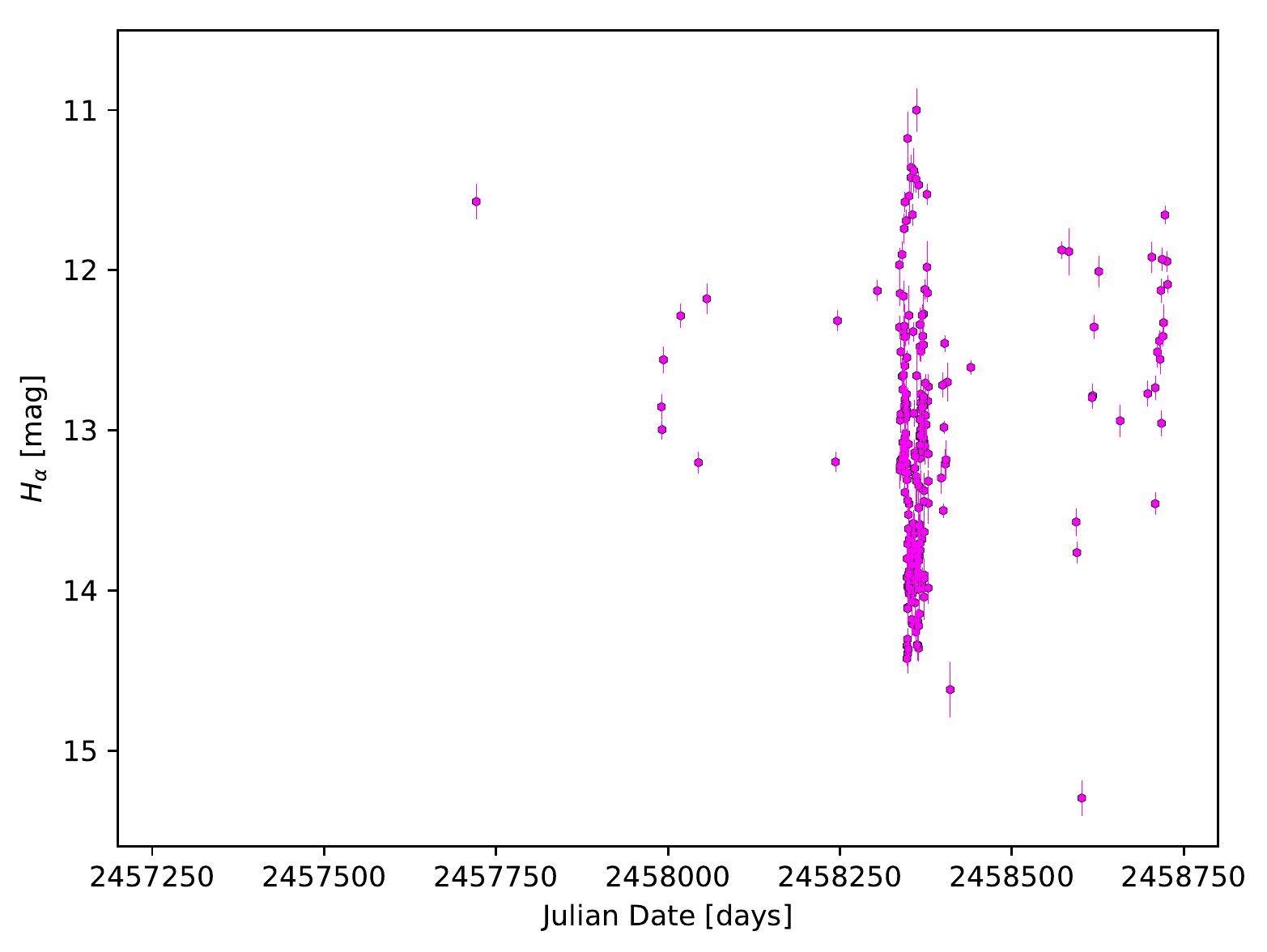} \\
\includegraphics[width=0.75\textwidth]{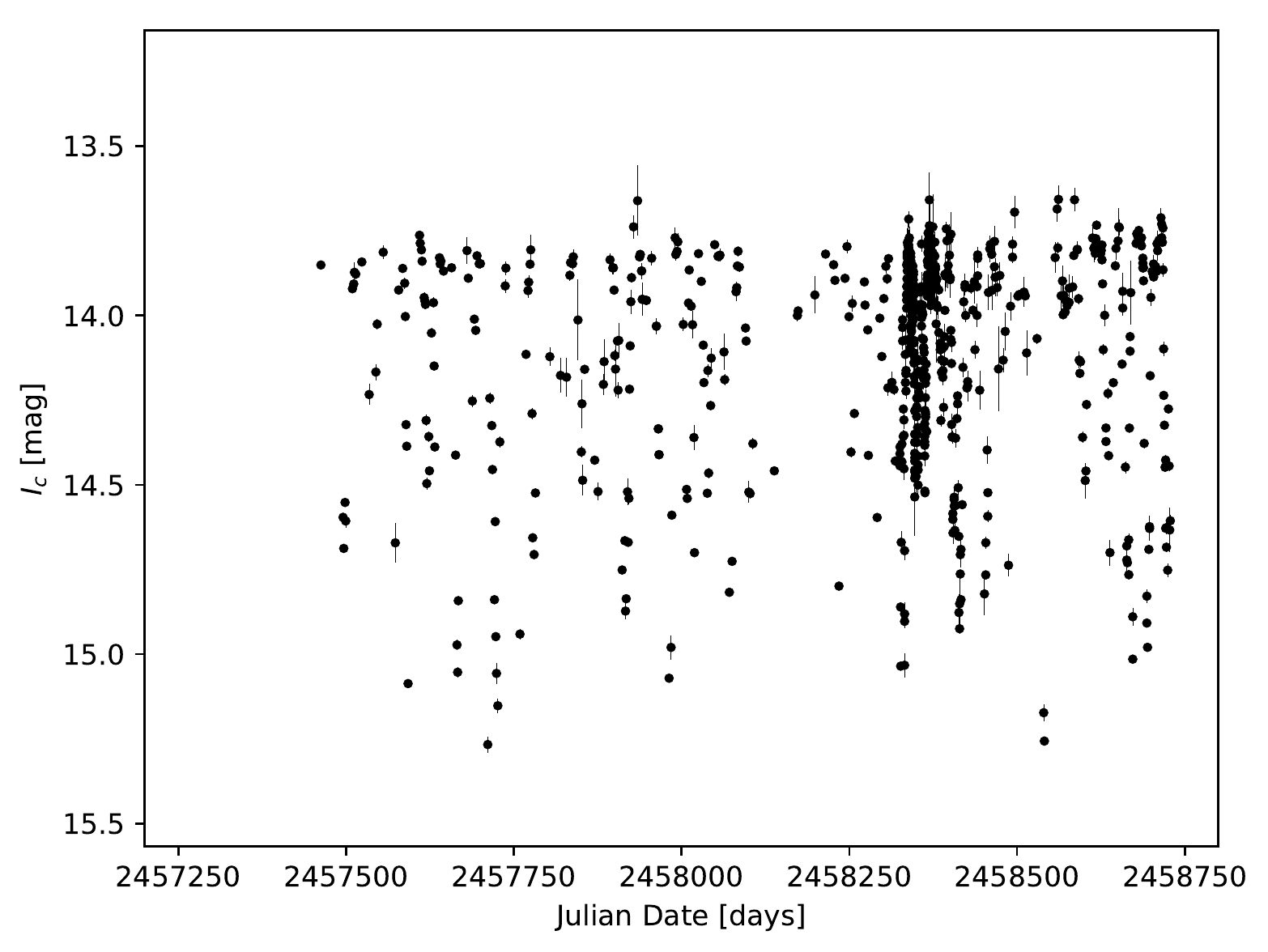} \\
\caption{\label{HAI_lc} Long term \hc\ \ha\ and I$_c$ light curves of \vc.}
\end{figure}

\clearpage\newpage

\section{Lomb-Scarcle Periodograms}\label{lombscarcle}

\begin{figure}
\centering
\vspace{-10cm}
\includegraphics[width=0.75\textwidth]{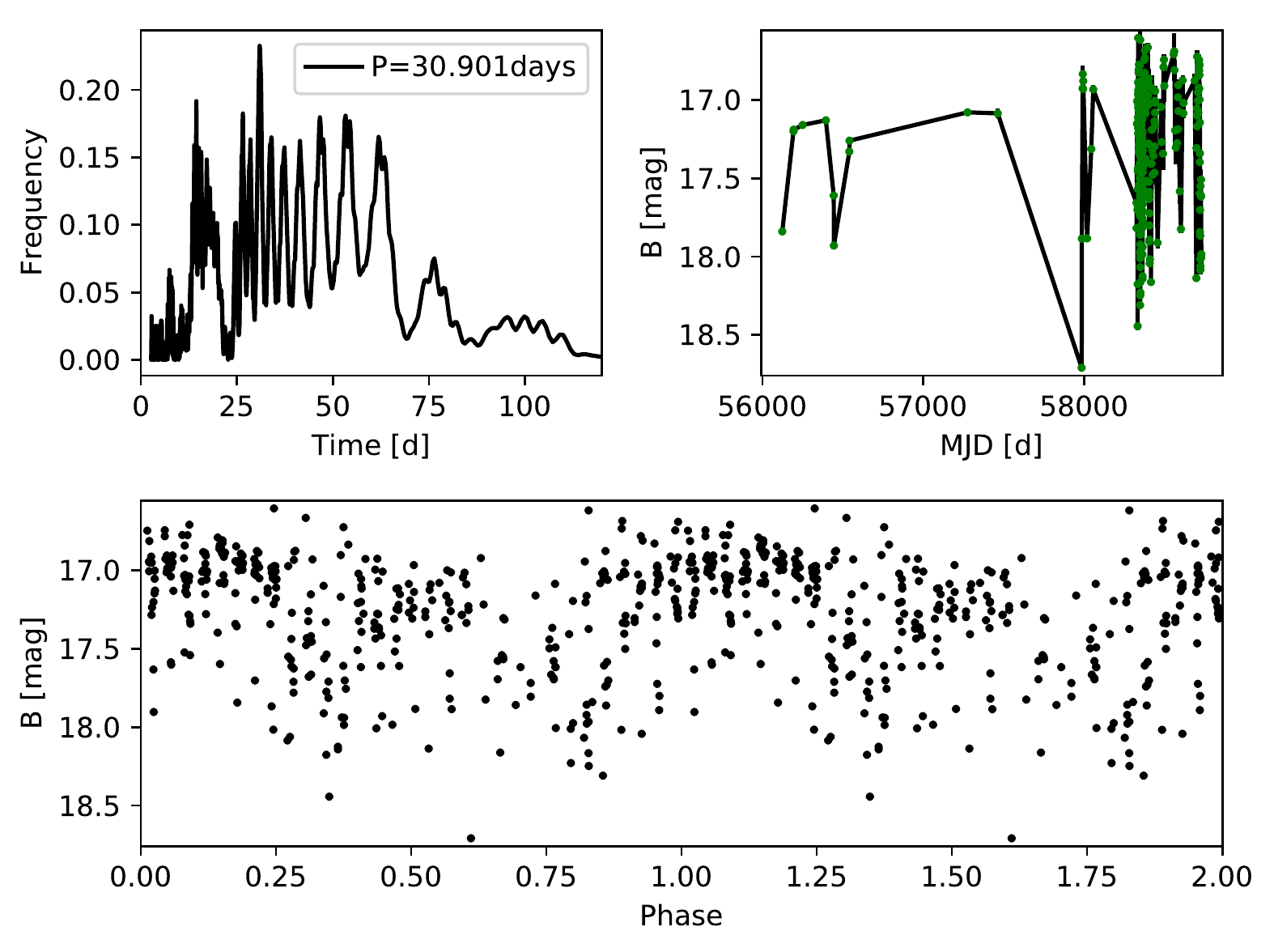}
\includegraphics[width=0.75\textwidth]{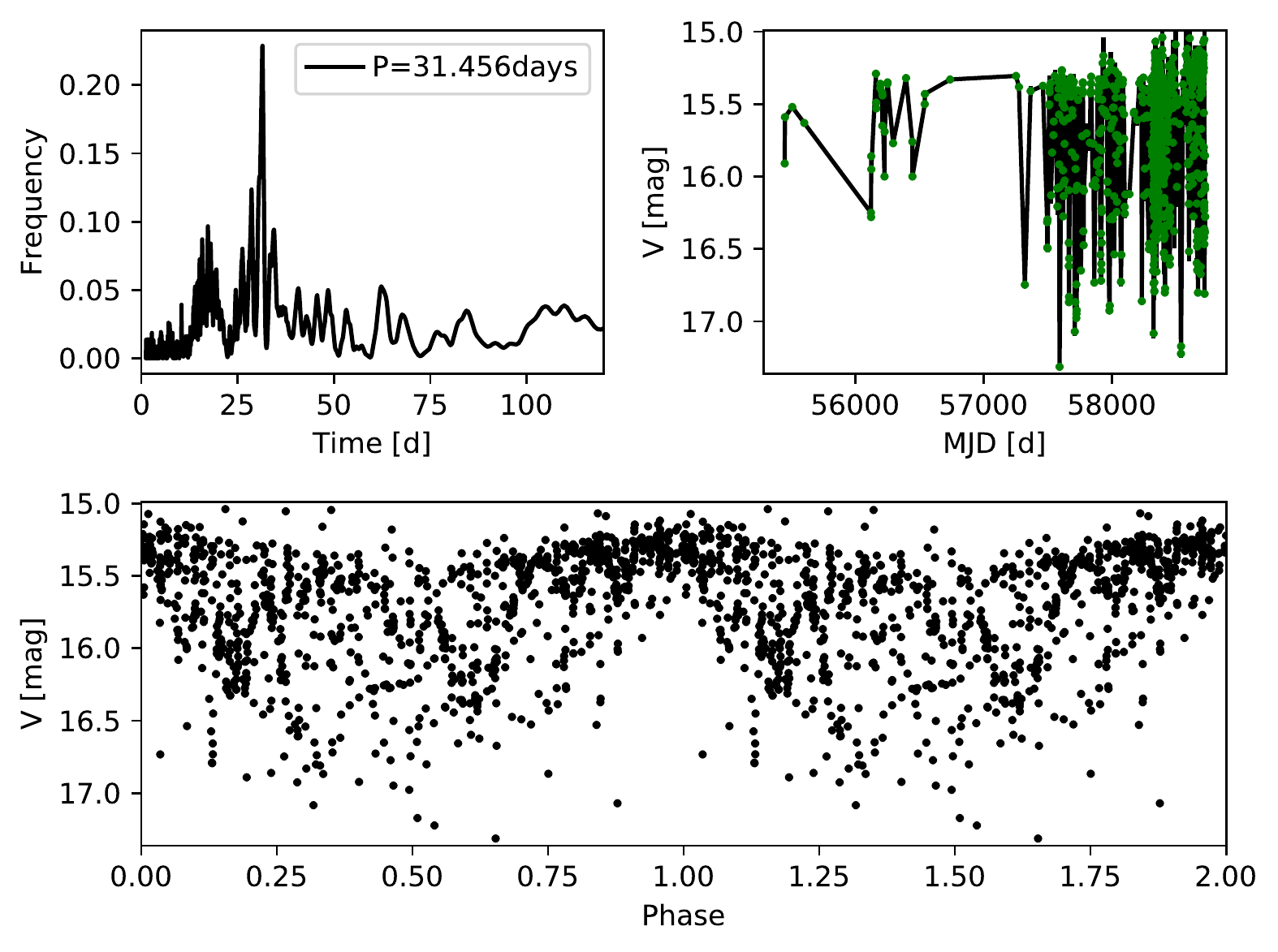} \\
\caption{\label{period_bv} {\bf Top Left:} Shown is a Lomb-Scargle periodogram \citep{1982ApJ...263..835S} of the \vc\ light curve in the B (top) and V (bottom) filter. Only data a with a magnitude uncertainty of less than $0.2$\,mag has been included, as well as the data from \citet{2018RAA....18..137I}. {\bf Top right:} The full light curve of \vc. {\bf Bottom:} The phase plot of \vc\ showing two full periods.}
\end{figure}
        
\begin{figure}
\centering
\includegraphics[width=0.75\textwidth]{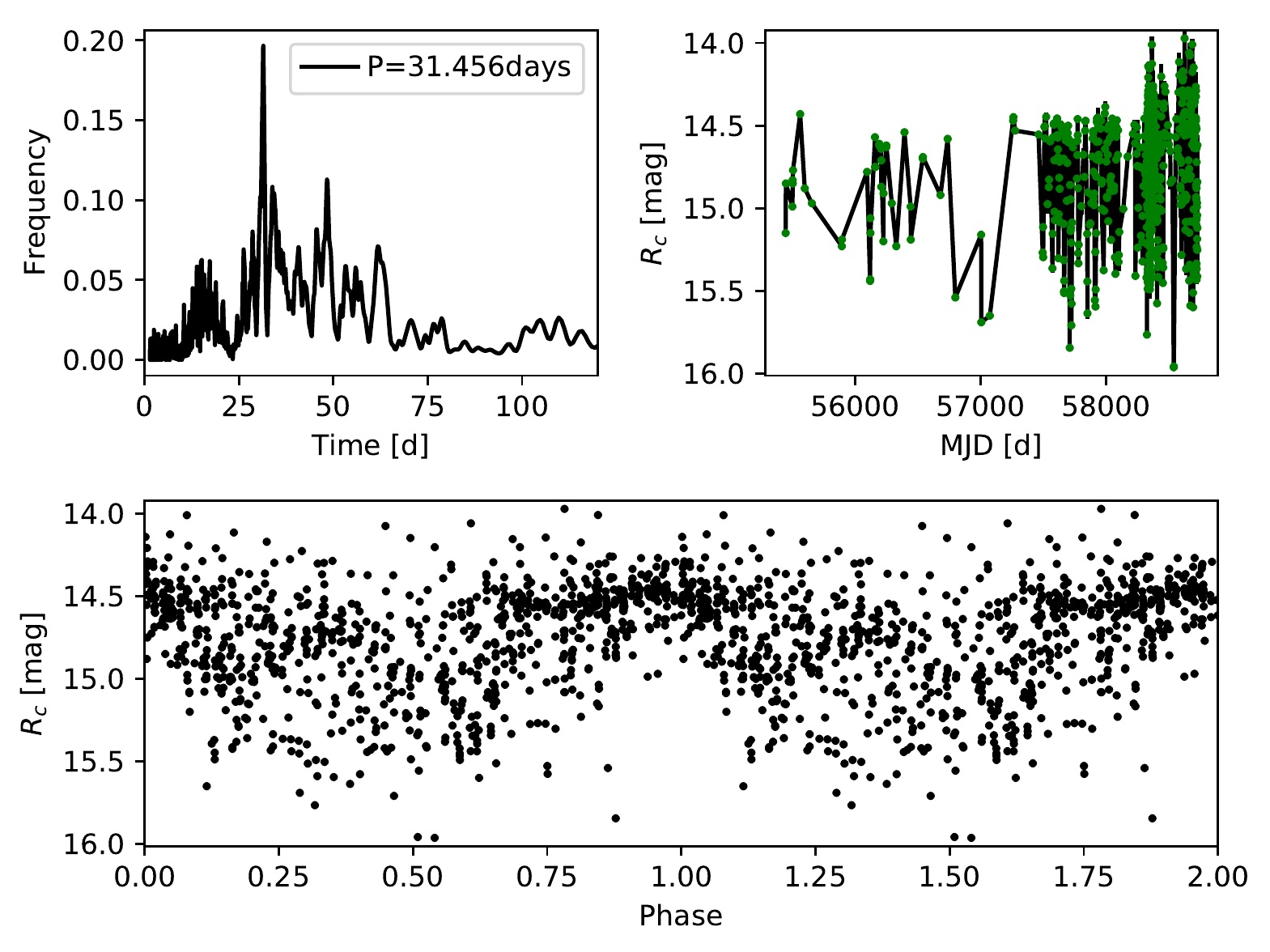} \\
\includegraphics[width=0.75\textwidth]{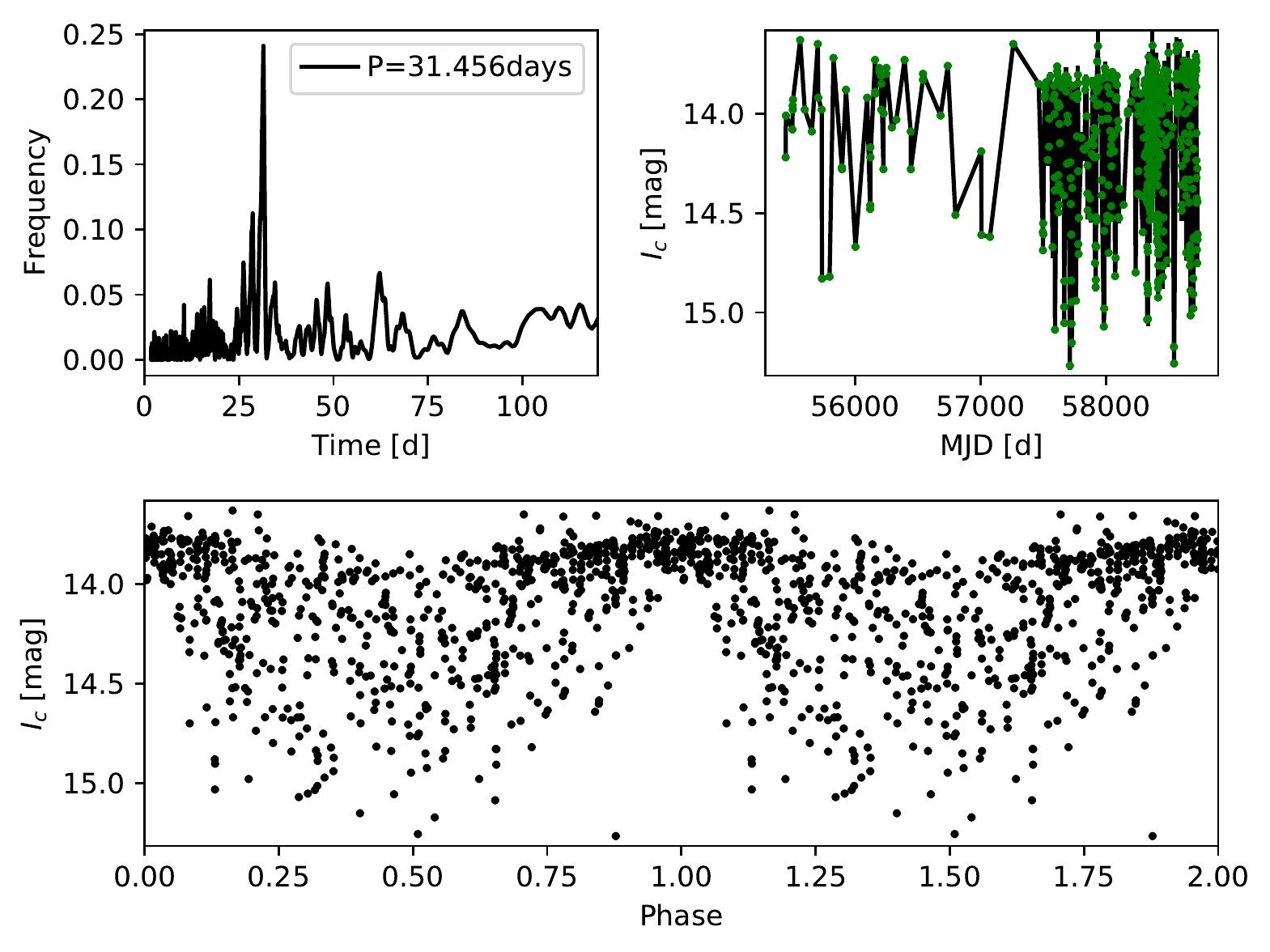} 
\caption{\label{period_ri} {\bf Top Left:} Shown is a Lomb-Scargle periodogram \citep{1982ApJ...263..835S} of the \vc\ light curve in the R$_c$ (top) and I$_c$ (bottom) filter. Only data a with a magnitude uncertainty of less than $0.2$\,mag has been included, as well as the data from \citet{2018RAA....18..137I}. {\bf Top right:} The full light curve of \vc. {\bf Bottom:} The phase plot of \vc\ showing two full periods.}
\end{figure}
        
\clearpage\newpage

\section{Gaia Parallax vs. Proper Motion in IC\,5070 \vc}\label{gaiapmparallax}

\begin{figure}
\centering
\vspace{-10cm}
\includegraphics[width=0.75\textwidth]{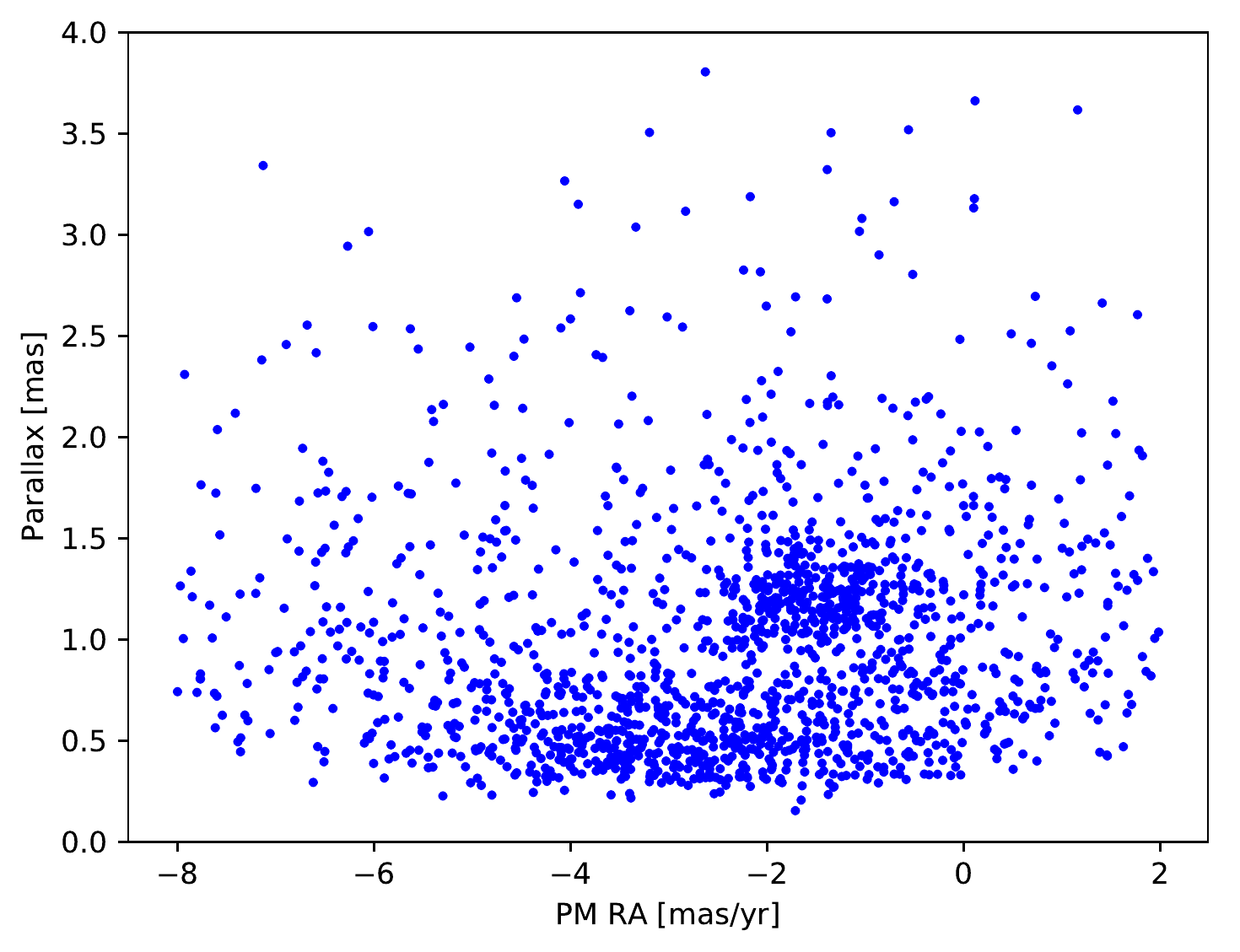} \hfill
\includegraphics[width=0.75\textwidth]{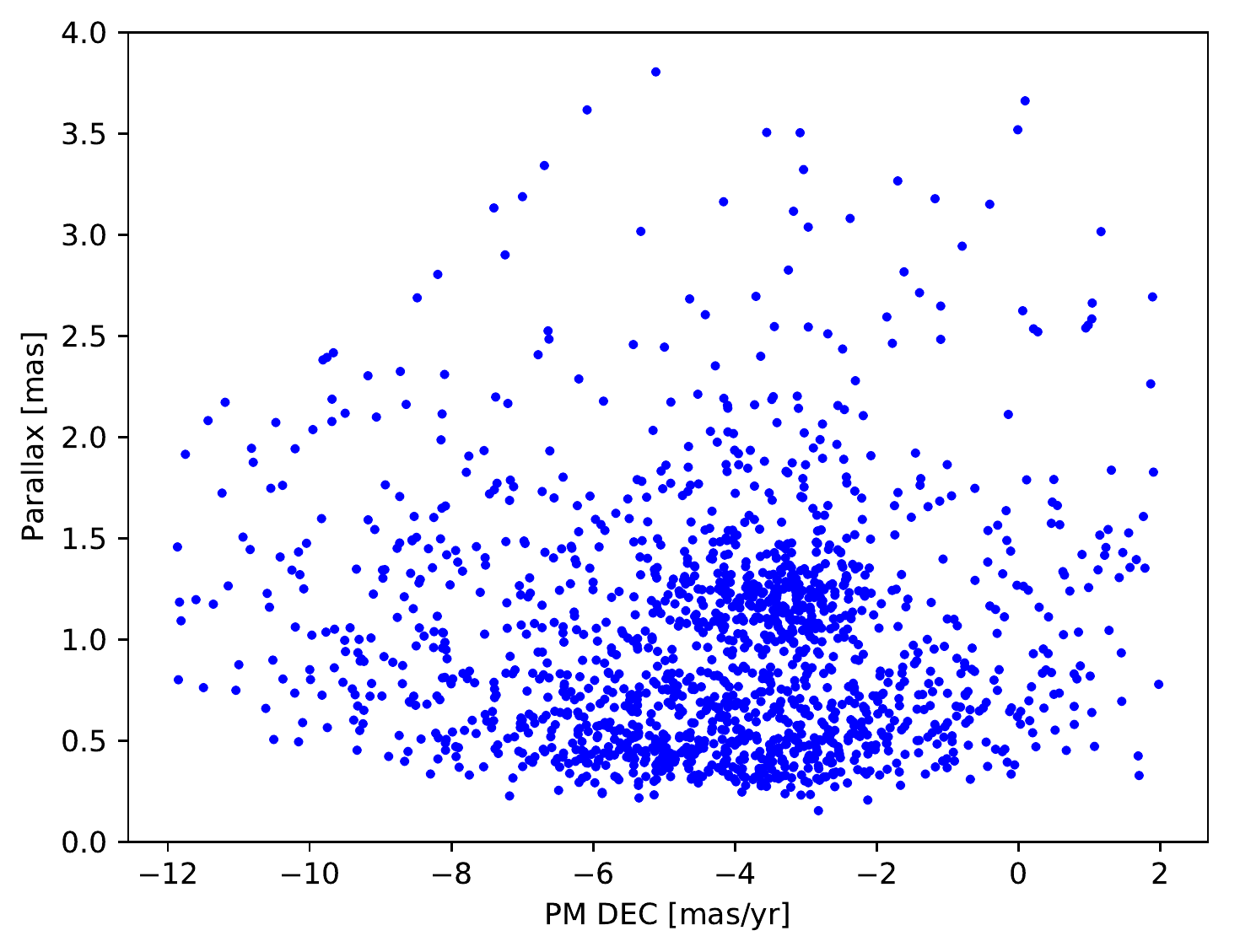}
\caption{\label{gaiafig} Gaia DR2 parallax vs. proper motion in RA (left) and DEC (right) of stars in the IC\,5070 region with a parallax S/N of better than 3. IC\,5070 is identifiable as cluster of points with distinct proper motion and distance.}
\end{figure}

\bsp	
\label{lastpage}
\end{document}